\documentclass[3p,times]{elsarticle}

\usepackage[normalem]{ulem}  
\newcommand\redsout{\bgroup\markoverwith{\textcolor{red}{\rule[0.4ex]{2pt}{1.5pt}}}\ULon}  

\usepackage{url}
\usepackage{amsmath,amsfonts,amssymb,amsthm}
\usepackage{nccmath} 
\usepackage{xfrac}
\usepackage{fancyvrb}
\usepackage{stmaryrd}
\usepackage[inline]{enumitem}
\usepackage{etoolbox}
\usepackage[ruled,vlined,linesnumbered]{algorithm2e}
\SetKwRepeat{Do}{do}{while}
\SetKwComment{tcp}{$\rhd\;$}{}%
\makeatletter
\patchcmd{\@algocf@start}
{-1.5em}
{0pt}
{}{}
\makeatother

\usepackage{pifont}
\newcommand{\cmark}{\textcolor{ForestGreen}{\ding{51}}}%
\newcommand{\xmark}{\textcolor{Maroon}{\ding{55}}}%
\newcommand{\namark}{\textcolor{Maroon}{\textbf{?}}}%

\usepackage[font=small,labelfont=bf]{caption}
\usepackage{comment}
\usepackage[long]{optidef}
\usepackage{algpseudocode}
\usepackage{float}
\usepackage{graphicx}
\usepackage{adjustbox}
\usepackage{booktabs}
\usepackage[usenames,dvipsnames]{xcolor}

\usepackage{makecell}
\usepackage{tikz}
\usetikzlibrary{calc,arrows,shapes,snakes,automata,backgrounds,petri,positioning}
\tikzset{mynode/.style={draw,solid,circle,inner sep=1pt}}

\usepackage{nicematrix}
\NiceMatrixOptions{cell-space-top-limit = 1pt, cell-space-bottom-limit = 1pt}

\AtBeginEnvironment{pmatrix}{\setlength{\arraycolsep}{4pt}}

\usepackage{array}
\usepackage{multirow}
\usepackage{bm}
\usepackage{marvosym}
\usepackage{fontawesome}

\usepackage{graphicx}
\graphicspath{{./figures/}}
\DeclareGraphicsExtensions{.pdf,.eps}
\usepackage{wrapfig}
\usepackage{subfig}

\usepackage{appendix}

\usepackage{scalerel} 

\patchcmd{\paragraph}{\itshape}{\bfseries\boldmath}{}{} 

\patchcmd\subequations
{\theparentequation\alph{equation}}
{\subequationsformat}
{}{}

\makeatletter
\def\thanks#1{\protected@xdef\@thanks{\@thanks
		\protect\footnotetext{#1}}}
\makeatother

\usepackage{hyperref}
\hypersetup{
	colorlinks=true
}
\newcommand{\subequationsformat}{\theparentequation.\arabic{equation}}

\include{def}

\journal{Information and Computation}

\begin{document}
	
\hypersetup{
	linkcolor=RedViolet,
	anchorcolor=black,
	citecolor=blue,
	urlcolor=blue
}

\begin{frontmatter}

\tnotetext[mytitlenote]{
		This work has been partially funded by the NSFC under grant No.~62192732, 61625206, 61732001, 61872341, and 61836005, by the ERC Advanced Project FRAPPANT under grant No.~787914, by the European Union's Horizon 2020 research and innovation programme under the Marie Sk\l{}odowska-Curie grant agreement No.~101008233, and by the CAS Pioneer Hundred Talents Program.}
\title{Encoding inductive invariants as barrier certificates:\\ synthesis via difference-of-convex programming\tnoteref{mytitlenote}}
%



\author[ISCAS,UCAS]{Qiuye Wang}
\ead{wangqye@ios.ac.cn}

\author[RWTH]{Mingshuai Chen\corref{correspondingauthor}}
\ead{chenms@cs.rwth-aachen.de}

\author[ISCAS,UCAS]{Bai Xue}
\ead{xuebai@ios.ac.cn}

\author[ISCAS,UCAS,ISCAS2]{Naijun Zhan\corref{correspondingauthor}}
\cortext[correspondingauthor]{Corresponding authors}
\ead{znj@ios.ac.cn}

\author[RWTH]{Joost-Pieter~Katoen}
\ead{katoen@cs.rwth-aachen.de}

\address[ISCAS]{State Key Laboratory of Computer Science, Institute of Software, CAS, Beijing, China}
\address[UCAS]{University of Chinese Academy of Sciences, Beijing, China}
\address[RWTH]{RWTH Aachen University, Aachen, Germany}
\address[ISCAS2]{Science and Technology on Integrated Information System Laboratory, Institute of Software, CAS, Beijing, China}

%
%

\begin{abstract}
    A barrier certificate often serves as an inductive invariant 
    that isolates an unsafe region from the reachable set of states,
    and hence is widely used in proving safety of hybrid systems
    possibly over an infinite time horizon. 
    We present a novel condition on barrier certificates, 
    termed the \emph{invariant barrier-certificate condition}, 
    that witnesses unbounded-time safety of differential dynamical systems. 
    The proposed condition is 
	the weakest possible one to attain inductive invariance. 
    We show that discharging the invariant barrier-certificate condition 
    ---thereby synthesizing invariant barrier certificates--- 
    can be encoded as solving an \emph{optimization problem subject to bilinear matrix inequalities} (BMIs).
    We further propose a synthesis algorithm based on difference-of-convex programming, 
    which approaches a local optimum of the BMI problem 
    via solving 
    \emph{a series of convex optimization problems}.
    This algorithm is incorporated in a branch-and-bound framework 
    that searches for the global optimum in a divide-and-conquer fashion. 
    We present a weak completeness result of our method, 
    namely, a barrier certificate is guaranteed to be found (under some mild assumptions) 
    whenever there exists an inductive invariant (in the form of a given template) 
    that suffices to certify safety of the system. 
    Experimental results on benchmarks demonstrate the effectiveness and efficiency of our approach. 
\end{abstract}

\begin{keyword} 
    Barrier certificates \sep 
    Inductive invariants \sep 
    Bilinear matrix inequalities \sep 
    Difference-of-convex programming \sep 
    Semidefinite programming
\end{keyword}

\end{frontmatter}

\section{Introduction}\label{sec:introduction}

Hybrid systems are mathematical models that capture the interaction 
between continuous physical dynamics and discrete switching behaviors, 
and hence are widely used in modelling cyber-physical systems (CPS). 
These CPS may be complex and safety-critical, 
with sensitive variables of the environment in its sphere of control. 
Everyday examples include process control at all scales, 
ranging from household appliances to nuclear power plants, 
or embedded systems in transportation domain, 
such as autonomous driving maneuvers in automotive, 
aircraft collision-avoidance protocols in avionics, 
or automatic train control applications, 
as well as a broad range of devices in health technologies, such as cardiac pacemakers.

The safety-critical feature of these CPS, with increasingly complex behaviors, has initiated automatic safety 
or, dually, reachability verification of hybrid systems~\cite{ACH95,DBLP:journals/siglog/Fraenzle19}. 
The problem of reachability verification is undecidable in general~\cite{ACH95}, 
albeit with decidable families of sub-classes 
(see, e.g.,~\cite{LPY01,DBLP:conf/hybrid/AnaiW01,Gan15,DBLP:conf/eucc/GanCLXZ16,Gan18}) 
identified in the literature. 
The hard core of the verification problem lies in 
reasoning about the continuous dynamics, 
which are often characterized by ordinary differential equations (ODEs). 
In particular, when nonlinearity arises in the ODEs, 
the explicit computation of the exact reachable set is usually intractable 
even for purely continuous dynamics~\cite{WDSmith06}.

Therefore in the literature, a plethora of approximation schemes, 
as surveyed in~\cite{DBLP:journals/siglog/Fraenzle19}, 
for reachability analysis of hybrid systems has been developed, 
including an invariant-style reasoning scheme known as \emph{barrier certificate}~\cite{Prajna04}. 
A barrier certificate often serves as an inductive invariant that isolates an unsafe region from the reachable set, 
thereby witnessing safety of hybrid (polynomial) systems possibly over an infinite time horizon. 
A common way to synthesize barrier certificates is 
to reduce the condition defining barrier certificates 
to a numerical optimization or constraint solving problem. 
There is, however, a trade-off between the expressiveness of the barrier-certificate condition 
and the efficiency in discharging the reduced constraints. 
Hence, to enable efficient algorithmic synthesis of barrier certificates 
via, e.g., linear programming (LP), second-order cone programming (SOCP), semidefinite programming (SDP) 
and interval analysis~\cite{djaballah2017construction,DBLP:conf/cav/KongSG18}, 
the general condition on inductive invariance 
(that a barrier certificate defines an invariant, see~\cite{Platzer18FM,Gan17}) 
has been strengthened into a spectrum of different shapes, 
e.g., \cite{Kong13, yang2015exact, zeng2016darboux, Gan17, Platzer18FM}. 
It has been, nevertheless, a long-standing challenge 
\emph{to find a barrier-certificate condition that is as weak as possible 
while admitting efficient synthesis algorithms}.

In this paper, we present a new condition on barrier certificates, 
termed the \emph{invariant barrier-certificate condition}, 
based on the sufficient and necessary condition on being an inductive invariant~\cite{LZZ11}. 
Our invariant barrier-certificate condition is 
the weakest possible condition on barrier certificates to attain inductive invariance. 
We show, by leveraging Putinar's Positivstellensatz~\cite{lasserre2010moments}, 
that discharging the invariant barrier-certificate condition 
---thereby synthesizing invariant barrier certificates--- 
can be encoded as solving an optimization problem subject to 
\emph{bilinear matrix inequalities} (BMIs).
It is known that general BMI problems are NP-hard and non-convex~\cite{toker1995np}. 
Existing solvers for BMI problems, e.g.,~\cite{kocvara2005penbmi, orsi2005lmirank},
are thus considerably less efficient than solvers for (linear) SDP problems.
We show that general bilinear matrix-valued functions 
can be decomposed as a difference 
of two 
convex (matrix-valued) functions 
using 
matrix decomposition,
thus resulting in a synthesis algorithm 
as per 
\emph{difference-of-convex programming} (DCP)~\cite{tao1986algorithms,le2018dc}, 
which solves a series of convex sub-problems (in the form of 
\emph{linear matrix inequalities} (LMIs)) 
that approaches (arbitrarily close to) a local optimum of the BMI problem. 
This algorithm is incorporated in a branch-and-bound framework
that searches for the global optimum in a divide-and-conquer fashion. 
We present a weak completeness result of our method: 
a barrier certificate is guaranteed to be found (under some mild assumptions) 
whenever there exists an inductive invariant (in the form of a given template) 
that suffices to certify the system's safety. 
A similar result on completeness is previously provided only by symbolic approaches, 
yet to the best of our knowledge, 
not by methods based on numerical constraint solving, e.g.,~\cite{yang2015exact, yang2016linear, CAV20BMI}. 
Experiments on a collection of examples suggested that 
our invariant barrier-certificate condition recognizes more barrier certificates than existing conditions, 
and that our DCP-based algorithm is more efficient than directly solving the BMIs via off-the-shelf solvers.


Our main contributions in this paper can be summarized as follows.
\begin{itemize}\itemsep2pt
	\item We present the invariant barrier-certificate condition, which is the weakest possible condition on barrier certificates to attain inductive invariance.
	\item We show that synthesizing invariant barrier certificates can be encoded as solving a BMI optimization problem.
	\item We propose a locally-convergent synthesis algorithm based on difference-of-convex programming.
	\item We present a weak completeness result by augmenting the local algorithm with a branch-and-bound framework.
	\item Experimental results suggested that our condition recognizes more barrier certificates than existing ones, 
	and that our DCP-based algorithm is more efficient than directly solving the BMIs.
\end{itemize}

This article is an extended version of the conference paper~\cite{DBLP:conf/cav/WangCXZK21}. Major extensions include 
\begin{itemize}\itemsep2pt
    \item two alternative matrix decomposition methods (besides eigendecomposition, cf.~Section~\ref{subsec:dc-decomposition}) that better exploit matrix sparsity to accelerate various matrix operations;
    \item a convex relaxation-based method for pruning branches in the branch-and-bound framework (see Algorithm~\ref{alg:bbAlgorithm} and Section~\ref{subsec:upperBound}) to mitigate the effect of exponential blow-up;
  \item complexity analysis of the DCP iterative procedure (cf.~Section~\ref{subsec:complexity}) and potential solutions to circumvent numerical errors in SDP solving (cf.~Section~\ref{subsec:numerical}); and 
  \item generalization to hybrid systems (in Section~\ref{subsec:invBC}), additional experimental results, and all the technical proofs.
\end{itemize}

\paragraph*{\it Paper structure}
The rest of this paper is structured as follows. 
Section~\ref{sec:overview} gives an overview of our approach through a simple example. 
Section~\ref{sec:preliminaries} introduces the necessary mathematical preliminaries. 
Section~\ref{sec:formulation} presents the invariant barrier-certificate condition 
and shows how to encode it as a BMI optimization problem. 
Section~\ref{sec:algorithm} elucidates an algorithm for solving general BMI optimizations via DCP. 
Section~\ref{sec:bbframework} shows how to incorporate the BMI-solving algorithm 
into a branch-and-bound framework to attain weak completeness. 
Section~\ref{sec:experiments} demonstrates our method on a collection of examples. 
After discussing related work in Section~\ref{sec:related}, 
we conclude the paper in Section~\ref{sec:conclusion}. 

\section{A bird's-eye perspective}\label{sec:overview}

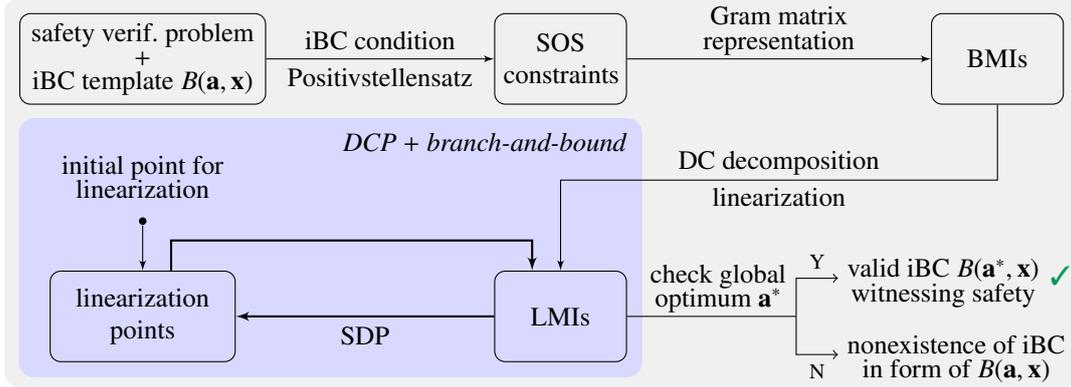
\begin{figure}[h]
	\centering
	\begin{tikzpicture}[
		box/.style args = {#1/#2}{draw, align=flush center, rounded corners, text width=#1, minimum height=#2},
		box/.default = 1.4cm/12mm,
		unframedbox/.style args = {#1/#2}{align=flush center, text width=#1, minimum height=#2},
		unframedbox/.default = 1.4cm/12mm
		]
		\tikzstyle{line} = [draw, -latex']
		
		\begin{scope}
		\node[box=3cm/12mm]          (problem)     {safety verif.~problem\\[-1mm]$+$\\[-1mm]iBC template $B(\aaa,\xx)$};
		\node[box=1.5cm/12mm,right=3cm of problem]          (sos)			       {SOS\\constraints};
		\node[box=1.5cm/12mm,right=4cm of sos]          (bmi)			       {BMIs};
		\node[box=1.5cm/12mm,below=2.2cm of sos]          (lmi)			       {LMIs};
		\node[box=2.2cm/12mm]          (lp)		at  (problem |- lmi)	       {linearization\\points};
		\node[circle,fill,inner sep=1pt,above=.6cm of lp,label=\makecell{initial point for\\[-1mm]linearization}] (ip)  {};
		\coordinate[above =1.2cm of lmi]          (lmighost);
		\coordinate[above right=0cm and -.85cm of lp]          (lpcorner0);
		\coordinate[above right=.4cm and -.85cm of lp]          (lpcorner1) ;
		\coordinate[above left=.4cm and -.5cm of lmi]          (lpcorner2);
		\coordinate[above left=0cm and -.5cm of lmi]          (lpcorner3);
		
		\coordinate[right=3.1cm of $(lmi.north)!0.5!(lmi.south)$] (ghost);
		\node[anchor=east,above right=.07cm and .55cm of ghost]               (Y)	     {\makecell{valid iBC $B(\aaa^*,\xx)$\\[-1mm]witnessing safety\\[-1.2mm]}~\,\cmark};
		\node[anchor=east,below right=.07cm and .55cm of ghost]              (N)   {\makecell{nonexistence of iBC\\[-1mm]in form of $B(\aaa,\xx)$\\[-1.2mm]}};
		
		\path [line] (problem) -- (sos) node [above,pos=0.5] {iBC condition} node [below,pos=0.5] {Positivstellensatz};
		\path [line] (sos) -- (bmi) node [above,pos=0.5] {\makecell{Gram matrix\\[-1mm]representation\\[-1.2mm]}};
		\path [line] (bmi) |- (lmighost) node [above,pos=0.75] {\makecell{DC decomposition\\[-1mm]}} node [below,pos=0.75] {\makecell{\\[-5mm]linearization}} -- (lmi) ;
		\path [line,thick] (lmi) -- (lp) node [below,pos=0.5] {SDP};
		\path [line] (ip) -- (lp);
		\path [line,thick] (lpcorner0) -- (lpcorner1) -- (lpcorner2) -- (lpcorner3);
		
		\draw[-]  (lmi) -- (ghost) node [above,pos=0.54] {\makecell{check global\\[-1mm]optimum $\aaa^*$\\[-1.2mm]}};
		\draw[->] (ghost) |- (Y) node [above,pos=0.75] {\footnotesize Y};
		\draw[->] (ghost) |- (N) node [below,pos=0.75] {\footnotesize N};
		
		\node [circle,white,minimum size = 0mm,above right=1.7cm and .8cm of lmi] (tedge) {};
		\node [circle,white,minimum size = 0mm,above left=1.7cm and -.09cm of lp] (dedge) {};
		
		\node [circle,white,minimum size = 0mm,above left=-1.3cm and .1cm of lp] (llcorner) {};
		\end{scope}
	
		\begin{pgfonlayer}{background}
			\filldraw [line width=4mm,join=round,black!6]
			(bmi.north  -| bmi.east)  rectangle (llcorner.south  -| llcorner.west);
			
			\filldraw [line width=4mm,join=round,blue!15]
			(tedge.east  -| lmi.east)  rectangle (lp.south  -| dedge.west);
			\node[above right=1.45cm and -3.86cm of lmi] (dcbab)  {\emph{DCP $+$ branch-and-bound}};
		\end{pgfonlayer}
	\end{tikzpicture}
	\caption{A sketch of our method for unbounded-time safety verification via invariant barrier certificates (iBC, for short).}\label{fig:method-diagram}
\end{figure}

The diagram in Fig.~\ref{fig:method-diagram} sketches out a bird's-eye view of our method for the unbounded-time safety verification of differential dynamical systems. We use the following example to demonstrate several core steps underneath.

\begin{example}[\expname{overview}~\textnormal{\cite{djaballah2017construction}}]\label{exmp:overview}
Consider the following continuous-time dynamical system modelled by an ordinary differential equation:
\begin{equation*}
	\dot{\xx} \seq
    \begin{pmatrix}
        \dot{x}_1 \\
        \dot{x}_2
    \end{pmatrix} 
    \seq
    \begin{pmatrix}
        x_1 + x_2 \\
        x_1 x_2 - 0.5 x_2^2 + 0.1 
    \end{pmatrix}
	~.
\end{equation*}%
The verification obligation is to show that 
the system trajectory originating from any state in the initial set 
$\init = \{ \xx \mid \initBound(\xx) \leq 0 \}$ with $\initBound(\xx) = x_1^2 + (x_2 - 2)^2 -1$ 
will never enter the unsafe set $\unsafe = \{ \xx \mid \unsafeBound(\xx)\leq 0 \}$ 
with $\unsafeBound(\xx) = x_2 + 1$.
\qedTT
\end{example}

A barrier certificate satisfying our invariant barrier-certificate condition (cf.~Definition~\ref{def:invBc}) 
serves as an inductive invariant that suffices to isolate the unsafe region $\unsafe$ 
from the set of reachable states from $\init$, 
thereby proving safety of the system over an infinite time horizon. 
To this end, we proceed in the following steps.

\paragraph*{\bf 1) Encode as sum-of-squares (SOS) constraints}
We first set a (polynomial) barrier-certificate template, for example, $B(\aaa, \xx) = a x_2$ 
with unknown coefficient $a \in \mathbb{R}$. 
According to Theorem~\ref{thm:invariantCondition}, 
we only need to consider Lie derivatives up to order $\LieBound = 1$, 
i.e., $\mathcal{L}_{\ff}^0 B(\aaa, \xx) = a x_2$ 
and $\mathcal{L}_{\ff}^1 B(\aaa, \xx) = a (x_1 x_2 - 0.5 x_2^2 + 0.1)$.

We show that
$B(\aaa, \xx)$ is an invariant barrier certificate 
if there exists a polynomial $v(\xx)$, SOS polynomials (i.e., polynomials that can be written as a finite sum of squares of polynomials) $\sigma(\xx), \sigma'(\xx)$ 
and a constant $ \epsilon > 0$ such that
\begin{subequations}\label{eqn:expInvCond}
\begin{align}
    &-\underbrace{a x_2}_{B} +\ \sigma(\xx) \underbrace{\left(x_1^2 + (x_2 - 2)^2 - 1\right)}_{\initBound}~, \owntag[expInvCond1]{initial} \\[.1cm]
    &-\underbrace{a \left(x_1 x_2 - 0.5 x_2^2 + 0.1\right)}_{\mathcal{L}_{\ff}^1 B} + \,v(\xx) \underbrace{a x_2}_{\mathcal{L}_{\ff}^0 B}, \owntag[expInvCond2]{Lie consecution} \\[.1cm]
    & \underbrace{a x_2}_{B} +\ \sigma'(\xx) \underbrace{(x_2 + 1)}_{\unsafeBound} - \epsilon \owntag[expInvCond3]{separation}
\end{align}
\end{subequations}%
are SOS polynomials. 

\paragraph*{\bf 2) Reduce to a BMI optimization problem}
Observe that the above SOS constraints can be formulated as BMI constraints (via the Gram matrix representation, as formalized later). 
For instance, let us assume that \eqref{eqn:expInvCond2} is an SOS polynomial of degree at most 2 
and $v(\sss, \xx) = s_0 + s_1 x_1 + s_2 x_2$ is a template polynomial with unknown coefficients $\sss$. 
Then constraint \eqref{eqn:expInvCond2} is equivalent to the BMI constraint 
\begin{equation*}
    \mathcal{F}_2(\aaa, \sss) \seq -
	    \begin{pmatrix}
	        -0.1 a & 0 & 0.5a s_0 \\
	        0 & 0 & 0.5(a s_1 - a) \\
	        0.5a s_0 & 0.5(a s_1 - a) & a s_2 + 0.5a 
	    \end{pmatrix}
    \spreceq 0
\end{equation*}%
meaning that the bilinear matrix (the LHS of $\preceq$) is negative semidefinite. 
Note that the bilinearity arises due to the coupling of the unknown coefficients $\aaa$ and $\sss$.

Constraints \eqref{eqn:expInvCond1} and \eqref{eqn:expInvCond3} can be reduced 
to BMI constraints in an analogous way\footnote{
Despite that no bilinearity is involved in constraints \eqref{eqn:expInvCond1} and \eqref{eqn:expInvCond3}, 
they can be processed in the same way as \eqref{eqn:expInvCond2}, yielding LMI constraints.}, 
yielding $\mathcal{F}_1$ and $\mathcal{F}_3$. 
It then follows that, to solve the SOS constraints, 
we need to find a feasible solution $(\aaa, \sss)$ such that\footnote{
Extra constraints on $\sigma(\xx)$ and $\sigma'(\xx)$ being SOS polynomials 
can be encoded analogously in the feasibility problem, yet are omitted here for the sake of simplicity.}
\begin{equation}\label{eqn:bmiExpFeasible}
    \mathcal{F}_1(\aaa, \sss) \,\preceq\, 0 
    \sland \mathcal{F}_2(\aaa, \sss) \,\preceq\, 0 
    \sland \mathcal{F}_3(\aaa, \sss) \,\preceq\, 0~.
\end{equation}%

To exploit well-developed optimization techniques, 
the feasibility problem \eqref{eqn:bmiExpFeasible} is transformed 
to an optimization problem subject to BMI constraints:
\begin{maxi}
    {\lambda, \aaa, \sss}
    {\lambda}
    {\label{eqn:bmiExp}}
    {}
    \addConstraint{\mathcal{B}_i(\lambda, \aaa, \sss) \sdefine \mathcal{F}_i(\aaa, \sss) + \lambda I}{\spreceq 0,\quad}{i=1, 2, 3}
\end{maxi}%
where $I$ is the identity matrix with compatible dimensions. 
Note that problem \eqref{eqn:bmiExpFeasible} has a feasible solution 
if and only if the optimal value $\lambda^*$ in \eqref{eqn:bmiExp} is non-negative.

\paragraph*{\bf 3) Decompose as difference-of-convex problems}
The problem \eqref{eqn:bmiExp} contains non-convex constraints 
and hence does not admit efficient (polynomial-time) algorithms tailored for convex optimizations. 
However, using our DCP-based technique, 
a non-convex function $\mathcal{B}_i(\lambda, \aaa, \sss)$ can be decomposed 
as the difference of two (positive semidefinite) convex matrix-valued functions: 
\begin{equation}\label{eq:diff-of-psd-convex}
    \mathcal{B}_i(\lambda, \aaa, \sss) \seq
    \mathcal{B}_i^+(\lambda, \aaa, \sss) - \mathcal{B}_i^-(\lambda, \aaa, \sss)~.
\end{equation}%
The decomposition of $\mathcal{B}_2(\lambda, \aaa, \sss)$ (via eigendecomposition), 
for instance, gives
\begin{align*}
    \mathcal{B}_2^+(\lambda, \aaa, \sss) &= 
    \frac{1}{8}\,\begin{pNiceMatrix}
        8 \lambda + 0.08 a + a^2 + 0.408 s_0^2 & 
        0.408 s_0 s_1 & 
        -2 a s_0 + 0.816 s_0 s_2 \\
        0.408 s_0 s_1 & 
        8 \lambda + a^2 + 0.408 s_1^2 & 
        4 a - 2 a s_1 + 0.816 s_1 s_2 \\
        -2 a s_0 + 0.816 s_0 s_2 & 
        4 a - 2 a s_1 + 0.816 s_1 s_2 & 
        8 \lambda - 4 a + 2.449 a^2 - 4 a s_2 + s_0^2 + s_1^2 + 1.632 s_2^2 
    \end{pNiceMatrix} \\
	\mathcal{B}_2^-(\lambda, \aaa, \sss) &= 
    \frac{1}{8}\,\begin{pNiceMatrix}
        a^2 + 0.408 s_0^2 & 
        0.408 s_0 s_1 & 
        2 a s_0 + 0.816 s_0 s_2 \\
        0.408 s_0 s_1 & 
        a^2 + 0.408 s_1^2 & 
        2 a s_1 + 0.816 s_1 s_2 \\
        2 a s_0 + 0.816 s_0 s_2 & 
        2 a s_1 + 0.816 s_1 s_2 & 
        2.449 a^2 + 4 a s_2 + s_0^2 + s_1^2 + 1.632 s_2^2 
    \end{pNiceMatrix} 
	~.
\end{align*}%

\paragraph*{\bf 4) Solve a series of convex sub-problems}
Now, we apply a standard iterative procedure 
in difference-of-convex programming~\cite{dinh2011combining} as follows. 
Given a feasible solution $\zz^k = (\lambda^k, \aaa^k, \sss^k)$ 
to the BMI optimization problem \eqref{eqn:bmiExp},
the concave part $-\mathcal{B}_i^-(\lambda, \aaa, \sss)$ in \eqref{eq:diff-of-psd-convex} 
is linearized around $\zz^k$, thus yielding a series of convex programs ($k = 0, 1, \ldots$): 
\begin{maxi}
    {\lambda, \aaa, \sss}
    {\lambda}
    {\label{eqn:bmiExpLinearized}}
    {}
    \addConstraint{\mathcal{B}_i^+(\zz) - \mathcal{B}_i^-\left(\zz^k\right) - \mathcal{DB}_i^-\left(\zz^k\right)\left(\zz -\zz^k\right)}{\spreceq 0,\quad}{i=1, 2, 3}
\end{maxi}%
where $\mathcal{DB}_i^-(\zz^k)(\cdot)$ denotes the derivative of the matrix-valued function $\mathcal{B}_i^-(\cdot)$ at $\zz^k$. 

The soundness of our approach 
asserts that \emph{the feasible set of the linearized program \eqref{eqn:bmiExpLinearized} 
under-approximates the feasible set of the original BMI program \eqref{eqn:bmiExp}}. 
Therefore, if $\lambda^k \geq 0$ after iteration $k$, 
we can safely claim that $(\aaa^k, \sss^k)$ is a feasible solution to \eqref{eqn:bmiExpFeasible}. 
A barrier certificate $B(\xx)$ is then obtained by substituting $\aaa^k$ in $B(\aaa, \xx)$. 
Moreover, if we take the optimum $\zz^{*, k}$ of \eqref{eqn:bmiExpLinearized} 
to be the next linearization point $\zz^{k+1}$, 
the solution sequence $\{ \zz^k \}_{k \in \NN}$ converges to a local optimum of \eqref{eqn:bmiExp}. 

\begin{wrapfigure}{r}{0.43\textwidth}
	\vspace*{-8mm}
	\begin{center}
		\includegraphics[width=0.43\textwidth]{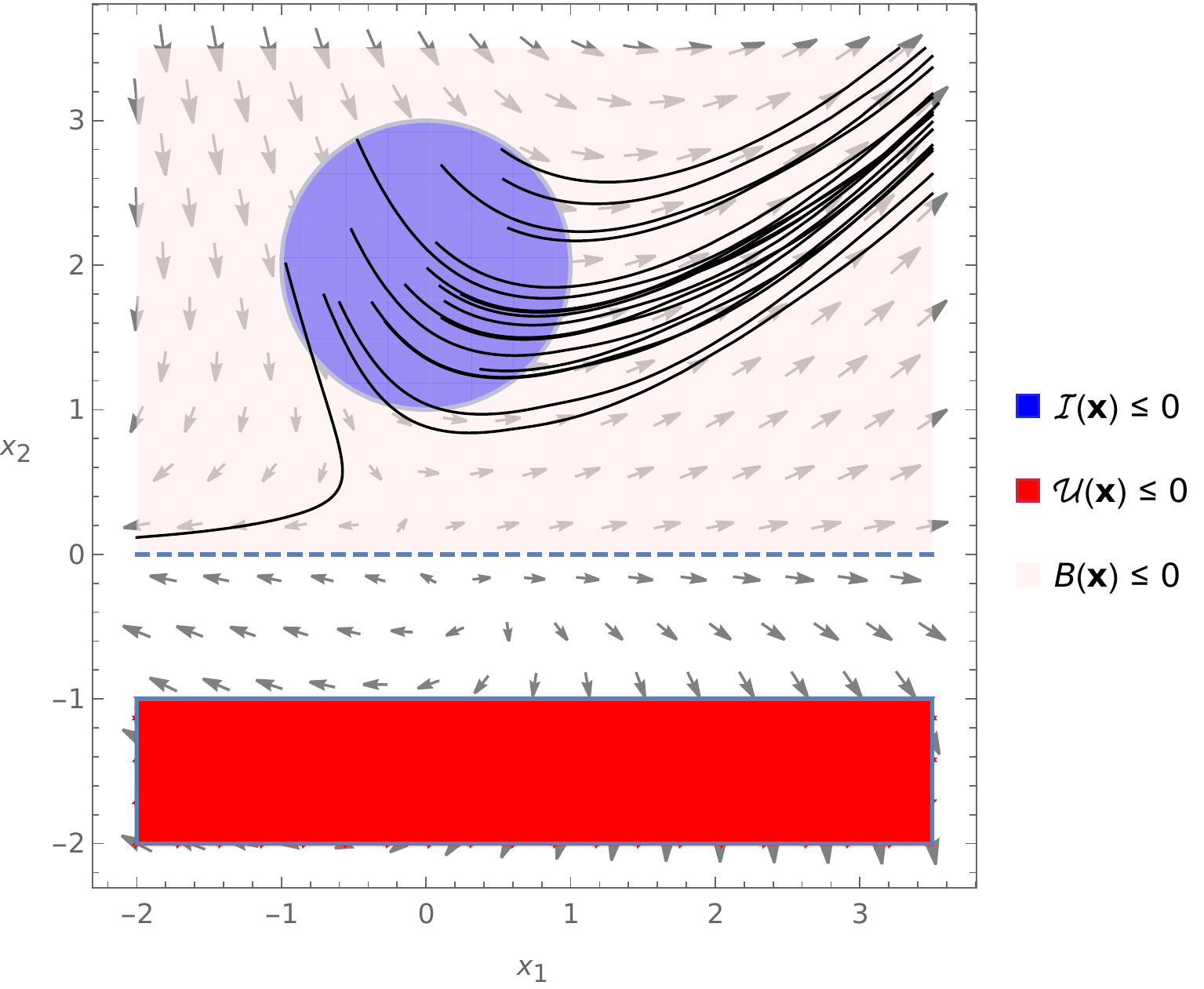}
	\end{center}
	\vspace*{-6mm}
	\caption{Phase portrait of the system in Example \ref{exmp:overview}. The arrows indicate the vector field and the solid curves are randomly sampled trajectories.}
	\label{fig:overview}
\end{wrapfigure}

We show that the linearized program \eqref{eqn:bmiExpLinearized} 
is equivalent to an LMI optimization problem 
admitting polynomial-time algorithms \cite{boyd1994linear}, 
say the well-known \emph{interior-point methods} supported by most off-the-shelf SDP solvers.
Our iterative procedure starts with a strictly feasible initial solution $\zz^0$ 
to program \eqref{eqn:bmiExp} and terminates after iteration $k=2$
with $\lambda^2 \ge 0$ (subject to numerical round-off) and $a^2 = -0.00363421$, 
yielding the barrier certificate
{
	\setlength{\abovedisplayskip}{8pt}
	\setlength{\belowdisplayskip}{8pt}
\[
    B(\aaa^2, \xx) \seq -0.00363421 x_2 \sleq 0~.
\]
}%
Fig.~\ref{fig:overview} depicts the system dynamics and the synthesized barrier certificate.


We remark that the aforementioned iterative procedure on solving a series of convex optimizations 
converges only to a local optimum of the BMI problem \eqref{eqn:bmiExp}. 
This means that, in some cases, it may miss the global optimum that induces a non-negative $\lambda^*$. 
We will present in Section~\ref{sec:bbframework} a solution to this problem 
by incorporating our iterative procedure into a branch-and-bound framework 
that searches for the global optimum in a divide-and-conquer fashion. 
%
%
%

\section{Mathematical foundations}\label{sec:preliminaries}

\paragraph*{\bf Notations}
Let $\NN$, $\NN^+$, $\mathbb{R}$, $\RR^+$ and $\RR^+_0$ be respectively the set of natural, positive natural, real, positive real and non-negative real numbers. For a vector $\xx \in \RR^n$, $x_i$ refers to its $i$-th component and $\normsqrt{\xx}$ denotes the $\ell^2$-norm; we write $\diag(\xx) \in \mathbb{R}^{n \times n}$ for a diagonal matrix with $x_i$ being the $i$-th diagonal element. For a matrix $A \in \mathbb{R}^{n \times m}$, $A(i, j)$ refers to its $(i, j)$-th element; for a square matrix $A \in \mathbb{R}^{n \times n}$, its trace is $\trace(A)=\sum^n_{i=1} A(i, i)$. Given two matrices $A \in \mathbb{R}^{a \times b}$ and $B \in \mathbb{R}^{c \times d}$, their \emph{Kronecker product} is $A \otimes B \define [A(1,1) B, \ldots, A(1,b) B; \text{\rotatebox[origin=c]{-15}{$\cdots$}}; A(a,1) B, \ldots, A(a,b) B]  \in \mathbb{R}^{a c \times b d}$. $\mathcal{S}^n$ denotes the space of $n \times n$ real, symmetric matrices. For $A \in \mathcal{S}^n$, $A \succeq 0$ means that $A$ is \emph{positive semidefinite} (\psd, for short),
i.e., $\forall \xx \in \mathbb{R}^n\colon \xx^\trans\! A \xx \ge 0$. More generally, for $A, B \in \mathcal{S}^n$, $A \preceq B$ indicates that $B - A$ is positive semidefinite.
A matrix-valued function $\mathcal{B}\colon \mathbb{R}^n \to \mathcal{S}^m$ is \emph{\psd-convex} on a convex set $\compactSet \subseteq \mathbb{R}^n$ if 
$
	\forall \xx_1, \xx_2 \in \compactSet \ldotp \forall \mu \in (0, 1)\colon
	\mathcal{B}(\mu \xx_1 + (1 - \mu) \xx_2) 
	\preceq \mu \mathcal{B}(\xx_1) + (1 - \mu) \mathcal{B}(\xx_2)
$.


\paragraph*{\bf SOS, LMIs, and BMIs}
Let $\mathbb{R}[\xx]$ be the polynomial ring in $\xx$ over the field $\mathbb{R}$. A polynomial $h \in \mathbb{R}[\xx]$ is \emph{sum-of-squares} (SOS) iff there exist polynomials $g_1, \ldots, g_k \in \mathbb{R}[\xx]$ such that $h = \sum_{i=1}^{k} g_i^2$. We denote by $\Sigma[\xx] \subset \mathbb{R}[\xx]$ the set of SOS polynomials over $\xx$.	A \emph{linear matrix inequality} (LMI) is a constraint of the form $\mathcal{L}(\xx) \define F + \sum_{i=1}^{m} x_i H_i \preceq 0$,
where $\xx \in \RR^m$ is a vector of variables and $F, H_i \in \mathcal{S}^p$ are constant symmetric matrices. 
LMIs are convex and hence admit polynomial-time algorithms to find feasible solutions (or prove the infeasibility) given the desired precision \cite{boyd1994linear}. 
A \emph{bilinear matrix inequality} (BMI) is a constraint of the form $\mathcal{B}(\xx, \yy) \define F + \sum_{i=1}^{m} x_i H_i + \sum_{j=1}^{n} y_j G_j + \sum_{i=1}^{m} \sum_{j=1}^{n} x_i y_j F_{i, j} \preceq 0$,
where $\xx, \yy \in \RR^m$ are vectors of variables and $F, H_i, G_j, F_{i,j} \in \mathcal{S}^p$ are constant symmetric matrices. Solving general BMIs is NP-hard due to the non-convex nature of the constraints \cite{toker1995np}.

\paragraph*{\bf Differential dynamical systems}
We consider a class of continuous dynamical systems modelled by ordinary differential equations of the autonomous type:
\begin{equation}
    \label{eqn:dynamics}
    \dot{\xx} \seq \ff(\xx)
\end{equation}%
where $\xx \in \RR^n$ is the \emph{state} vector, $\dot{\xx}$ denotes its temporal derivative ${\rm d}\xx/{\rm d}t$, with $t \in \RR^+_0$ modelling time, and $\ff\colon \RR^n \to \RR^n$ is a polynomial \emph{flow field} (or \emph{vector field}) that governs the evolution of the system. A polynomial vector field is local Lipschitz, and hence for some $T \in \RR^+ \cup \{\infty\}$, there exists a unique \emph{solution} (or \emph{trajectory}) $\sol_{\xx_0}\colon [0,T) \to \RR^n$ originating from any initial state $\xx_0 \in \RR^n$ such that (1) $\sol_{\xx_0}(0) = \xx_0$, and (2) $\forall \tau \in [0,T)\colon \frac{{\rm d}\sol_{\xx_0}}{{\rm d}t}\big\vert_{t=\tau} \!= \ff(\sol_{\xx_0}(\tau))$. We assume in the sequel that $T$ is the maximal instant up to which $\sol_{\xx_0}$ exists for all $\xx_0$.

\begin{remark}
Our techniques on synthesizing barrier certificates in this paper focus on differential dynamics of the form \eqref{eqn:dynamics}. However, we will show that there is no substantial difficulty in extending the results to multi-mode hybrid systems where extra constraints on the system evolution, e.g., guards, are present.
\end{remark}

\paragraph*{\bf Safety verification problem}
Given a domain set $\domain \subseteq \RR^n$ and an initial set $\init \subseteq \domain$, the \emph{reachable set} of a dynamical system of the form \eqref{eqn:dynamics} at time instant $t \in [0,T)$ is defined as $\reach_{\init}(t) \define \{\sol_{\xx_0}(t) \mid \xx_0 \in \init\}$.
We denote by $\reach_{\init}$ the aggregated reachable set, i.e., the union of $\reach_{\init}(t)$ over $t \in [0,T)$.
Given an unsafe set $\unsafe \subseteq \domain$, the system is said to be \emph{safe} iff $\reach_{\init} \cap \unsafe = \emptyset$, and \emph{unsafe} otherwise. For simplicity, we consider $\domain = \RR^n$ unless explicitly stated otherwise.

To avoid the explicit computation of the exact reachable set, which is usually intractable for nonlinear hybrid systems (cf., e.g., \cite{DBLP:journals/siglog/Fraenzle19}), barrier-certificate methods make use of a partial differential operator, termed the \emph{Lie derivative}, to capture the evolution of a barrier function along the vector field:
\begin{definition}[Lie derivative~\textnormal{\cite{kolar1993natural}}]\label{def:Lie-Der}
	Given a vector field $\ff\colon \RR^n \to \RR^n$ over $\xx$, the \emph{Lie derivative} of a polynomial $B \in \mathbb{R}[\xx]$ along $\ff$, $\mathcal{L}_{\ff}^k B\colon \RR^n \to \RR$ of order $k \in \NN$, is
	\begin{equation*}
		\mathcal{L}_{\ff}^k B(\xx) \sdefine
		\left\{
		\begin{array}{ll}
			B(\xx), \quad k=0~,\\
			\left\langle\frac{\partial}{\partial \xx} \mathcal{L}_{\ff}^{k-1} B(\xx), \ff(\xx)\right\rangle, \quad k>0
		\end{array}
		\right.
	\end{equation*}%
	where $\langle\cdot,\cdot\rangle$ is the inner product of vectors, i.e., $\langle \uu, \vv\rangle \define \sum_{i=1}^{n}u_i v_i$ for $\uu, \vv \in \RR^n$.
\end{definition}

The Lie derivative $\mathcal{L}_{\ff}^k B(\xx)$ is essentially the $k$-th temporal derivative of the (barrier) function $B(\xx)$, and thus captures the change of $B(\xx)$ over time. In fact, given a polynomial vector field, one can use (high-order) Lie derivatives to identify the tendency of its trajectories in terms of a polynomial function $B(\xx)$, as exemplified in Appendix~\ref{appendix_lie}. 

An \emph{inductive invariant} $\invt \subseteq \mathbb{R}^n$ of a dynamical system is a set of states such that all the trajectories starting from within $\invt$ remain in $\invt$:
\begin{definition}[Inductive invariant~\textnormal{\cite{PC08}}]\label{def:inv}
    Given a system \eqref{eqn:dynamics}, a set $\invt \subseteq \mathbb{R}^n$ is an \emph{inductive invariant} of system \eqref{eqn:dynamics} if and only if
    \begin{equation}
        \forall \xx_0 \in \invt \ldotp\, \forall t \in [0, T)\colon\ \sol_{\xx_0}(t) \in \invt~.
    \end{equation}%
\end{definition}

In the sequel, we refer to inductive invariants simply as invariants. In~\cite{LZZ11}, a sufficient and necessary condition on being a polynomial invariant is proposed:
\begin{theorem}[Invariant condition~\textnormal{\cite{LZZ11}}]
    \label{thm:invariantCondition}
    Given a polynomial $B \in \mathbb{R}[\xx]$, its \emph{zero sub-level set} $\{ \xx \mid B(\xx) \leq 0 \}$ is an invariant 
    of system \eqref{eqn:dynamics} if and only if\,\footnote{In \eqref{eqn:invariantCondition}, $\bigwedge_{j = 0}^{i-1} \mathcal{L}_{\ff}^j B = 0$ is $\true$ for $i = 0$ by default. This applies in the sequel. Moreover, the sub-level set of $B$ can be non-zero in general.}
    \begin{equation}
        \label{eqn:invariantCondition}
        B \,\leq\, 0 \simplies 
        \bigvee\nolimits_{i = 0}^{\LieBound} \left(
        \left(\bigwedge\nolimits_{j = 0}^{i-1} \mathcal{L}_{\ff}^j B \,=\, 0\right) 
        \,\land\, \mathcal{L}_{\ff}^i B \,<\, 0\right)
        \,\lor\, \bigwedge\nolimits_{i = 0}^{\LieBound} \mathcal{L}_{\ff}^i B \,=\, 0
    \end{equation}%
    where $\LieBound \in \NN^+$ is the completeness threshold, i.e., a positive integer that bounds the order of Lie derivatives.
\end{theorem}

\begin{remark}
	$\LieBound$ is the minimal index $i$ such that $\mathcal{L}_{\ff}^{i+1} B$ is in the polynomial ideal generated by $\mathcal{L}_{\ff}^0 B, \ldots, \mathcal{L}_{\ff}^i B$. The ideal membership can be decided by computing the Gr\"{o}bner basis of this ideal~\cite{LZZ11}. The complexity of computing $\LieBound$ will be discussed in the complexity analysis of our approach (see Section~\ref{subsec:complexity}).
\end{remark}

In contrast, a \emph{barrier certificate} is a function whose zero sub-level set isolates an unsafe region $\unsafe$ from the reachable set $\reach_{\init}$ w.r.t.~some initial set $\init$ (the sub-level set can be non-zero in general):
\begin{definition}[Semantic barrier certificate~\textnormal{\cite{Platzer18FM}}]
\label{def:semanticBC}
Given a system \eqref{eqn:dynamics}, an initial set $\init$ and an unsafe set $\unsafe$, a \emph{barrier certificate} of \eqref{eqn:dynamics} is a differentiable function $B\colon \RR^n \to \RR$ satisfying
\begin{equation}
\label{eqn:semanticBc}
    \forall \xx_0 \in \init \ldotp \, \forall t \in [0,T)\colon\ B\left(\sol_{\xx_0}(t)\right) \sleq 0 \quad \text{and} \quad 
    \forall \xx \in \unsafe\colon\ B(\xx) \sgt 0~.
\end{equation}%
\end{definition}
The existence of such a barrier certificate trivially implies safety of the system. Moreover, one may readily verify that if some set $\invt = \{ \xx \mid B(\xx) \leq 0 \}$ is an invariant and satisfies $(\init \subseteq \invt) \land (\invt \cap \unsafe = \emptyset)$, then $B(\xx)$ is a barrier certificate. 

As observed in \cite{Platzer18FM}, however, the semantic statement in Definition \ref{def:semanticBC} encodes merely the general \emph{principle of barrier certificates}~\cite{Gan17}, yet in itself is not that useful for safety verification because it explicitly involves the system solutions. Therefore, in order to enable efficient synthesis, the semantic condition on barrier certificates has been strengthened into a handful of different shapes (see, e.g., \cite{Prajna04, Kong13, yang2015exact, Gan17}) which all imply inductive invariance\footnote{\label{ftn-t-BC}An exception is known as the $t$-barrier certificate condition~\cite{DBLP:conf/adhs/Bak18},
which is a continuous analogy to $k$-induction, thus more general than (classical) inductive invariance. However, this condition also explicitly involves the system solutions, and hence does not admit efficient synthesis.}.
It has been yet a long-standing challenge \emph{to find a barrier-certificate condition that is as weak as possible while admitting efficient synthesis algorithms}.


Our BMI encoding of the invariant barrier-certificate condition 
roots in Putinar's Positivstellensatz, which characterizes positivity of polynomials on a semi-algebraic set defined by a system of polynomial inequalities:

\begin{theorem}[Putinar's Positivstellensatz~\textnormal{\cite{lasserre2010moments}}]
	Let $\mathcal{K} = \{ \xx \mid \bigwedge_{i=1}^{m} g_i(\xx) \geq 0\}$ be a compact semi-algebraic set defined by $g_1, \ldots, g_m \in \RR[\xx]$.
        Assume the \emph{Archimedean condition} holds,
        i.e., there exists $L \in \RR^+$ such that $L - \norm{\xx} =\eta_0(\xx) + \sum_{i=1}^{m} \eta_i(\xx) g_i(\xx)$ for some $\eta_0, \ldots, \eta_m \in \Sigma[\xx]$.
	If $h \in \RR[\xx]$ is strictly positive on $\mathcal{K}$, then
	\begin{equation*}
		h(\xx) \seq \sigma_0(\xx) + \sum\nolimits_{i=1}^m \sigma_i(\xx) g_i(\xx)
	\end{equation*}%
	holds for some SOS polynomials $\sigma_0, \ldots, \sigma_m \in \Sigma[\xx]$.
\end{theorem}%
\begin{remark}
The Archimedean condition can be met by adding a (redundant) constraint $g_{m+1}(\xx) = L_0 - \norm{\xx} \leq 0$, provided that a bound $L_0 \in \RR^+$ is known such that $\forall \xx \in \mathcal{K}\colon L_0 - \norm{\xx} \geq 0$. See~\cite[Chapter~2]{lasserre2010moments} for more details on the Archimedean condition.
\end{remark}

We now recall a key technique used in our reduction to semidefinite optimizations. Given a symmetric matrix $X \in \mathcal{S}^n$ partitioned as
$X = 
\begin{pmatrix}
	A & C \\
	C^\trans & D
\end{pmatrix}$
with invertible $A$, the \emph{Schur complement} of $A$ in $X$ is defined as
$X / A \define D - C^\trans A^{-1} C$.
An important property of the Schur complement $X / A$ is that it characterizes the positive semidefiniteness of the block matrix $X$ (which will be used later to transform nonlinear convex constraints into linear constraints):
\begin{theorem}[Schur complement~\textnormal{\cite{boyd2004convex}}]
    \label{thm:schurComplement}
    If $A \succ 0$, then $X \succeq 0$ iff $X / A \succeq 0$. 
\end{theorem}%

\section{Invariant barrier-certificate condition as BMIs}
\label{sec:formulation}
In this section, we present our \emph{invariant barrier-certificate condition} 
based on the necessary and sufficient condition on being an inductive invariant (cf.~Theorem~\ref{thm:invariantCondition}), and show how to encode it as BMI constraints.

\subsection{Invariant barrier-certificate condition}\label{subsec:invBC}

\begin{definition}[Invariant barrier certificate]
    \label{def:invBc}
    Given a system \eqref{eqn:dynamics}, an initial set $\init$ and an unsafe set $\unsafe$, a polynomial function $B\colon \RR^n \to \RR$ is an \emph{invariant barrier certificate} of system \eqref{eqn:dynamics} if and only if
    \begin{enumerate}
        \item (initial): $\forall \xx \in \init\colon\ B(\xx) \sleq 0$~;
        \item (consecution): $\forall \xx \in \RR^n\colon\ 
                \bigwedge_{i = 1}^{\LieBound} \left(
                \left(\bigwedge_{j = 0}^{i-1} \mathcal{L}_{\ff}^j B(\xx) \,=\, 0\right)
                \,\implies\, \mathcal{L}_{\ff}^i B(\xx) \,\leq\, 0\right)$~;
        \item (separation): $\forall \xx \in \unsafe\colon\ B(\xx) \sgt 0$~.
    \end{enumerate}
\end{definition}

Notice that the consecution constraint in Definition~\ref{def:invBc} involves 
Lie derivatives of orders up to $\LieBound \in \NN^+$, 
as is the case in Theorem~\ref{thm:invariantCondition}. 
Our invariant barrier-certificate condition 
hence generalizes existing conditions on barrier certificates, 
e.g., \cite{yang2015exact, zhang2018safety, CAV20BMI}, 
which consider Lie derivatives only up to the first order.

The following lemma states that \emph{the consecution condition in Definition~\refeq{def:invBc} 
is in fact equivalent to the invariant condition \eqref{eqn:invariantCondition} 
in Theorem~\refeq{thm:invariantCondition}}.
\begin{lemma}[Equivalence of Lie consecution]\label{lem:eqConsecution}
	The consecution condition in Definition~\refeq{def:invBc} holds 
        if and only if the invariant condition \eqref{eqn:invariantCondition} 
        in Theorem~\refeq{thm:invariantCondition} holds.
\end{lemma}

\begin{proof}
	We prove both the ``if'' and the ``only if'' part by contradiction.
	
	For the ``if'' part, suppose that the invariant condition \eqref{eqn:invariantCondition} holds but the consecution condition is invalid. The latter implies that for some $\xx_0 \in \RR^n$ and $1 \leq i_0 \leq \LieBound$,
	\begin{equation}
		\label{eqn:invBcCase}
		\left(\bigwedge\nolimits_{j = 0}^{i_0 - 1} \mathcal{L}_{\ff}^j (\xx_0) \,=\, 0\right) 
		\sland \mathcal{L}_{\ff}^{i_0} B(\xx_0) \,>\, 0~.
	\end{equation}%
	Note that \eqref{eqn:invBcCase} implies $B(\xx_0) = 0$. From \eqref{eqn:invariantCondition}, it follows that either
	\begin{equation}
		\label{eqn:invariantConditionCase1}
		\bigwedge\nolimits_{i = 0}^{\LieBound} \mathcal{L}_{\ff}^i B(\xx_0) \,=\, 0
	\end{equation}%
	holds, or there exists $0 \leq i_1 \leq \LieBound$ such that
	\begin{equation}
		\label{eqn:invariantConditionCase2}
		\left(\bigwedge\nolimits_{j = 0}^{i_1-1} \mathcal{L}_{\ff}^j B(\xx_0) \,=\, 0\right) 
		\sland \mathcal{L}_{\ff}^{i_1} B(\xx_0) \,<\, 0
	\end{equation}%
	holds. However, 
	\begin{itemize}
		\item \eqref{eqn:invariantConditionCase1} cannot hold as $\mathcal{L}_{\ff}^{i_0}B(\xx_0) = 0$ in \eqref{eqn:invariantConditionCase1} but $\mathcal{L}_{\ff}^{i_0}B(\xx_0) > 0$ in \eqref{eqn:invBcCase};
		\item for $i_1 \leq i_0$, \eqref{eqn:invariantConditionCase2} cannot hold as $\mathcal{L}_{\ff}^{i_1} B(\xx_0) < 0$ in \eqref{eqn:invariantConditionCase2} but $\mathcal{L}_{\ff}^{i_1} B(\xx_0) \geq 0$ in \eqref{eqn:invBcCase}; 
		\item for $i_1 > i_0$, \eqref{eqn:invariantConditionCase2} cannot hold as $\mathcal{L}_{\ff}^{i_0} B(\xx_0) = 0$ in \eqref{eqn:invariantConditionCase2} but $\mathcal{L}_{\ff}^{i_0} B(\xx_0) > 0$ in \eqref{eqn:invBcCase}. 
	\end{itemize}
	%
	
	For the ``only if'' direction, suppose that the consecution condition in Definition~\ref{def:invBc} holds but the invariant condition \eqref{eqn:invariantCondition} is invalid. The latter implies that there exists $\xx_0$ such that $B(\xx_0) \leq 0$ and
	\begin{equation}
		\label{eqn:invariantConditionNegCase}
		\neg\ \left(\left(\bigwedge\nolimits_{j = 0}^{i-1} \mathcal{L}_{\ff}^j B (\xx_0) \,=\, 0\right) 
		\sland \mathcal{L}_{\ff}^i B (\xx_0) \,<\, 0\right)
	\end{equation}%
	holds for any $0 \leq i \leq \LieBound$.
	
	For $i = 0$, \eqref{eqn:invariantConditionNegCase} yields that $B(\xx_0) \geq 0$. Together with the premise $B(\xx_0) \leq 0$, we have $B(\xx_0) = \mathcal{L}_{\ff}^{0} B(\xx_0) = 0$. 
	Now, by taking the case $i = 1$ in the consecution condition, we deduce $\mathcal{L}_{\ff}^{1} B(\xx_0) \leq 0$. Meanwhile, for $i = 1$, \eqref{eqn:invariantConditionNegCase} yields $\mathcal{L}_{\ff}^{1} B(\xx_0) \geq 0$. It thus follows that $\mathcal{L}_{\ff}^{1} B(\xx_0) = 0$. 
	Analogously, by taking $i = 2, \ldots, \LieBound$, we conclude $\mathcal{L}_{\ff}^{i} B(\xx_0) = 0$ for all $0 \leq i \leq \LieBound$. This is exactly encoded in \eqref{eqn:invariantCondition} (the rightmost conjunctive clause) and hence contradicts the assumption that \eqref{eqn:invariantCondition} is invalid. Therefore, the consecution condition implies \eqref{eqn:invariantCondition}.
\end{proof}

Lemma~\ref{lem:eqConsecution} reveals the relation 
between an inductive invariant and an invariant barrier certificate:

\begin{theorem}[Inductive invariance]\label{thm:inductiveInvariance}
    Given a system \eqref{eqn:dynamics}, an initial set $\init$ and an unsafe set $\unsafe$. (1) If polynomial $B(\xx)$ is an invariant barrier certificate, then $\invt = \{ \xx \mid B(\xx) \leq 0 \}$ is an invariant. Conversely, (2) if $\invt = \{ \xx \mid B(\xx) \leq 0 \}$ is an invariant satisfying $\init \subseteq \invt$ and $\invt \cap \unsafe = \emptyset$, then $B(\xx)$ is an invariant barrier certificate.
\end{theorem}

\begin{proof}
    The claim is an immediate consequence of Lemma~\ref{lem:eqConsecution}.
\end{proof}

%

It follows from Theorem~\ref{thm:inductiveInvariance} that \emph{our invariant barrier-certificate condition is the least conservative (and in fact the weakest possible) one on barrier certificates to attain inductive invariance}.

\begin{remark}
We do not employ the invariant condition \eqref{eqn:invariantCondition} in Theorem~\refeq{thm:invariantCondition} as the constraint on the consecution of Lie derivatives. This is because our consecution condition in Definition~\refeq{def:invBc} is simpler, and in particular, amenable to more straightforward transformations to SOS constraints via Putinar's Positivstellensatz, as shown later in Section~\ref{subsec:BMIEncoding}.
\end{remark}

\begin{remark}\label{remark:strengthening}
	For a fixed $0 < \mathfrak{N} < \LieBound$, the consecution condition in Definition~\ref{def:invBc} can be strengthened in the following way while preserving inductive invariance:
	\begin{equation*}
            \forall \xx \in \RR^n\colon\ 
		\bigwedge\nolimits_{i = 1}^{\mathfrak{N}-1} \left(
		\left(\bigwedge\nolimits_{j = 0}^{i-1} \mathcal{L}_{\ff}^j B(\xx) = 0\right) 
		\implies \mathcal{L}_{\ff}^i B(\xx) \leq 0\right) \sland 
		\left(\left(\bigwedge\nolimits_{j = 0}^{\mathfrak{N}-1} \mathcal{L}_{\ff}^j B(\xx) = 0\right) 
		\implies \mathcal{L}_{\ff}^\mathfrak{N} B(\xx) < 0\right)
	\end{equation*}%
        where for the $\mathfrak{N}$-th Lie derivative, one needs $\mathcal{L}_{\ff}^\mathfrak{N} B(\xx) < 0$ (rather than $\mathcal{L}_{\ff}^\mathfrak{N} B(\xx) \leq 0$). In practice, using
        such a strengthened consecution condition ---with less sub-constraints to solve--- may yield more efficient synthesis.
\end{remark}

%
\begin{wrapfigure}{r}{0.43\textwidth}
	\vspace*{-6mm}
	\begin{center}
		\adjustbox{width=0.43\textwidth}{
		\begin{tikzpicture}[align=center, node distance=5cm]
			\node[state, accepting, inner sep=2pt] (q0) 
			{
				$q_0\colon$\\[.2cm]
				$\dot{\xx} = \ffk{0}(\xx)$\\[.1cm]
				$\xx \in \domain_0$\\[-.3cm]
			};
			\node[state, right of=q0, inner sep=2pt] (q1) 
			{
				$q_1\colon$\\[.2cm]
				$\dot{\xx} = \ffk{1}(\xx)$\\[.1cm]
				$\xx \in \domain_1$\\[-.3cm]
			};
			\draw [-latex']
			(q0) edge[bend left] node[align=center]{\\[-.08cm]$\mathit{guard}_0(\xx) \leq 0$\\[.2cm]$\xx' = \mathit{reset}_0(\xx)$} (q1)
			(q1) edge[bend left] node[align=center]{$\xx' = \mathit{reset}_1(\xx)$\\[.2cm]$\mathit{guard}_1(\xx) \leq 0$\\[-.24cm]} (q0);
		\end{tikzpicture}
	}
	\end{center}
	\vspace*{-6mm}
	\caption{A simple (symbolic) hybrid automaton.}
	\vspace*{-2mm}
	\label{fig:hybrid}
\end{wrapfigure}
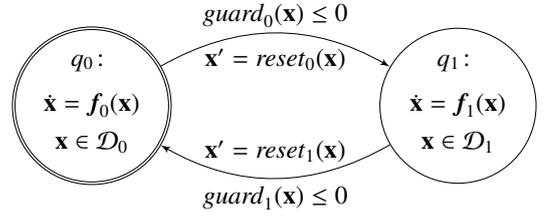	

\paragraph*{\bf Generalization to hybrid systems}
Our invariant barrier-certificate condition can be readily generalized to multi-mode hybrid systems exhibiting both continuous dynamics and discrete transitions in the same vein as in \cite{Prajna04,CAV20BMI}.
%
    We illustrate such generalization by a simple (symbolic) hybrid automaton~\cite{DBLP:journals/siglog/Fraenzle19} as depicted in Fig.~\ref{fig:hybrid}. 
    The system has two modes $q_0$ (initial mode) and $q_1$ governed respectively by polynomial flow fields $\ffk{0}(\xx)$ and $\ffk{1}(\xx)$ and mode domains $\domain_0$ and $\domain_1$. 
    The system may evolve continuously in mode $q_k$ (for $k = 0, 1$) within $\domain_k$ or jump to mode $q_{1-k}$ when $\mathit{guard}_k(\xx) \leq 0$ is satisfied. In the latter case, the system state will be set to  $\xx' = \mathit{reset}_k(\xx) \in \domain_{1-k}$ after the jump. 
    We aim to verify that no trajectory originating from an initial set $\init \subseteq \domain_0$ will ever visit states in the unsafe sets $\unsafek \subseteq \domain_k$. To this end, our invariant barrier-certificate condition (cf.\  Definition~\ref{def:invBc}) can be augmented to recognize an invariant barrier certificate $B_k(\xx)$ for each mode $q_k$:
    \begin{enumerate}
        \item (initial): $\forall \xx \in \init\colon\ B_0(\xx) \sleq 0$~;
        \item (consecution): $\forall k \in \{0, 1\} \ldotp\, \forall \xx \in \domain_k\colon\ 
                \bigwedge_{i = 1}^{\LieBoundBi} \left(
                \left(
                    \bigwedge_{j = 0}^{i-1} \mathcal{L}_{\ffk{k}}^j B_k(\xx) \,=\, 0
                \right)
                \,\implies\, \mathcal{L}_{\ffk{k}}^i B_k(\xx) \,\leq\, 0\right)$~;
        \item (transition): $\forall k \in \{0, 1\} \ldotp\, \forall \xx \in \domain_k\colon\ 
                \left(B_k(\xx) \,\leq\, 0 \land \mathit{guard}_k(\xx) \,\leq\, 0\right) 
                \,\implies\, B_{1-k}(\mathit{reset}_k(\xx)) \,\leq\, 0$~;
        \item (separation): $\forall k \in \{0, 1\} \ldotp\, \forall \xx \in \unsafek\colon\ B_k(\xx) \sgt 0$~.
    \end{enumerate}
	The existence of $B_k(\xx)$ satisfying the above constraints ensures safety of the hybrid system model. In fact, all these constraints (with polynomial guards and resets as well as domains described by polynomials) can be encoded in a BMI optimization problem and thereby solved by our DCP-based algorithm without substantial changes. For simplicity, however, we present our techniques for single-mode dynamical systems based on the invariant barrier-certificate condition given in Definition~\ref{def:invBc}.
%

\subsection{Encoding as BMI optimizations}
\label{subsec:BMIEncoding}

Next, we show how to encode the synthesis of an invariant barrier certificate 
as an optimization problem subject to BMIs. To this end, we first recast the invariant barrier-certificate condition into a collection of SOS constraints. For simplicity, we assume that $\init$ and $\unsafe$ are both captured by a single polynomial. Our formulations, however, apply also to cases with basic semi-algebraic $\init$ or $\unsafe$.
%

\begin{theorem}[Sufficient condition for invariant barrier certificate]
    \label{thm:invariantBcSosSufficient}
    Given a system \eqref{eqn:dynamics}, an initial set $\init = \{ \xx \mid \initBound(\xx) \leq 0 \}$ and an unsafe set $\unsafe = \{ \xx \mid \unsafeBound(\xx) \leq 0 \}$. A polynomial $B \in \mathbb{R}[\xx]$ is an invariant barrier certificate of \eqref{eqn:dynamics} if for some $\epsilon \in \RR^+$, there exist polynomials $v_{i, j} \in \mathbb{R}[\xx]$ and SOS polynomials $\sigma(\xx), \sigma'(\xx) \in \Sigma[\xx]$ s.t.
    \begin{enumerate}
        \item $-B(\xx) + \sigma(\xx) \initBound(\xx)$~,
        \item for all $1 \leq i \leq \LieBound$, $-\mathcal{L}_{\ff}^i B(\xx) + \sum\nolimits_{j = 0}^{i - 1} v_{i, j}(\xx) \mathcal{L}_{\ff}^j B(\xx)$~,
        \item $B(\xx)+ \sigma'(\xx) \unsafeBound(\xx) - \epsilon$
    \end{enumerate}
    are SOS polynomials in $\Sigma[\xx]$.
\end{theorem}

\begin{proof}
    It can be shown that the $k$-th condition in Theorem~\ref{thm:invariantBcSosSufficient} 
    implies the $k$-th condition in Definition~\ref{def:invBc}, for $k =1, 2, 3$. 
    For instance, the second condition in Theorem~\ref{thm:invariantBcSosSufficient} requires that 
    $-\mathcal{L}_{\ff}^i B(\xx) + \sum\nolimits_{j = 0}^{i - 1} v_{i, j}(\xx) \mathcal{L}_{\ff}^j B(\xx)$
    is an SOS polynomial (and thus non-negative) for all $1 \leq i \leq \LieBound$, 
    we therefore have 
    $\mathcal{L}_{\ff}^i B(\xx) \leq \sum\nolimits_{j = 0}^{i - 1} v_{i, j}(\xx) \mathcal{L}_{\ff}^j B(\xx)$ 
    for all $1 \leq i \leq \LieBound$.
    It follows that for all $\xx$, when $\mathcal{L}_{\ff}^j B(\xx) = 0$ with $0 \leq j \leq i-1$, 
    we have $\mathcal{L}_{\ff}^i B(\xx) \leq 0$, 
    which is the consecution condition in Definition~\ref{def:invBc}. 
    A similar argument applies to the other two conditions.
\end{proof}

By enforcing the Archimedean condition and applying Putinar's Positivstellensatz, 
we further derive a necessary condition of invariant barrier certificate:

\begin{theorem}[Necessary condition for invariant barrier certificate]
    \label{thm:invariantBcSosNecessary}
    Given a system \eqref{eqn:dynamics}, an initial set $\init = \{ \xx \mid \initBound(\xx) \leq 0 \}$ and an unsafe set $\unsafe = \{ \xx \mid \unsafeBound(\xx) \leq 0 \}$. If $B \in \mathbb{R}[\xx]$ is an invariant barrier certificate of \eqref{eqn:dynamics}, then for some $\epsilon \in \RR^+$, there exist polynomials $v_{i, j}  \in \mathbb{R}[\xx]$ and SOS polynomials $\sigma(\xx), \sigma'(\xx), \rho(\xx), \rho'(\xx), \rho_i''(\xx) \in \Sigma[\xx]$ s.t.~for any $L \in \RR^+$,
    \begin{enumerate}
        \item $-B(\xx) + \rho(\xx) (\normDisplay{\xx} - L) + \sigma(\xx) \initBound(\xx) + \epsilon$~,
        \item for all $1 \leq i \leq \LieBound$, $ -\mathcal{L}_{\ff}^i B(\xx) + \rho_i''(\xx) (\normDisplay{\xx} - L) + \sum\nolimits_{j = 0}^{i - 1} v_{i, j}(\xx) \mathcal{L}_{\ff}^j B(\xx) + \epsilon$~,
        \item $B(\xx) + \rho'(\xx) (\normDisplay{\xx} - L) + \sigma'(\xx) \unsafeBound(\xx)$
    \end{enumerate}
    are SOS polynomials in $\Sigma[\xx]$. 
\end{theorem}

\begin{proof}
    The invariant barrier-certificate condition in Definition~\ref{def:invBc} 
    characterizes positivity of polynomials over certain sets. 
    By adding a ``ball'' constraint $\norm{\xx} - L \leq 0$ to those sets 
    (thus achieving the Archimedean condition), 
    we can apply Putinar's Positivstellensatz to rewrite those polynomials into SOS forms. 
%
    
    For instance, the consecution condition in Definition~\ref{def:invBc} implies that        
    $-\mathcal{L}_{\ff}^i B(\xx) + \epsilon$ is strictly positive on 
    $\mathcal{K} = \{ \xx \mid 
        (\bigwedge_{j = 0}^{i-1} \mathcal{L}_{\ff}^0 B(\xx) = 0) \land
        - (\norm{\xx} - L) \geq 0
    \}$ 
    for all $1 \leq i \leq \LieBound$. 
    Putinar's Positivstellensatz can then be applied to show that 
    $-\mathcal{L}_{\ff}^i B(\xx) + \epsilon = 
        \sigma_i(\xx) 
        - \rho_i''(\xx) (\norm{\xx} - L) 
        - \sum_{j = 0}^{i-1} v_{i, j} (\xx) \mathcal{L}_{\ff}^j B(\xx)
    $
    holds for some SOS polynomials $\sigma_i(\xx)$, $\rho_i''(\xx)$ and some polynomials $v_{i, j}(\xx)$
    for $1 \leq i \leq \LieBound$ and $0 \leq j \leq i-1$. 
    The second condition in Theorem~\ref{thm:invariantBcSosNecessary} then follows immediately.

    A similar argument applies to the other two conditions.
\end{proof}

Notice that a polynomial $B(\xx)$ satisfying the sufficient condition in Theorem~\ref{thm:invariantBcSosSufficient} suffices as an invariant barrier certificate that witnesses safety of the system. In contrast, a polynomial $B(\xx)$ satisfying the necessary condition in Theorem~\ref{thm:invariantBcSosNecessary} may serve as a candidate invariant barrier certificate, and safety of the system can be concluded via a posterior check 
of $B(\xx)$ per Definition~\ref{def:invBc}.
Such a check inherits decidability of the first-order theory over real-closed fields~\cite{Tarski51}.
%

Next \emph{we show how to encode an SOS constraint of the shape ``$h(\xx) \in \Sigma[\xx]$'' in Theorems~\refeq{thm:invariantBcSosSufficient} and~\refeq{thm:invariantBcSosNecessary} as a BMI constraint}. To this end, we first set a \emph{template polynomial}
$B(\aaa, \xx)$ parameterized by unknown real coefficients $\aaa$ as the barrier certificate (required to be linear in its parameters $\aaa$). We then proceed by setting templates for the remaining unknown polynomials (e.g., $v_{i, j}(\xx)$) and SOS polynomials (e.g., $\sigma(\xx)$ and $\rho(\xx)$) in $h(\xx)$, with all the parameters in these templates grouped in $\sss$. Observe that the parameterized SOS polynomial $h(\aaa,\sss,\xx)$ is of a bilinear form on the parameter spaces, i.e., $h(\aaa,\sss,\xx)$ is linear in $\aaa$ and $\sss$ separately. However, nonlinearity arises in the combined parameter space $(\aaa,\sss)$ due to the product couplings of $\aaa$ and $\sss$, i.e., $v_{i, j}(\sss_{i, j}, \xx) \mathcal{L}_{\ff}^j B(\aaa, \xx)$ in the consecution constraint.

Now the problem of synthesizing an invariant barrier certificate boils down to searching for an instantiation of the parameters $\aaa$ and $\sss$ such that the sufficient condition in Theorem~\ref{thm:invariantBcSosSufficient} holds (or alternatively, the necessary condition in Theorem~\ref{thm:invariantBcSosNecessary} holds and the posterior check of Definition~\ref{def:invBc} passed). Such an instantiation of $\aaa$ (making $B(\aaa,\xx)$ an invariant barrier certificate) will be called \emph{valid} in the sequel.

Suppose that a parameterized SOS polynomial $h(\aaa,\sss,\xx)$ is of degree at most $2d$, with user-specified $d \in \NN$. Then $h(\aaa,\sss,\xx)$ can always be written in \emph{quadratic form} as $h(\aaa,\sss,\xx) = \mathbf{b}^\trans Q(\aaa,\sss) \mathbf{b}$, where $\mathbf{b} = (1, x_1, x_2, x_1 x_2, \ldots, x^d_n)$ is the \emph{basis vector} of size $p = \tbinom{n+d}{n}$ containing all monomials of degree up to $d$, and $Q(\aaa,\sss) \in \mathcal{S}^p$ is a parameterized real symmetric matrix known as the \emph{Gram matrix}~\cite{choi1995sums}\footnote{Extracting the Gram matrix amounts to solving a system of linear equations resulting from coefficient matching. The derived Gram matrix may contain extra unknowns if the system of linear equations admits multiple solutions, which nevertheless can be encoded in our subsequent workflow by enumerating the basis of its null space.}. An important fact states that \emph{$h(\aaa,\sss,\xx)$ is SOS if and only if $Q(\aaa,\sss) \succeq 0$}.

Let $\mathcal{F}(\aaa, \sss) = - Q (\aaa,\sss)$. As per $h(\aaa,\sss,\xx)$, the matrix-valued function $\mathcal{F}(\aaa, \sss)$ is bilinear in $(\aaa,\sss)$. Observe that \emph{$h(\aaa,\sss,\xx)$ is SOS if and only if the BMI constraint $\mathcal{F}(\aaa, \sss) \preceq 0$ holds}. See Example~\ref{exmp:overview} for an illustration of this BMI encoding.

In general, $\mathcal{F}(\aaa, \sss)$ can be flattened in an expanded bilinear form as
\[
    \mathcal{F}(\aaa, \sss) \seq F + \sum\nolimits_{i=1}^m a_i H_i + \sum\nolimits_{j=1}^n s_j G_j + \sum\nolimits_{i=1}^m \sum\nolimits_{j=1}^n a_i s_j  F_{i, j}
\]%
where $m$ and $n$ are the size of $\aaa$ and $\sss$, respectively; $F, H_i, G_j, F_{i, j} \in \mathcal{S}^p$ are constant matrices.
Discharging the conditions of invariant barrier certificates hence amounts to solving the BMI feasibility problem of finding $\aaa$ and $\sss$ s.t.
\begin{equation}\label{eqn:bmiFeasiblity}
	\mathcal{F}_\iota(\aaa, \sss) \spreceq 0, \quad \iota=1, 2, \ldots, l~.
\end{equation}%
Here $\mathcal{F}(\aaa, \sss)$ is indexed by $\iota$ and $l$ is the number of SOS constraints involved.

To exploit well-developed techniques in optimization, the feasibility problem \eqref{eqn:bmiFeasiblity} is transformed to an optimization problem subject to BMI constraints:
\begin{maxi}
	{\lambda, \aaa, \sss}
	{\lambda}
	{\label{eqn:bmiBc}}
	{}
	\addConstraint{\mathcal{F}_\iota(\aaa, \sss) + \lambda I}{\spreceq 0,\quad}{\iota=1, 2, \ldots, l~.}
\end{maxi}%
A solution $(\lambda, \aaa, \sss)$ to~\eqref{eqn:bmiBc} is \emph{feasible} if it satisfies the BMIs in~\eqref{eqn:bmiBc}, and \emph{strictly feasible} if all the BMIs are satisfied with strict inequalities. We sometimes drop the $\lambda$ component in the solution when it is clear from the context. Notice that \emph{problem~\eqref{eqn:bmiFeasiblity} has a feasible solution if and only if the optimal value $\lambda^*$ in the BMI optimization problem~\eqref{eqn:bmiBc} is non-negative}.

To achieve (weak) completeness of our method in subsequent sections on solving the BMI optimization problem, we make the following assumption on the boundedness of the search space $(\aaa, \sss)$ of the optimization.

\begin{assumption}[Boundedness on the parameters]
    \label{ass:compactness}
    Every feasible solution $(\aaa, \sss)$ to the BMI problem~\eqref{eqn:bmiBc} is in a compact set with non-empty interior, i.e.,
    \begin{equation*}
        (\aaa, \sss) \ \in\  \compactSet_{\aaa} \times \compactSet_{\sss} \seq \left\{(\aaa,\sss) \mathrel{\big|} \normDisplay{\aaa} \leq L_{\aaa}, \normDisplay{\sss} \leq L_{\sss}\right\}
    \end{equation*} 
    for some known bounds $L_{\aaa}, L_{\sss} \in \RR^+$. 
\end{assumption}

\begin{remark}
	The boundedness on $\aaa$ in Assumption~\ref{ass:compactness} makes sense in practice since we usually prefer barrier certificates with bounded coefficients. Moreover, when the bilinear functions $\mathcal{F}_\iota(\aaa, \sss)$ in~\eqref{eqn:bmiBc} are affine in $\aaa$ and $\sss$, i.e., with a zero constant matrix $F$, the parameters $\aaa$ and $\sss$ can be scaled independently by any positive factor. Therefore in this case, w.l.o.g., one may simply set $L_{\aaa} = L_{\sss} = 1$.
\end{remark}

\section{Solving BMI optimizations via DCP}
\label{sec:algorithm}

The BMI optimization problem~\eqref{eqn:bmiBc}, derived from the synthesis problem, is known to be NP-hard and contains non-convex constraints~\cite{toker1995np}, and hence is not amenable to efficient (polynomial-time) algorithms in contrast to convex optimization. In this section, we present an algorithm for solving general BMI optimizations via difference-of-convex programming~\cite{tao1986algorithms,le2018dc}, which solves a series of convex sub-problems that approaches a local optimum of~\eqref{eqn:bmiBc}.

For brevity, we consider optimization problems with a single BMI constraint (whereas multiple BMI constraints can be joined as a single BMI in a block-diagonal fashion):
\begin{maxi}
	{\zz=(\xx, \yy)}
	{g(\zz)}
	{\label{eqn:bmip}}
	{}
	\addConstraint{}{\mathcal{B}(\xx, \yy) \sdefine F + \sum_{i=1}^m x_i H_i + \sum_{j=1}^n y_j G_j + \sum_{i=1}^m \sum_{j=1}^n x_i y_j  F_{i, j}}{\spreceq 0}
\end{maxi}%
where the objective function $g\colon \RR^{m+n} \to \RR$ is linear in $\zz = (\xx$, $\yy)$; $F, H_i, G_j, F_{i, j} \in \mathcal{S}^p$ are constant symmetric matrices.

\subsection{Difference-of-convex decomposition}
\label{subsec:dc-decomposition}
The key challenge in solving the BMI problem \eqref{eqn:bmip} is its non-convexity, that is, the matrix-valued function $\mathcal{B}(\xx, \yy)$ is, in general, not \psd-convex.

There have been attempts, most pertinently in~\cite{dinh2011combining}, to decompose a bilinear \mbox{function} as a difference between two \psd-convex functions, known as the \emph{difference-of-convex} (DC) \emph{decomposition}, such that the optimization in its decomposed form enjoys well-established techniques in difference-of-convex programming~\cite{tao1986algorithms,le2018dc}. The DC decomposition in~\cite{dinh2011combining}, however, is confined to BMIs of a specific structure, namely, $X^\trans Y + Y^\trans X \preceq 0$, where $X$ and $Y$ are matrix variables containing variables $x_i$ and $y_j$, respectively. The more general bilinear function $\mathcal{B}(\xx, \yy)$ in \eqref{eqn:bmip} does unfortunately not admit straightforward forms of decomposition such as those in~\cite[Lemma~3.1]{dinh2011combining}.

In this subsection, we first show how to formulate a difference-of-convex decomposition of the matrix-valued function $\mathcal{B}(\xx, \yy)$ using matrix decomposition (inspired by~\cite{wang2016feasibility}), and then present three different ways to obtain such a matrix decomposition. These decomposition methods compete with each other in terms of theoretical simplicity, generality, and the exploitation of matrix sparsity.
%
%

First, observe that the function $\mathcal{B}(\xx, \yy)$ can be written as 
\begin{equation}
    \label{eqn:bmiKronecker}
    \mathcal{B}(\xx, \yy) \seq 
    \begin{pmatrix}
        \xx \otimes I \\
        \yy \otimes I
    \end{pmatrix}^\trans
    \begin{pmatrix}
        0 & \Gamma \\
        \Gamma^\trans & 0
    \end{pmatrix}
    \begin{pmatrix}
        \xx \otimes I \\
        \yy \otimes I
    \end{pmatrix} 
    \mathbin{+} 
    \begin{pmatrix}
        \Omega_1 & \Omega_2 
    \end{pmatrix} 
    \begin{pmatrix}
        \xx \otimes I \\
        \yy \otimes I
    \end{pmatrix}
    + F
\end{equation}%
where $0$ represents the zero matrices with compatible dimensions 
and
\begin{equation*}
    \label{eqn:defGammaOmega}
    \Gamma \seq \frac{1}{2} 
    \begin{pmatrix}
        F_{1, 1} & \dots & F_{1, n} \\
        \vdots & \ddots & \vdots \\
        F_{m, 1} & \dots & F_{m, n}
    \end{pmatrix},\quad
    \Omega_1 \seq 
    \begin{pmatrix}
        H_1 & \dots & H_m
    \end{pmatrix},\quad
    \Omega_2 \seq 
    \begin{pmatrix}
        G_1 & \dots & G_n
    \end{pmatrix}~.
\end{equation*}%

The form of~\eqref{eqn:bmiKronecker} implies that 
\emph{$\mathcal{B}(\xx, \yy)$ is \psd-convex if the matrix
$
    M = \begin{pmatrix}
            0 & \Gamma \\
            \Gamma^\trans & 0
        \end{pmatrix}
$
is positive semidefinite}. Unfortunately, as \cite[Theorem~1]{wang2016feasibility} points out, 
for a non-trivial bilinear function $\mathcal{B}(\xx, \yy)$, 
$M$ may not be positive semidefinite.

Nevertheless, the matrix $M$ can always be decomposed 
as $M = M_1 - M_2$ with $M_1, M_2 \succeq 0$, 
i.e., a difference between two \psd-matrices. 
This, in turn, leads to a DC decomposition of $\mathcal{B}(\xx, \yy)$:
\begin{theorem}[DC decompostion by matrix decomposition]
    \label{thm:psdConvexity}
    Suppose $M = M_1 - M_2$ with $M_1, M_2 \succeq 0$. Then, the form
    \begin{equation}
    	\label{eqn:dc}
    	\mathcal{B}(\xx, \yy) \seq 
    	\mathcal{B}^+(\xx, \yy) - \mathcal{B}^-(\xx, \yy)
    \end{equation}%
    where
    \begin{equation*}
    	\begin{split}
    		\mathcal{B}^+(\xx, \yy) &\seq     
    		\begin{pmatrix}
    			\xx \otimes I \\
    			\yy \otimes I
    		\end{pmatrix}^\trans
    		M_1
    		\begin{pmatrix}
    			\xx \otimes I \\
    			\yy \otimes I
    		\end{pmatrix}
    		\mathbin{+} 
    		\begin{pmatrix}
    			\Omega_1 & \Omega_2 
    		\end{pmatrix} 
    		\begin{pmatrix}
    			\xx \otimes I \\
    			\yy \otimes I
    		\end{pmatrix}
    		+ F \\
    		\mathcal{B}^-(\xx, \yy) &\seq
    		\begin{pmatrix}
    			\xx \otimes I \\
    			\yy \otimes I
    		\end{pmatrix}^\trans
    		M_2
    		\begin{pmatrix}
    			\xx \otimes I \\
    			\yy \otimes I
    		\end{pmatrix}
    	\end{split}
    \end{equation*}%
    is a difference-of-convex decomposition of $\mathcal{B}(\xx, \yy)$, i.e., the matrix-valued functions $\mathcal{B}^+(\xx, \yy)$ and $\mathcal{B}^-(\xx, \yy)$ 
    are \psd-convex on $\RR^{m+n}$.
\end{theorem}

\begin{proof}
	We first show the \psd-convexity of $\mathcal{B}^+(\xx, \yy)$. Let $\zz = (\xx, \yy) \in \RR^{m+n}$.	According to~\cite[Proposition 1]{DBLP:journals/mp/Shapiro97}, 
        $\mathcal{B}^+(\zz) = \mathcal{B}^+(\xx, \yy)$ is \psd-convex if (and only if) 
        for any $\vv \in \RR^{p}$, the function
	$
	\phi_{\vv}(\zz) = \vv^\trans \mathcal{B}^+(\zz) \vv
	$
	is convex. Note that
	\begin{equation*}
		\begin{split}
			\phi_{\vv}(\zz) 
			&\seq \vv^\trans 
			\begin{pmatrix}
				\zz \otimes I
			\end{pmatrix}^\trans
			M_1
			\begin{pmatrix}
				\zz \otimes I
			\end{pmatrix}
			\vv
			+
			\vv^\trans
			\begin{pmatrix}
				\Omega_1 & \Omega_2
			\end{pmatrix}
			\begin{pmatrix}
				\zz \otimes I
			\end{pmatrix}
			\vv
			+
			\vv^\trans F \vv  \\
			&\seq (\zz \otimes \vv)^\trans
			M_1
			(\zz \otimes \vv)
			+
			\vv^\trans
			\begin{pmatrix}
				\Omega_1 & \Omega_2
			\end{pmatrix}
			(\zz \otimes \vv)
			+
			\vv^\trans F \vv~.
		\end{split}
	\end{equation*}%
	Then, for any $\mu_1 \in (0, 1)$ and $\mu_2 = 1 - \mu_1$, 
        we have, for any $\zz_1, \zz_2 \in \RR^{m+n}$,
	\begin{align*}
		&\phi_{\vv}(\mu_1 \zz_1 + \mu_2 \zz_2) - 
		(\mu_1 \phi_{\vv}(\zz_1) + \mu_2 \phi_{\vv}(\zz_2)) \\
		\seq\ &
		(\mu_1 (\zz_1 \otimes \vv) + \mu_2 (\zz_2 \otimes \vv))^\trans M_1 
		(\mu_1 (\zz_1 \otimes \vv) + \mu_2 (\zz_2 \otimes \vv)) - 
		\mu_1 (\zz_1 \otimes \vv)^\trans M_1 (\zz_1 \otimes \vv)
		- \mu_2 (\zz_2 \otimes \vv)^\trans M_1 (\zz_2 \otimes \vv) \\
		\seq\ &
		\mu_1 \mu_2 (\zz_2 \otimes \vv)^\trans M_1 (\zz_1 \otimes \vv) + 
		\mu_1 \mu_2 (\zz_1 \otimes \vv)^\trans M_1 (\zz_2 \otimes \vv) - 
		\mu_1 \mu_2 (\zz_1 \otimes \vv)^\trans M_1 (\zz_1 \otimes \vv) - 
		\mu_1 \mu_2 (\zz_1 \otimes \vv)^\trans M_1 (\zz_1 \otimes \vv) \\
		\seq\ &
		- \mu_1 \mu_2 ((\zz_1 - \zz_2) \otimes \vv)^\trans M_1 
		((\zz_1 - \zz_2) \otimes \vv) \\
                \sleq & 0 \tag{positive semidefiniteness of $M_1$} 
	\end{align*}%
	which means that $\phi_{\vv}(\zz)$ is convex. Thus, $\mathcal{B}^+(\xx, \yy)$ is \psd-convex.
	
	The \psd-convexity of $\mathcal{B}^-(\xx, \yy)$ can be shown in an analogous way. 
\end{proof}

It remains to find a matrix decomposition of $M$. In what follows, we present three different ways to decompose the
matrix $M\in \mathcal{S}^{(m+n)p}$ as a difference between two \psd-matrices. Notice that $M$ is a real symmetric matrix and thus only has real eigenvalues.

\subsubsection{Decompose \texorpdfstring{$M$}{M} via eigendecomposition}
\label{sec:DcEigendecomposition}
A (real symmetric) matrix is positive semidefinite if and only if all of its eigenvalues are non-negative.
Although the matrix $M$ may have both non-negative and negative eigenvalues, we can ``group'' them respectively in \psd-matrices $M_1$ and $M_2$ such that $M = M_1 - M_2$. 

One way to do so is to use the \emph{eigendecomposition} of $M$. That is,
$
    M = V^\trans D V
$,
where the orthogonal matrix $V$ contains the eigenvectors of $M$, and $D$ is a diagonal matrix whose diagonal elements are the eigenvalues of $M$.

Let $D^+$ be the matrix obtained by setting all negative elements of $D$ to zero, and $D^- = D^+ - D$. Then,
\begin{equation}
    \label{eqn:DcEigendecomposition}
    M \seq \underbrace{V^\trans D^+ V}_{M_1} - \underbrace{V^\trans D^- V}_{M_2}~.
\end{equation}%
It follows from construction that $M_1, M_2 \succeq 0$, and therefore, by Theorem~\ref{thm:psdConvexity}, we obtain a DC decomposition of $\mathcal{B}(\xx, \yy)$. 

\subsubsection{Decompose \texorpdfstring{$M$}{M} via bounds on eigenvalues}
The eigendecomposition-based DC decomposition is theoretically simple, yet does not benefit from the sparsity nature of $M$:
The matrix 
$
    M = \begin{pmatrix}
            0 & \Gamma \\
            \Gamma^\trans & 0
        \end{pmatrix}
    \in \mathcal{S}^{(m+n)p}
$
in~\eqref{eqn:bmiKronecker} is often highly sparse, 
which is potentially a useful feature in accelerating many matrix operations. 
However, sparsity is of little value when \emph{all} of the eigenvalues and eigenvectors are needed, 
which typically takes time cubic in the matrix size~\cite{pan1999complexity}. 
In particular, the decomposed matrices $M_1$ and $M_2$ may not be as sparse as $M$ is, thus slowing down almost all the subsequent matrix manipulations. 

The key observation here is that, \emph{to obtain a DC decomposition, 
one does not need to compute all the eigenvalues}. 
In fact, \emph{it suffices to find a bound on the eigenvalues}:
%
    \label{prop:DcEigenvalue}
    Let $\lambda_u \in \RR^+_0$ be an upper-bound 
    on all the eigenvalues of $M$ (the symbol $\lambda$ shall not be confused with those used in optimization problems).
    We have
    \begin{equation}
    	\label{eqn:DcEigenbound}
    	M \seq \underbrace{\lambda_u I}_{M_1} - \ \underbrace{\left(\lambda_u I - M\right)}_{M_2}~.
    \end{equation}%
%
%
    Here, $M_1 \succeq 0$ trivially holds as $\lambda_u \ge 0$. The positive semidefiniteness of $M_2 = \lambda_u I - M$ can be shown by considering the eigendecomposition of $M$:
    \begin{equation*}
        M_2 \seq \lambda_u I - V^\trans D V  \seq V^\trans \left(\lambda_u I - D\right) V
    \end{equation*}
    where the diagonal matrix $\lambda_u I - D$ contains the eigenvalues of $M_2$. Since $\lambda_u$ upper-bounds all the eigenvalues of $M$ (diagonal elements in $D$), $\lambda_u I - D$ contains only non-negative values, and thus we conclude that $M_2 \succeq 0$. 

In order to obtain the upper-bound $\lambda_u$, it suffices to compute only the \emph{largest eigenvalue} of $M$, 
which can be done substantially more efficient than conducting the full eigendecomposition, especially for sparse $M$~\cite[Chapter VI]{trefethen1997numerical}. Moreover, the decomposed matrices $M_1$ and $M_2$ given in~\eqref{eqn:DcEigenbound} 
are guaranteed to be as sparse as $M$ is.

We remark, however, that the derived matrices $M_1$ and $M_2$ in~\eqref{eqn:DcEigenbound} have inevitably larger entries than those built from eigendecomposition. 
In practice, larger entries in $M_2$ may increase the linearization error (in the transformation to convex sub-problems, cf.~Section~\ref{sec:DcAlgorithm}), thereby slowing down the convergence of the iterative DCP procedure.

\begin{remark}
    Apart from using an upper-bound $\lambda_u \geq 0$ on the eigenvalues of $M$, 
    a DC decomposition can also be obtained 
    by using a lower bound $\lambda_l \leq 0$ on the eigenvalues of $M$. 
    In that case, we have $M_1 = M - \lambda_l I$ and $M_2 = -\lambda_l I$. 
\end{remark}

\subsubsection{Decompose \texorpdfstring{$M$}{M} via SDP}
The problem of decomposing the matrix $M$
as a difference between two \psd-matrices can alternatively be modelled as an SDP problem: 
\begin{mini}|l|
    {M_2}
    {\trace\left(M_2\right)}
    {\label{eqn:DcSdpNaive}}
    {}
    \addConstraint{M + M_2}{\ssucceq 0~}{}
    \addConstraint{M_2}{\ssucceq 0~.}{}
\end{mini}%
A feasible solution to~\eqref{eqn:DcSdpNaive} clearly induces a matrix decomposition (with $M_1 = M + M_2$) as required in Theorem~\ref{thm:psdConvexity}. The objective function (i.e., the trace of $M_2$) in~\eqref{eqn:DcSdpNaive} intuitively measures the magnitude of the (undesired) ``concave part'' $-\mathcal{B}^-(\xx, \yy)$ in~\eqref{eqn:dc}. As argued previously, minimizing such an objective may reduce the linearization error 
and thus expedite the DCP procedure\footnote{A good DC decomposition should make the concave part (locally) ``as affine as possible''. Such ``affineness'' can be measured by the Hessian matrix for scalar-valued functions (see~\cite{ahmadi2018dc}). For matrix-valued functions, the Hessian is in fact a $4$-th rank tensor, but its norm can still be bounded by the norm of a certain matrix (cf.~\cite{wang2009practical}). That matrix, in our case, is exactly the matrix $M_2$.}.

Although it would seem to be more time-consuming to solve an SDP problem than to perform the eigendecomposition, the specific SDP instance~\eqref{eqn:DcSdpNaive} can often be solved rather efficiently by exploiting the sparsity pattern of $M$, e.g., the chordal sparsity~\cite{zhang2018sparse}. Alternatively, one can improve the performance by imposing a certain sparsity structure (e.g., to be diagonal) on $M_1$ or $M_2$. For instance, one possible formulation using diagonal matrix $M_1 = \diag(\mathbf{c})$ is
\begin{mini*}
    {\mathbf{c}}
    {\trace\left(\diag\left(\mathbf{c}\right) - M\right)}
    {\label{eqn:DcSdpDiagnoal}}
    {}
    \addConstraint{c_i}{\sgeq 0, \quad}{i=1, 2, \ldots, (m+n)p~}
    \addConstraint{\diag(\mathbf{c}) - M}{\ssucceq 0}{}
\end{mini*}%
which can be further rewritten as a (sparse) LMI problem: 
\begin{mini}
    {\mathbf{c}}
    {\sum\nolimits_{i} c_i}
    {\label{eqn:DcSdpLmi}}
    {}
    \addConstraint{c_i}{\sgeq 0, \quad}{i=1, 2, \ldots, (m+n)p~}
    \addConstraint{\sum\nolimits_{i} c_i \,\mathbf{e}^{\trans}_i \mathbf{e}_i - M}{\ssucceq 0}{}
\end{mini}
where $\mathbf{e}_i$ denotes a row vector with $1$ in its $i$-th column and 0's elsewhere. 
When $M$ admits a specific sparsity pattern, the LMI problem~\eqref{eqn:DcSdpLmi} can be solved extremely efficiently (see, e.g., \cite{zhang2018efficient}, for solving LMIs with thousands of variables in minutes).

In a nutshell, the eigendecomposition-based method is theoretically simple, yet does not benefit from the sparsity nature of $M$. Decomposing $M$ via bounds on eigenvalues exploits the sparsity nature of $M$ ---thereby yielding considerably faster matrix operations, but may slow down the convergence of the iterative DCP procedure. The SDP-based decomposition may expedite the DCP procedure, but is theoretically more involved and stands out only when $M$ admits specific sparsity patterns. 
We will compare these different DC decomposition methods empirically in Section~\ref{sec:experiments}.

\subsection{Reduction to LMIs}
\label{sec:DcAlgorithm}
On top of a DC decomposition (cf.~Theorem~\ref{thm:psdConvexity}), we can now apply a standard iterative procedure 
in difference-of-convex programming~\cite{dinh2011combining} to solve the BMIs.

The core idea of the procedure is to iteratively solve a series of convex sub-problems. More specifically, given a feasible solution $\zz^k = (\xx^k, \yy^k)$ to the BMI optimization problem~\eqref{eqn:bmip}, the ``concave part'' $-\mathcal{B}^-(\xx, \yy)$ in~\eqref{eqn:dc} is linearized around $\zz^k$, thereby yielding a series of convex programs ($k = 0, 1, \ldots$):

\begin{maxi}
	{\zz=(\xx, \yy)}
	{g(\zz) + \frac{1}{2} \delta \normDisplay{\zz - \zz^k}}
	{\label{eqn:bmipLinearized}}
	{}
	\addConstraint{}{\mathcal{B}^+(\zz) - \mathcal{B}^-\left(\zz^k\right) - \mathcal{DB}^-\left(\zz^k\right)\left(\zz -\zz^k\right)}{\spreceq 0}
\end{maxi}%
where $\mathcal{DB}^-(\zz)\colon \RR^{m+n} \to \mathcal{S}^p$ is the derivative of the matrix-valued function $\mathcal{B}^-$ at $\zz$, i.e., a linear mapping from a vector $\uu \in \RR^{m+n}$ to a matrix in $\mathcal{S}^p$:
\begin{equation*}
	\mathcal{DB}^-(\zz) (\uu) \sdefine \sum\nolimits_{i = 1}^{n+m} u_i \frac{\partial \mathcal{B}^-}{\partial z_i}(\zz)~.
\end{equation*}%
An extra regularization term $\frac{1}{2} \delta \norm{\zz - \zz^k}$ with $\delta < 0$ is added in~\eqref{eqn:bmipLinearized} to enforce that $g(\zz)$ strictly increases after each iteration until it stabilizes, which can be encoded as a second-order cone constraint and embedded in SDP solving.

Note that the linearized problem~\eqref{eqn:bmipLinearized} is convex and therefore can be solved efficiently
(see, e.g., \cite{zhang2013alternating}). 
Furthermore, Theorem~\ref{thm:schurComplement} can also be used 
to reformulate \eqref{eqn:bmipLinearized} as an LMI problem:

\begin{theorem}[Reduction to LMIs]
    \label{thm:bmiToLmi}
    The quadratic matrix inequality (QMI) constraint
    \begin{equation*}
       \mathcal{B}^+(\zz) - \mathcal{B}^-\left(\zz^k\right) - \mathcal{DB}^-\left(\zz^k\right)\left(\zz -\zz^k\right) \spreceq 0
    \end{equation*} 
    in \eqref{eqn:bmipLinearized} is equivalent to the LMI constraint (of the size $(m+n+1)p$)
    \begin{equation*}
        \begin{pmatrix}
            -I \quad & N (\zz \otimes I) \\
            (\zz \otimes I)^\trans N^\trans \quad  & - \mathcal{B}^-\left(\zz^k\right) - \mathcal{DB}^-\left(\zz^k\right)\left(\zz -\zz^k\right) + \Omega (\zz \otimes I) + F
        \end{pmatrix}
        \spreceq 0 
    \end{equation*}
    where $N$ is the square root matrix of $M_1$, i.e., 
    $M_1$ = $N^\trans N$, and $\Omega = \begin{pmatrix} \Omega_1 & \Omega_2 \end{pmatrix}$.  
\end{theorem}

\begin{proof}
    Note that the square root matrix $N$ of $M_1$ exists since $M_1 \succeq 0$\footnote{In case we have $M_1 = V^\trans D^+ V$ (with only non-negative eigenvalues in $D^+$) from the eigendecomposition of $M$, the matrix $N$ can be computed as $N = V^\trans (D^+)^{1/2} V$, where $(D^+)^{1/2}$ is the diagonal matrix whose diagonal elements are square roots of those in $D^+$.
	For the other decomposition methods as presented in Section~\ref{subsec:dc-decomposition}, $N$ can be obtained via Cholesky decomposition of $M_1$.}. 
    The claim then follows immediately by applying the Schur complement in Theorem~\ref{thm:schurComplement}. 
\end{proof}


Theorem~\ref{thm:bmiToLmi} entails that the series of linearized convex sub-problems of the form~\eqref{eqn:bmipLinearized} can be solved alternatively by most off-the-shelf SDP solvers designated for discharging LMIs via polynomial-time algorithms \cite{boyd1994linear}, say the interior-point methods.
Furthermore, by taking the optimum of the $k$-th sub-problem to be the next linearization point $\zz^{k+1}$, we obtain an iterative procedure for solving general BMIs, as depicted in Algorithm~\ref{alg:localBMI}.

\begin{algorithm}[t]
	\caption{\toolname{BMI-DC}: solving BMIs based on DC decomposition}
	\label{alg:localBMI}
	\SetKwInput{Input}{input}\SetKwInOut{Output}{output}\SetNoFillComment
	\Input{A BMI optimization problem~\eqref{eqn:bmip} with a strictly feasible initial solution $\zz^0$.}
	\Output{A sequence of feasible solutions $S = \left\{\zz^0, \ldots, \zz^k\right\}$ to the BMI optimization.}
    $k \gets 0$;~
    $S \gets \left\{ \zz^0 \right\}$\; 
    $M \gets \text{reformulation of~\eqref{eqn:bmip} as~\eqref{eqn:bmiKronecker}}$\;
    $(M_1, M_2) \gets \text{matrix decomposition of } M$ as in Theorem~\ref{thm:psdConvexity}\;
    \Repeat{$\normDisplaysqrt{\zz^k - \zz^{k-1}} < \varepsilon~\text{\upshape for a given tolerance } \varepsilon \in \RR^+_0$}{
    	 Construct the convex sub-problem~\eqref{eqn:bmipLinearized} out of $(M_1, M_2)$ linearized around $\zz^k$\;
         $\zz^{k+1} \gets \text{optimum of the program~\eqref{eqn:bmipLinearized}}$\;
         $S \gets S \cup \left\{ \zz^{k+1} \right\}$\tcp*{$S\;\mathtt{keeps\;track\;of\;visited\;points}$}
         $k \gets k+1$\;
    }
	\Return $S$\;
\end{algorithm}

Algorithm~\ref{alg:localBMI} falls into the DCP framework~\cite{dinh2011combining} and thus enjoys useful properties, e.g., soundness, termination and convergence as follows.

\begin{theorem}[Soundness]
    \label{thm:soundnessLocal}
    Every solution $\zz^i  = (\xx^i, \yy^i) \in S$ with $i = 0, \ldots, k$ returned by Algorithm~\ref{alg:localBMI} is a feasible solution to the original BMI problem~\eqref{eqn:bmip}.
\end{theorem}

\begin{proof}
	We prove by induction on $i$.
        The base case holds as $\zz^0$ is assumed to be a feasible solution to~\eqref{eqn:bmip}.
        For the induction step, we show that $\zz^{i+1}$ is a feasible solution 
        to~\eqref{eqn:bmip} if $\zz^i$ is a feasible solution to~\eqref{eqn:bmip}. 
        Since $\zz^{i+1}$ is a feasible solution to~\eqref{eqn:bmipLinearized} linearized at $\zz^i$, 
        it suffices to show that the feasible set of~\eqref{eqn:bmipLinearized} 
        is a subset (or, an under-approximation) of the feasible set of~\eqref{eqn:bmip}. 
	
	Theorem~\ref{thm:psdConvexity} shows that $\mathcal{B}^-(\zz)$ is \psd-convex, then by~\cite[Lemma~2.2~(b)]{dinh2011combining}, we have 
	\begin{equation}\label{eq:soundnessProof1}
		\mathcal{B}^-(\zz) - \mathcal{B}^-\left(\zz^i\right) \ssucceq \mathcal{DB}^-\left(\zz^i\right)\left(\zz - \zz^i\right)~.
	\end{equation}%
	In the meantime, $\zz^i$ is a feasible solution to~\eqref{eqn:bmipLinearized} and thus fulfils
	\begin{equation}\label{eq:soundnessProof2}
		\mathcal{B}^+(\zz) - \mathcal{B}^-\left(\zz^i\right) - \mathcal{DB}^-\left(\zz^i\right)\left(\zz - \zz^i\right) \spreceq 0~.
	\end{equation}%

	Combining~\eqref{eq:soundnessProof1} and~\eqref{eq:soundnessProof2}, we have
	$
	\mathcal{B}(\xx, \yy) = 
	\mathcal{B}^+(\zz) - \mathcal{B}^-(\zz) \preceq 0
	$
	which is exactly the BMI constraint of~\eqref{eqn:bmip}. This completes the proof.
\end{proof}

The result below states termination and convergence of Algorithm~\refeq{alg:localBMI} in terms of \emph{KKT points} of~\eqref{eqn:bmip}, i.e., solutions fulfilling the 
KKT conditions~\cite{boyd2004convex} of~\eqref{eqn:bmip}. The KKT conditions, short for Karush-Kuhn-Tucker conditions, are used to determine the optimality of a solution to a constrained nonlinear optimization problem. Addressing these conditions in detail falls outside the scope of this paper.
%

\begin{theorem}[Termination and convergence]
	\label{thm:convergenceLocal}
	If~\eqref{eqn:bmip} has finitely many KKT points, then (1) for $\varepsilon \in \RR^+$, Algorithm~\refeq{alg:localBMI} terminates; (2) for $\varepsilon = 0$, Algorithm~\refeq{alg:localBMI} visits an infinite sequence of solutions converging to a KKT point.
\end{theorem}

\begin{proof}
	Let $\bar{S} = \{\zz^i\}_{i \in \NN}$ be the infinite sequence of visited points for $\varepsilon = 0$.
	
	We first show that (2) implies (1).
	Assume that (2) holds, i.e., $\bar{S}$ converges (to a KKT point of~\eqref{eqn:bmip}), then by Cauchy's criterion for convergence, we have $\forall \varepsilon \in \RR^+ \ldotp \exists k \in \NN^+\colon \normsqrt{\zz^k - \zz^{k-1}} < \varepsilon$ (with $\zz^k, \zz^{k-1} \in \bar{S}$). Algorithm~\refeq{alg:localBMI} thus terminates.
	
	It then remains to show that $\bar{S}$ converges to a KKT point of~\eqref{eqn:bmip} if the set of KKT points of~\eqref{eqn:bmip} is finite. This is in fact a straightforward corollary of~\cite[Theorem~4.3]{dinh2011combining}, by noticing that the assumptions thereof can be readily verified. For simplicity, we highlight the validity of only a few of these assumptions: Since $\zz^0$ in Algorithm~\refeq{alg:localBMI} is a strictly feasible solution to~\eqref{eqn:bmip}, the relative interior of the feasible set of~\eqref{eqn:bmip} is non-empty and thus Assumption~A1 in~\cite{dinh2011combining} holds; Our Assumption~\ref{ass:compactness} on the boundedness of the search space ensures that $g(\zz)$ in~\eqref{eqn:bmip} is bounded from above over a bounded feasible set, and therefore the boundedness assumptions in~\cite[Theorem~4.3]{dinh2011combining} holds.
\end{proof}


We remark that, under some sufficient KKT conditions and regularity conditions~\cite{boyd2004convex}, a KKT point suffices as a local optimum. In this case, the infinite sequence $\{\zz^i\}_{i \in \NN}$ of points visited by Algorithm~\refeq{alg:localBMI} (for $\varepsilon = 0$) converges to a local optimum of~\eqref{eqn:bmip}.

It is also worth noting that, in~\cite{DBLP:conf/atva/CubuktepeJJKT18}, the authors presented a DC-based approach to synthesizing parameters in parametric Markov decision processes, which integrates (probabilistic) model checking into the DCP procedure, thereby yielding possibly earlier termination and numerically more stable results in practice. It is our future interest to investigate a similar idea in the context of barrier-certificate synthesis for hybrid systems.

\subsection{Complexity of Algorithm~\refeq{alg:localBMI}}\label{subsec:complexity}
We discuss ingredients for establishing the time complexity of Algorithm~\refeq{alg:localBMI}, which concerns (1) computing the DC decomposition; (2) performing a single iteration; and (3) conducting a number of iterations (up to a desired precision).

Recall that the matrix $M$ to be decomposed (cf.~Theorem~\ref{thm:psdConvexity}) is of the size $(m+n)p$, where $m$ and $n$ are the number of parameters in the template barrier certificate (i.e., size of $\aaa$) and other template polynomials (i.e., size of $\sss$), respectively; $p= \tbinom{r+d}{r}$ bounds the size of the basis vector $\mathbf{b}$ (where $r$ is the system dimension and the SOS polynomial is of degree at most $2d$). All the three DC decomposition methods in Section~\ref{subsec:dc-decomposition} can be done in polynomial time, e.g., $O((m+n)^3 p^3)$ for the eigendecomposition of $M$~\cite{pan1999complexity}.
%

Performing a single iteration in Algorithm~\refeq{alg:localBMI} amounts to solving an LMI instance with $k+2$ constraints (derived from Definition~\ref{def:invBc}) where $k$ is the order of Lie derivatives considered (bounded by $\LieBound$). Computing $\LieBound$ is non-elementary in theory (described in terms of the fast-growing hierarchy~\cite{DBLP:conf/lics/FigueiraFSS11} or an explicit Ackermannian function~\cite{wang2017generating,https://doi.org/10.48550/arxiv.1510.05201}), yet it is relatively small in practice and can be obtained offline. Each LMI constraint involves matrices in $\mathcal{S}^{(m+n+1)p}$ (see Theorem~\ref{thm:bmiToLmi}), which can be solved in $O(((m+n) p)^{6.5})$~\cite{nemirovski2004interior}. Note that, in practice, the computation time is often significantly less than this theoretical bound especially when the matrices in the LMI instance admit specific sparsity patterns (see, e.g., \cite{zhang2018efficient}, for solving LMIs with thousands of variables in minutes).
%
%

Bounding or even estimating the number of iterations required to achieve a desired precision is non-trivial:
one needs to determine the \emph{rate of convergence} of the sequence of solutions produced by the iterative procedure. 
Since Algorithm~\refeq{alg:localBMI} essentially builds first-order approximations of the original BMI optimization problem, 
one may reasonably assume that it is at least linearly convergent. However, to the best of our knowledge, proving linear convergence for general difference-of-convex algorithms remains an open problem~\cite{sriperumbudur2009convergence}, 
albeit with some known results on typical subclasses~\cite{le2018convergence}. In practice, nonetheless,
difference-of-convex algorithms often converge to a local optimum within a few number of iterations, as can be observed in our experiments in Section~\ref{sec:experiments}.

\subsection{Finding the initial solution}\label{subsec:initialSolution}

The iterative procedure in Algorithm~\refeq{alg:localBMI} starts with a fed-by-oracle strictly feasible initial solution $\zz^0$ to the BMI problem~\eqref{eqn:bmip}. Finding such an initial solution, however, is non-trivial in general due to the non-convexity of~\eqref{eqn:bmip}. We argue though, that a strictly feasible initial solution can be obtained for the BMI problem of the form~\eqref{eqn:bmiBc} induced by the barrier-certificate synthesis problem.

Recall that in the BMI problem~\eqref{eqn:bmiBc}, bilinearity arises from the multiplication of 
$B(\aaa, \xx)$ with some unknown multiplier polynomials parameterized by $\sss$. One way to reduce the BMI constraints to LMIs is to fix every multiplier polynomial to be a non-negative constant, thereby yielding a linear program:

\begin{maxi}
	{\lambda, \aaa}
	{\lambda}
	{\label{eqn:lmiExponetialBc}}
	{}
	\addConstraint{\mathcal{F}_\iota(\aaa, \sss)\big\vert_{\sss = \left(c_\iota, 0, \ldots, 0\right)} + \lambda I}{\spreceq 0,\quad}{\iota=1, 2, \ldots, l}
\end{maxi}%
where $\sss$ in $\mathcal{F}_\iota(\aaa, \sss)$ is substituted by $(c_\iota, 0, \ldots, 0)$ with $c_\iota \in \RR^+_0$, which encodes a non-negative constant multiplier polynomial. Observe that no $\sss$-variable is involved in~\eqref{eqn:lmiExponetialBc} and the constraints therein are linear in $\aaa$.

Evidently, a strictly feasible solution $(\lambda, \aaa)$ to~\eqref{eqn:lmiExponetialBc} induces a strictly feasible solution $(\lambda, \aaa, (c_\iota, 0, \ldots, 0))$ to~\eqref{eqn:bmiBc} as well. Moreover, we have
\begin{lemma}\label{lem:LMIsolution}
    The LMI program~\eqref{eqn:lmiExponetialBc} always has a strictly feasible solution.
\end{lemma}

\begin{proof}
	Let $\Lambda_\aaa \define \min_{1 \le \iota \le l} -\rho\left(\mathcal{F}_\iota(\aaa, \sss)\big\vert_{\sss = \left(c_\iota, 0, \ldots, 0\right)}\right)$, where $\rho(A)$ denotes the spectral radius of matrix $A$, i.e., the largest absolute value of the eigenvalues of $A$. It follows that program~\eqref{eqn:lmiExponetialBc} has a strictly feasible solution if $\lambda < \Lambda_\aaa$.
	
	Furthermore, under Assumption~\ref{ass:compactness} on the boundedness of parameter $\aaa \in \compactSet_{\aaa}$, $\Lambda_\aaa$ can be shown to be bounded by the well-known Gershgorin circle theorem. 
	
	Therefore, by taking an interior point of $\compactSet_{\aaa}$ as $\tilde{\aaa}$, and $\tilde{\lambda} = \Lambda_{\tilde{\aaa}} - \epsilon$ for some $\epsilon \in \RR^+$, we obtain a strictly feasible solution $(\tilde{\lambda}, \tilde{\aaa})$ to program~\eqref{eqn:lmiExponetialBc}. 
\end{proof}

As a consequence, \emph{a strictly feasible solution to the BMI problem~\eqref{eqn:bmiBc} can be obtained by solving the LMI problem~\eqref{eqn:lmiExponetialBc}}. In fact, when considering Lie derivatives only up to the first order, solving (the feasibility counterpart of)~\eqref{eqn:lmiExponetialBc} is exactly the procedure to synthesize either an \emph{exponential barrier certificate}~\cite{Kong13} (with $c_\iota \in \RR^+$) or a \emph{convex barrier certificate}~\cite{Prajna04} (with $c_\iota = 0$). Algorithm~\refeq{alg:localBMI} therefore subsumes existing synthesis techniques in the sense that any valid barrier certificate synthesized by methods in~\cite{Kong13,Prajna04} can also be discovered by Algorithm~\refeq{alg:localBMI}. Moreover, an alternative way to reduce the BMI constraints to LMIs is to fix the multipliers to be some given non-trivial (SOS) polynomials~\cite{zeng2016darboux}.

\begin{remark}
	Different choices of the multiplier constants $c_\iota$ in~\eqref{eqn:lmiExponetialBc} may lead to different initial solutions fed to Algorithm~\refeq{alg:localBMI}, thereby considerably different numbers of iterations until termination. In practice, techniques like randomization are worth exploring when choosing these multiplier constants.
\end{remark}
%
%
%

\subsection{Numerical errors in SDP solving and potential solutions}\label{subsec:numerical}

Most of the existing off-the-shelf SDP solvers are based on numerical computations. The underlying numerical errors caused by, e.g., floating-point computations, may hence lead to unsound results in SDP-based verification or synthesis. To circumvent this issue, three different types of solution have been presented in the literature:
\begin{itemize}\itemsep2pt
    \item Validated SDP solving: In \cite{RVS16}, Roux et al.~presented  verified SDPs, where the basic idea is to compute a suitable bound $\epsilon \in \RR^+$ 
          and replace all matrix-inequality constraints of the form $A \preceq 0$ by the corresponding $\epsilon$-strengthened versions $A + \epsilon I \preceq 0$. 
          In \cite{Gan2020}, the authors further developed this idea to guarantee the soundness of SDP-based synthesis of nonlinear Craig interpolants. 
    \item Posterior check by symbolic methods:
    	  The soundness of numerical SDP-based approaches can be retrieved by performing a posterior check via symbolic methods, e.g., quantifier elimination \cite{Collins1975} and SMT solving \cite{DBLP:series/faia/BarrettSST09}.
    \item Exact SDP solving: Henrion et al.~presented in \cite{DBLP:journals/jsc/HenrionND21} 
          an exact algorithm based on symbolic homotopy for solving SDP problems. 
          This algorithm, as noted by the authors, can solve SDP instances only of small sizes.
\end{itemize}
In this article, we exploit the second approach to perform a posterior verification of the synthesized candidate barrier certificates via both the quantifier-elimination procedure in Wolfram \textsc{Mathematica} and the SMT solver \textsc{Z3}~\cite{z3}.

\section{Incorporating in a branch-and-bound framework}
\label{sec:bbframework}

The aforementioned iterative procedure on solving a series of convex optimizations converges only to a local optimum of the BMI problem~\eqref{eqn:bmiBc} (or more generally,~\eqref{eqn:bmip}). This means that, in some cases, it may miss the global optimum that induces a non-negative $\lambda^*$. We present in this section a solution to this problem by incorporating the iterative procedure into a branch-and-bound framework that searches for the global optimum in a divide-and-conquer fashion, as is a common technique in non-convex optimizations.
\subsection{The branch-and-bound algorithm}

The basic idea is as follows. We first try to solve the BMI problem~\eqref{eqn:bmiBc}
by Algorithm~\ref{alg:localBMI} over the compact parameter space $(\compactSet_{\aaa}, \compactSet_{\sss})$. If a valid solution, (i.e, a solution that contains a valid parameter $\bar{\aaa} \in \compactSet_{\aaa}$ such that $B(\bar{\aaa}, \xx)$ is an invariant barrier certificate) is found, then the corresponding barrier certificate can be obtained. Otherwise, we keep bisecting $\compactSet_{\aaa}$ and apply Algorithm~\ref{alg:localBMI} over each bisection (note that the validity of $\bar{\aaa} \in \compactSet_{\aaa}$ does not depend on $\sss$, thus we do not partition $\compactSet_{\sss}$).
The procedure, as depicted in Algorithm~\ref{alg:bbAlgorithm} in a recursive manner, terminates when a valid parameter is found or the partition is fine enough.

\begin{algorithm}[t]
\caption{\toolname{Branch-and-Bound}: searching for a valid parameter $\bar{\aaa}$}
\label{alg:bbAlgorithm}
\SetKwInput{Input}{input}\SetKwInOut{Output}{output}\SetNoFillComment
\Input{A BMI optimization problem of the form~\eqref{eqn:bmiBc} with $\compactSet_{\aaa} = \{\aaa \mid \norm{\aaa} \leq L_{\aaa}\}$.}
\Output{A valid parameter $\bar{\aaa}$, or otherwise $\bot$ indicating a failure.}
    \lIf(\tcp*[f]{$\mathtt{abort\;on\;fine\mbox{-}enough\;partitions\;(}\eta \in \RR^+\mathtt{)}$}){$L_{\aaa} < \eta$}{\Return $\bot$}
   $\hat{\lambda} \gets \text{an upper-bound on the objective value } \lambda \text{ of~\eqref{eqn:bmiBc} over } (\compactSet_{\aaa}, \compactSet_{\sss})$\;\label{lin:startBound}
   \If(\tcp*[f]{$\mathtt{skip\;branches\;inducing\;only\;negative\;objective\;values}$}){$\hat{\lambda} < 0$}
   {
   	{\DontPrintSemicolon \Return $\bot$\tcp*{$\mathtt{if\;Theorem\;\ref{thm:invariantBcSosNecessary}\;is\;used}$}}
   	$\parallel$~\textbf{goto} Line~\ref{lin:startPartition}\tcp*{$\mathtt{if\;Theorem\;\ref{thm:invariantBcSosSufficient}\;is\;used}$}\label{lin:endBound}
   }
    \tcc{$\mathtt{sample\mbox{-}and\mbox{-}check\;(Line\;\ref{lin:startSample}\!-\!\ref{lin:endSample})\;is\;not\;necessary\;if\;Theorem\;\ref{thm:invariantBcSosNecessary}\;is\;used}$}
    $\bar{\aaa} \gets \text{a randomly-sampled point in } \compactSet_{\aaa}$\;\label{lin:startSample}
    \lIf(\tcp*[f]{$\mathtt{check\;validity\;(inductive\;invariance)}$}){$\bar{\aaa}$ \text{\upshape is valid}}{\Return $\bar{\aaa}$\label{lin:endSample}}
    \If(\tcp*[f]{$S_{\mathit{glb}}\;\mathtt{contains\;a\;global\;set\;of\;visited\;points}$}){$\textit{proj}_{\aaa}(S_{\textit{glb}}) \cap \compactSet_{\aaa} = \emptyset$\label{lin:checkS}}
    {
        $S \gets \text{apply } \toolname{BMI-DC}$ in Algorithm~\ref{alg:localBMI} \text{to}~\eqref{eqn:bmiBc} with initial solution in $(\compactSet_{\aaa}, \compactSet_{\sss})$\;
        $S_{\textit{glb}} \gets S_{\textit{glb}} \cup S$\;
        \tcc{$\mathtt{checking\;validity\;is\;not\;necessary\;if\;Theorem\;\ref{thm:invariantBcSosSufficient}\;is\;used
        }$}\label{lin:comment}
        \lIf{\text{\upshape a valid parameter} $\bar{\aaa} \in \textit{proj}_{\aaa}(S)$ \text{\upshape is found}}{\Return $\bar{\aaa}$}\label{lin:checkForNec}
    }
    $(\compactSet_{\aaa}^1, \compactSet_{\aaa}^2) \gets \textit{bisect}(\compactSet_{\aaa})$\tcp*{$\mathtt{partition\;the\;parameter\;space}$}\label{lin:startPartition}
    $\bar{\aaa} \gets \toolname{Branch-and-Bound}(\compactSet_{\aaa}^1)$\;
    \lIf{$\bar{\aaa} \neq \bot$}{\Return $\bar{\aaa}$}
    \lElse{\Return $\toolname{Branch-and-Bound}(\compactSet_{\aaa}^2)$}\label{lin:endPartition}
\end{algorithm}

Algorithm~\ref{alg:bbAlgorithm} takes as input a BMI problem of the form~\eqref{eqn:bmiBc} that encodes either the sufficient condition in Theorem~\ref{thm:invariantBcSosSufficient} or the necessary condition in Theorem~\ref{thm:invariantBcSosNecessary} for invariant barrier certificates. In the former case, a sample-and-check process (Line~\ref{lin:startSample}--\ref{lin:endSample}) is necessary to attain (weak) completeness (see Theorem~\ref{thm:bbCompleteness}). The conditional statement in Line~\ref{lin:checkS} rules out parameter (sub-)spaces that have already been explored, which is the case when the projection of some visited point in $S_{\textit{glb}}$ (a global set that keeps track of visited points by Algorithm~\ref{alg:localBMI}, initialized as $\emptyset$) onto $\aaa$ is in the current parameter space.
%
%

To further improve the performance, Algorithm~\ref{alg:bbAlgorithm} is complemented by an operation (Line~\ref{lin:startBound}--\ref{lin:endBound}) that prunes branches inducing only negative objective values. This is witnessed by a negative upper-bound on the objective value of~\eqref{eqn:bmiBc} over the current parameter space. We defer the computation of such an upper-bound to Section~\ref{subsec:upperBound}. When Theorem~\ref{thm:invariantBcSosSufficient} is used to form~\eqref{eqn:bmiBc}, however, the partition of the parameter space (Line~\ref{lin:startPartition}--\ref{lin:endPartition}) is still necessary to attain completeness, as a negative objective value of~\eqref{eqn:bmiBc} encoding the sufficient condition for invariant barrier certificate may still induce a valid parameter. In practice, one may choose to preferentially explore (partition) branches with larger $\hat{\lambda}$.

%

The following theorem claims a weak completeness result: our method guarantees to find a barrier certificate when there exists an inductive invariant (in the form of a given template) that suffices to certify safety of the system.
%

\begin{theorem}[Weak completeness]
    \label{thm:bbCompleteness}
    Algorithm~\refeq{alg:bbAlgorithm} returns a valid parameter $\bar{\aaa} \in \compactSet_{\aaa}$, if (1) the partition granularity is fine enough (i.e., small enough $\eta \in \RR^+$), (2) the degrees of multiplier polynomials and SOS polynomials used to form~\eqref{eqn:bmiBc} are large enough, and (3) there exists, for the given template $B(\aaa,\xx)$, a strictly valid parameter $\hat{\aaa} \in \compactSet_{\aaa}$ (i.e., any parameter in some neighborhood of $\hat{\aaa}$ is valid).
\end{theorem}

\begin{proof}
    When the assumptions (1)--(3) hold, Algorithm~\ref{alg:bbAlgorithm} will eventually visit a branch wherein any parameter is valid (in case a valid parameter has not been found yet). If the necessary condition in Theorem~\ref{thm:invariantBcSosNecessary} is used to form the BMI problem~\eqref{eqn:bmiBc}, Line~\ref{lin:checkForNec} ensures to return a valid parameter $\bar{\aaa} \in \compactSet_{\aaa}$; Otherwise if the BMI problem~\eqref{eqn:bmiBc} encodes the sufficient condition in Theorem~\ref{thm:invariantBcSosSufficient} which strengthens the invariant barrier-certificate condition in Definition~\ref{def:invBc}, a valid parameter $\bar{\aaa}$ may not induce a non-negative objective value of~\eqref{eqn:bmiBc}. In this case, however, any parameter sampled and returned by Line~\ref{lin:startSample}--\ref{lin:endSample} in the branch is valid, as it contains only valid parameters.
\end{proof}
%

\subsection{Computing an upper-bound \texorpdfstring{$\hat{\lambda}$}{} by convex relaxation}\label{subsec:upperBound}
The bisection operation in Algorithm~\ref{alg:bbAlgorithm} incurs ---in the worst case--- an exponential blow-up in the number of branches. In practice, however, one can prune branches inducing only negative objective values, which can be evidenced by a negative upper-bound $\hat{\lambda}$ on the objective value of~\eqref{eqn:bmiBc} over the current parameter space (Line~\ref{lin:startBound}--\ref{lin:endBound} in  Algorithm~\ref{alg:bbAlgorithm}). Such an upper-bound can be computed by \emph{over-approximating the BMI problem} (in contrast to under-approximations pursued by Algorithm~\ref{alg:localBMI}) via, e.g., convex relaxation~\cite{kheirandishfard2018convex}. Moreover, the efficiency of Algorithm~\ref{alg:bbAlgorithm} greatly depends on the tightness of the upper-bound.

%
In this subsection, we show how to obtain a preferably tight upper-bound (on the objective value) of a BMI program by a classical semidefinite relaxation. Interested readers may refer to~\cite{kheirandishfard2018convex} for more established results on this topic.

To better illustrate the idea, we stick to the BMI optimization problem of the general form \eqref{eqn:bmip}. 
As the non-convexity comes from the quadratic terms $x_i y_i F_{i, j}$, 
a straightforward convex relaxation is 
\begin{maxi}
	{\substack{\zz=(\xx, \yy),\\ \quadraticVar=(\quadraticVar(i,j))_{m \times n}}}
	{g(\zz)}
	{\label{eqn:bmipNaiveRelaxation}}
	{}
	\addConstraint{}{F + \sum_{i=1}^m x_i H_i + \sum_{j=1}^n y_j G_j + \sum_{i=1}^m \sum_{j=1}^n \quadraticVar(i, j)  F_{i, j}}{\spreceq 0~.}
\end{maxi}%
%
%
That is, we replace each quadratic term $x_i y_i$ with a new variable $\quadraticVar(i, j)$, which constitutes a matrix $\quadraticVar=(\quadraticVar(i,j))_{m \times n}$ of fresh variables. The resulting constraint in~\eqref{eqn:bmipNaiveRelaxation} becomes an LMI that can be solved by SDP. 

Notice that the convex program~\eqref{eqn:bmipNaiveRelaxation} may lead to excessively coarse over-approximations, as the relation $\quadraticVar(i, j) = x_i y_j$ is completely abstracted away in the relaxation. However, by adding extra convex constraints, one can obtain better over-approximations of the feasible set and thereby tighter upper-bounds (despite the fact that finitely many convex constraints can never precisely capture a non-convex constraint):
%
%
The classical SDP relaxation replaces the non-convex constraints $\quadraticVar(i, j) = x_i y_j$, with $i=1,\ldots,m$; $j=1,\ldots,n$ by
\begin{equation}
    \label{eqn:relaxedBmiConstriant}
    \begin{pmatrix}
        0 & \quadraticVar\\ 
        \quadraticVar^\trans & 0 
    \end{pmatrix}
    - \zz^\trans \zz 
    \spreceq 0~.
\end{equation}%
Schur complement in Theorem~\ref{thm:schurComplement} implies that constraint~\eqref{eqn:relaxedBmiConstriant} is equivalent to the LMI constraint
\begin{equation}
	\label{eqn:relaxedBmiConstriant-LMI}
    \begin{pmatrix}
        1 & \xx & \yy \\
        \xx^\trans & 0 & \quadraticVar \\
        \yy^\trans & \quadraticVar^\trans & 0 
    \end{pmatrix} 
    \spreceq 0~.
\end{equation}%
By adding the LMI~\eqref{eqn:relaxedBmiConstriant-LMI} as an additional constraint to~\eqref{eqn:bmipNaiveRelaxation} and solving the consequent LMI optimization problem, one obtains an upper-bound (on the objective value) of the BMI program~\eqref{eqn:bmip}.
%

\section{Experimental results}\label{sec:experiments}

We have carried out a prototypical implementation\footnote{Available at~\faGithub~\url{https://github.com/Chenms404/BMI-DC}.} of our synthesis techniques in Wolfram \textsc{Mathematica}, which was selected due to its built-in primitives for SDP, polynomial algebra and matrix operations.  Given a safety verification problem as input, our implementation works toward discovering an invariant barrier certificate (in the form of a given template) that witnesses unbounded-time safety of the system. A collection of benchmark examples (detailed in Appendix~\ref{appendix_examples}) has been evaluated on a 2.10GHz 
Xeon processor with 376GB RAM running 64-bit CentOS Linux 7.

\begin{table}[h]
	\caption{Empirical results on benchmark examples (time in seconds)} 
	\label{tab:results}
	\vspace*{-.4cm}
	\begin{center}
		\adjustbox{width=1\textwidth}{
			\begin{tabular}{lcccc>{\centering}m{1mm}rcrc>{\centering}m{1mm}rc>{\centering}m{1mm}rc}
				\toprule
				\multirow{2}{*}{Example name} & &
				\multirow{2}{*}{$n_{\mathsf{sys}}$} &
				\multirow{2}{*}{$d_{\mathsf{flow}}$} &
				\multirow{2}{*}{$d_{\mathsf{BC}}$} & &
				\multicolumn{4}{c}{\toolname{BMI-DC}}& &
				\multicolumn{2}{c}{\toolname{PENLAB}}&  & 
				\multicolumn{2}{c}{\toolname{SOSTOOLS}} \\
				\cmidrule(lr){7-10} \cmidrule(lr){12-13} \cmidrule(lr){15-16}
				& & & & & & \#iter. & & time & validity & & time & validity & & time & validity \\
				\midrule	
				\expname{overview}~\cite{djaballah2017construction}        &   &  2   &  2        & 1   & &   2 &  & \textbf{0.03}  & \cmark & & 0.31  &  \cmark& & 0.07     &      \cmark      \\ 	
				\expname{contrived}          &  &2   &  1        & 2   &  &  0& & \textbf{0.01}  &  \cmark & & 0.48     &     \cmark    &  &   0.75     &    \cmark      \\ 
				\expname{lie-der}~\cite{LZZ11}        &   &  2   &  2        & 1  & &     0 & & \textbf{0.01}   &    \cmark   &    & 0.22        &    \cmark   &  &  0.04    &        \cmark        \\ 
				\expname{lorenz}~\cite{djaballah2017construction}      &     &  3   &  2        & 2  &  &    8 & & \textbf{2.37}  &     \cmark   & & 75.11 &   \xmark   &  &   1.47     &     \xmark          \\ 
				\expname{lti-stable}~\cite{DBLP:conf/cav/GaoKDRSAK19}    &       &  2   &  1        & 2  &   &   0 & & \textbf{0.01}      &   \cmark &   & 0.23               &   \cmark  &   &    0.14    &      \cmark         \\ 
				\expname{lotka-volterra}~\cite{goubault2014finding}       &    &  3   &  2        & 1  &  &    3 & & \textbf{0.07}      &   \cmark  & & 0.36   &    \cmark  &  &    0.21    &       \cmark        \\ 
				
				\expname{clock}~\cite{RatschanS05}      &     &  2   &  3        & 1 &   &    0 & & \textbf{0.01}       &   \cmark   &   & 0.88  &   \xmark   & &    0.18    &        \xmark        \\ 
				\expname{lyapunov}~\cite{ratschan2010providing}              &   &  3   &  3        & 2 &  &     4 & & 1.25       &    \cmark   &   & 56.98  &   \xmark  &  &   \textbf{0.35}     &        \cmark        \\ 
				\expname{arch1}~\cite{sogokon2016non}       &          &  2   &  5        & 2   & &    0 & & \textbf{0.01}      &    \cmark   &   & 33.76  &   \xmark  &   &   0.31     &      \cmark         \\ 
				\expname{arch2}~\cite{sogokon2016non}       &          &  2   &  2        & 2   & &   5 & & \textbf{0.37}       &     \cmark   &  & 0.38  &   \xmark  &   &     0.17   &       \xmark        \\ 
				\expname{arch3}~\cite{sogokon2016non}        &         &  2   &  3        & 2   &  &   1 & & \textbf{0.07}   &    \cmark    &      & 0.54  &  \cmark &  &     0.18   &        \cmark          \\ 
				\expname{arch4}~\cite{sogokon2016non}        &         &  2   &  2        & 1    &  &  2 & & 0.09       &    \cmark   &  & 0.49 &   \xmark  &  &    \textbf{0.06}    &        \cmark        \\ 
				
				\expname{barr-cert1}~\cite{Prajna04}     &      &  2   &  3        & 2 &   &   12  & & \textbf{0.85}   &     \cmark  &  & 2.53    &   \xmark  &  &    0.09    &        \xmark        \\ 
				\expname{barr-cert2}~\cite{djaballah2017construction}     &       &  2   &  2        & 2 &  &     6 & & \textbf{1.57}    &    \cmark  &   & 1.16       &    \xmark &   &    0.15    &      \cmark         \\ 
				\expname{barr-cert3}~\cite{zhang2018safety}        &         &  2   &  2        & 1 &  &     0 & & \textbf{0.01}       &   \cmark    &  & 0.20 &   \cmark &  &    0.11    &         \xmark        \\ 
				\expname{barr-cert4}~\cite{zhang2018safety}        &         &  2   &  3        & 2  & &     13 & & \textbf{0.96}     &      \cmark  &    & 0.89  &  \xmark &  &   0.23     &        \xmark          \\ 	
				
				\expname{fitzhugh-nagumo}~\cite{DBLP:conf/cdc/SassiGS14}        &        &  2  &  3        & 2 &   &    2 & & \textbf{0.16}      &      \cmark  &    &1.24 &  \cmark   &  &    0.25    &     \xmark           \\ 	
				\expname{stabilization}~\cite{DBLP:conf/hybrid/SassiS15}        &        &  3  &  2        & 2 &   &    9 & & 2.88     &      \cmark    &  & 55.22 &   \cmark   & &    \textbf{0.11}    &      \cmark          \\ 
				\expname{lie-high-order}       &   &  2   &  1        & 2  & &     32 & & \textbf{4.12}   &    \cmark   &    &1.56        &   \xmark   &  & 0.25    &       \xmark        \\ 
				\expname{raychaudhuri}~\cite{ferragut2015seeking}        &        &  4  &  2        & 2 &   &    34 & & \textbf{9.51}      &      \cmark &     & 33.64 &   \xmark  &  &   0.14     &      \xmark          \\ 
				
				\expname{focus}~\cite{ratschan2006constraints}         &       &  2  &  1        & 4  & &    100 & & 54.89      &      \xmark    &  & 0.95 &   \xmark  &  &     0.48   &        \xmark        \\ 
				
				\expname{sys-bio1}~\cite{klipp2008systems}         &        &  7   &  2        & 2 &   &    2& &  73.22    &     \namark   &     & 101.95 &  \namark  &  &   1.35     &      \namark           \\ 
				\expname{sys-bio2}~\cite{klipp2008systems}       &         &  9   &  2        & 1  & &     1&  & 1.03       &    \namark  &  & 15.54 &    \namark  &  &      0.16  &       \namark        \\ 
				\expname{quadcopter}~\cite{DBLP:conf/cav/GaoKDRSAK19}       &         &  12  &  1        & 1  &  &    0 & & 0.03      &      \namark   &   & 65.42 &   \namark  &  &   0.36     &     \namark           \\ 
				
				\bottomrule
			\end{tabular}
		}
	\end{center}
	\vspace*{-2mm}
	\scriptsize{$n_{\mathsf{sys}}$: system dimension;  $d_{\mathsf{flow}}$: maximal flow-field degree; $d_{\mathsf{BC}}$: degree of the template barrier certificate.\\
		\#iter.: number of DCP iterations. 0 means that the initial solution (cf.~Section~\ref{subsec:initialSolution}) is valid.\\
		validity: the synthesized barrier certificate is valid (\cmark), invalid (\xmark), or inconclusive within 15 minutes (\namark, beyond the capability of quantifier elimination in \textsc{Mathematica} and nonlinear reasoning in \textsc{Z3}).\\
		time: CPU-time, excluding that for casting the BMIs/LMIs. Boldface marks the winner among \cmark's.
	}
\end{table}

Table~\ref{tab:results} 
reports the empirical results. \toolname{BMI-DC} concerns our locally-convergent Algorithm~\ref{alg:localBMI} for solving BMIs (encoding the sufficient condition in Theorem~\ref{thm:invariantBcSosSufficient}) via the eigendecomposition-based DC decomposition (a comparison to other decomposition methods will be presented later).
We compare our approach with \toolname{PENLAB}~\cite{fiala2013penlab} ---an off-the-shelf solver in \textsc{Matlab} for directly discharging the same BMI problems (with no guarantee on convergence)--- and \toolname{SOSTOOLS}~\cite{sostools2013} ---for solving LMIs derived from Prajna and Jadbabaie's original barrier-certificate condition~\cite{Prajna04}. The comparison is performed under the same problem configurations\footnote{For \toolname{PENLAB} and \toolname{SOSTOOLS}, we use their optimized, built-in criteria for termination and finding initial solutions.}. Due to numerical errors caused by floating-point computations and the fact that reaching the local/global optimum does not necessarily yield a valid barrier certificate, we additionally perform a posterior check, via both the quantifier-elimination procedure in \textsc{Mathematica} and the SMT solver \textsc{Z3}~\cite{z3}, of the synthesized candidate barrier certificate per Definition~\ref{def:invBc}.

\begin{figure}[!t]
	\centering
	\resizebox{\textwidth}{!}{
		\begin{tabular}{ccc}
			\subfloat[\expname{lti-stable}]{~~~~\includegraphics[scale=0.38]{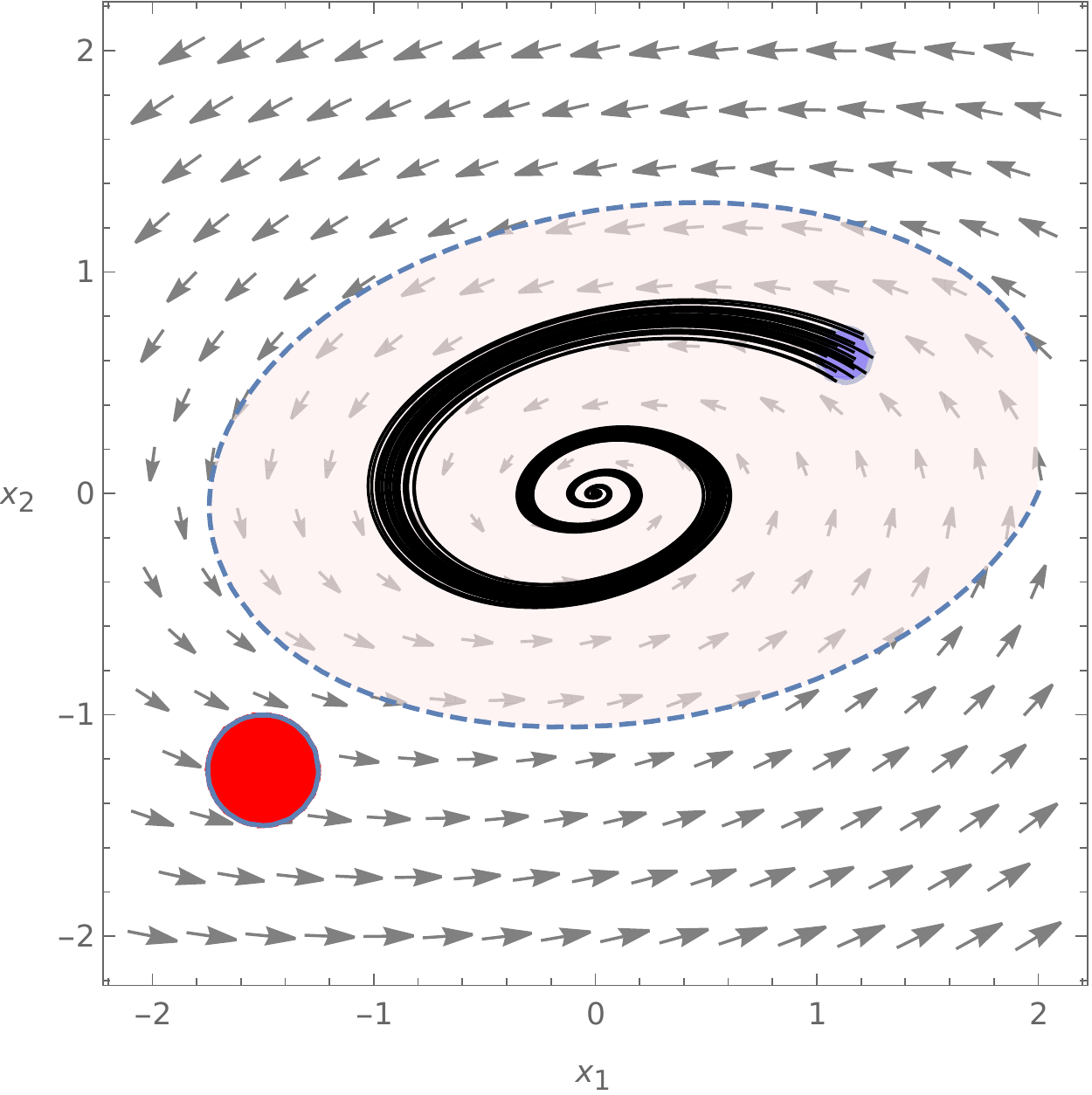}~~~~\label{fig:lti-stable}}&
			\hspace*{.1cm}
			\subfloat[\expname{lyapunov}]{~~~~\includegraphics[scale=0.28]{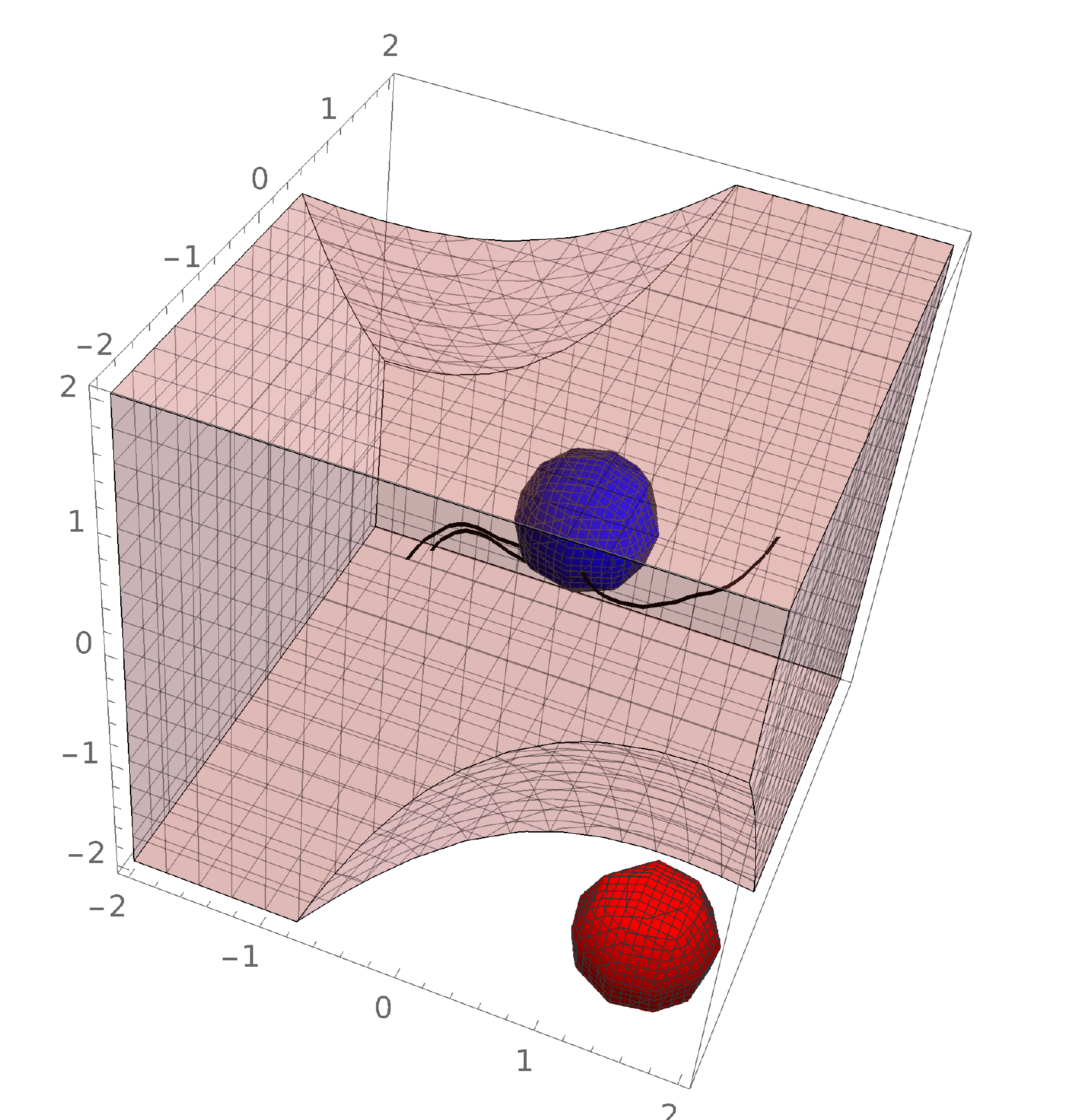}~~~~\label{fig:C2}}
			\hspace*{-.1cm}& 
			\subfloat[\expname{barr-cert2}]{~~~~\includegraphics[scale=0.382]{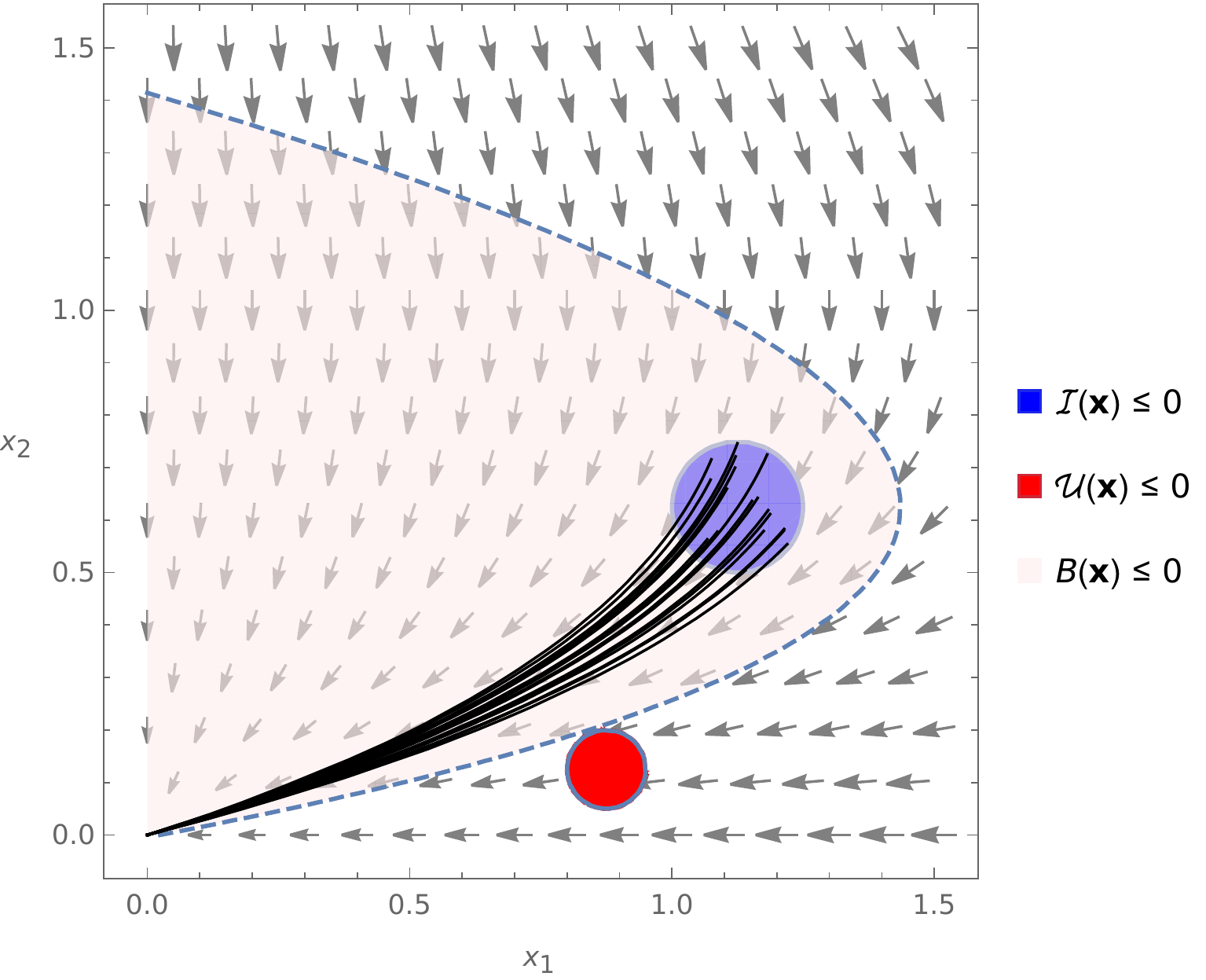}~~~~\label{fig:exmp3}}\\[-.1cm]
			\subfloat[\expname{barr-cert1}]{~~~~\includegraphics[scale=0.38]{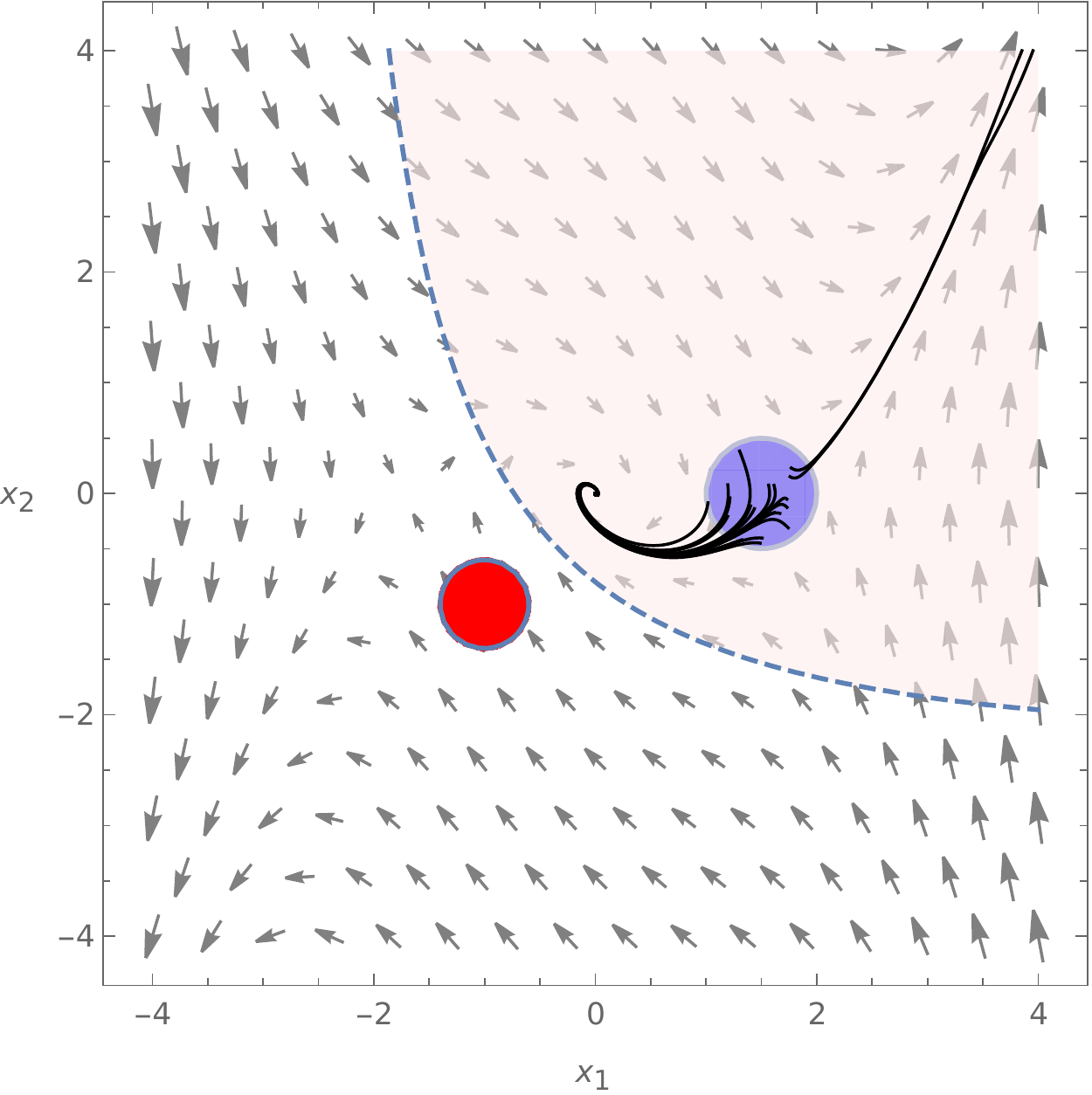}~~~~\label{fig:exmp1}}& 
			\subfloat[\expname{lie-der}]{~~~~\includegraphics[scale=0.38]{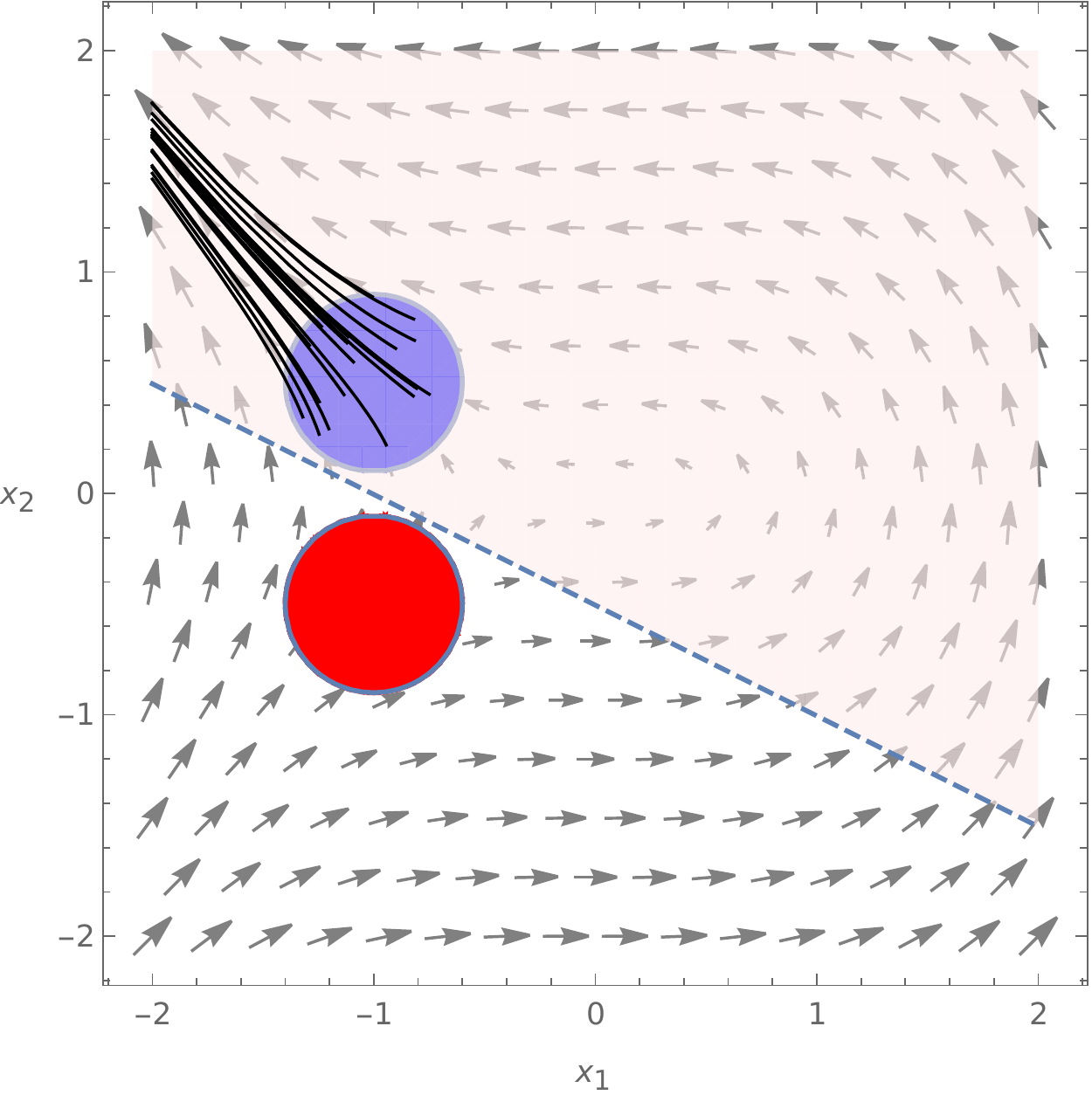}~~~~\label{fig:exmp8}}& 
			\subfloat[\expname{stabilization}]{~~~~\includegraphics[scale=0.382]{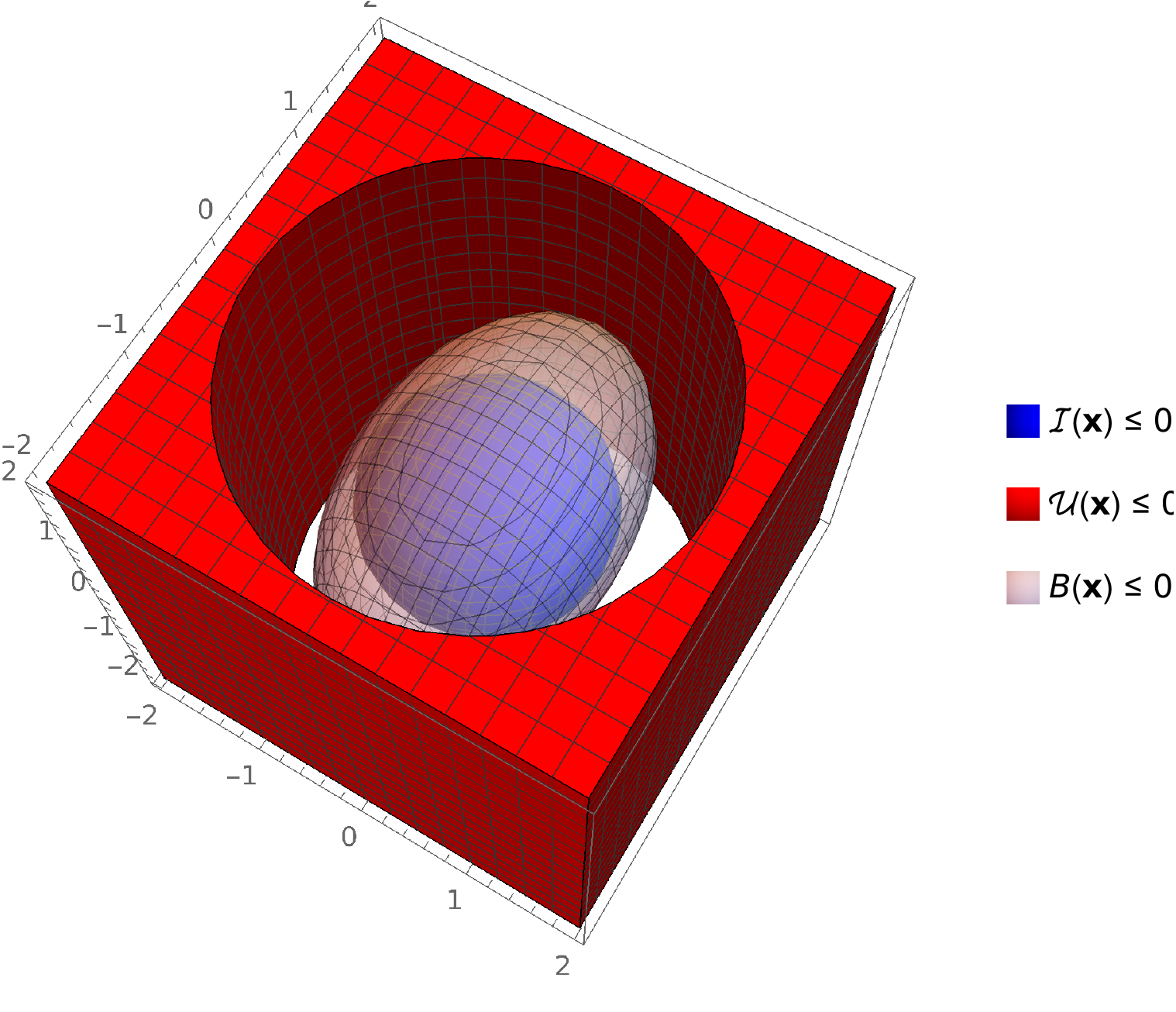}~~~~\label{fig:C14}}
			\\[-.1cm]
			\subfloat[\expname{clock}]{~~~~\includegraphics[scale=0.38]{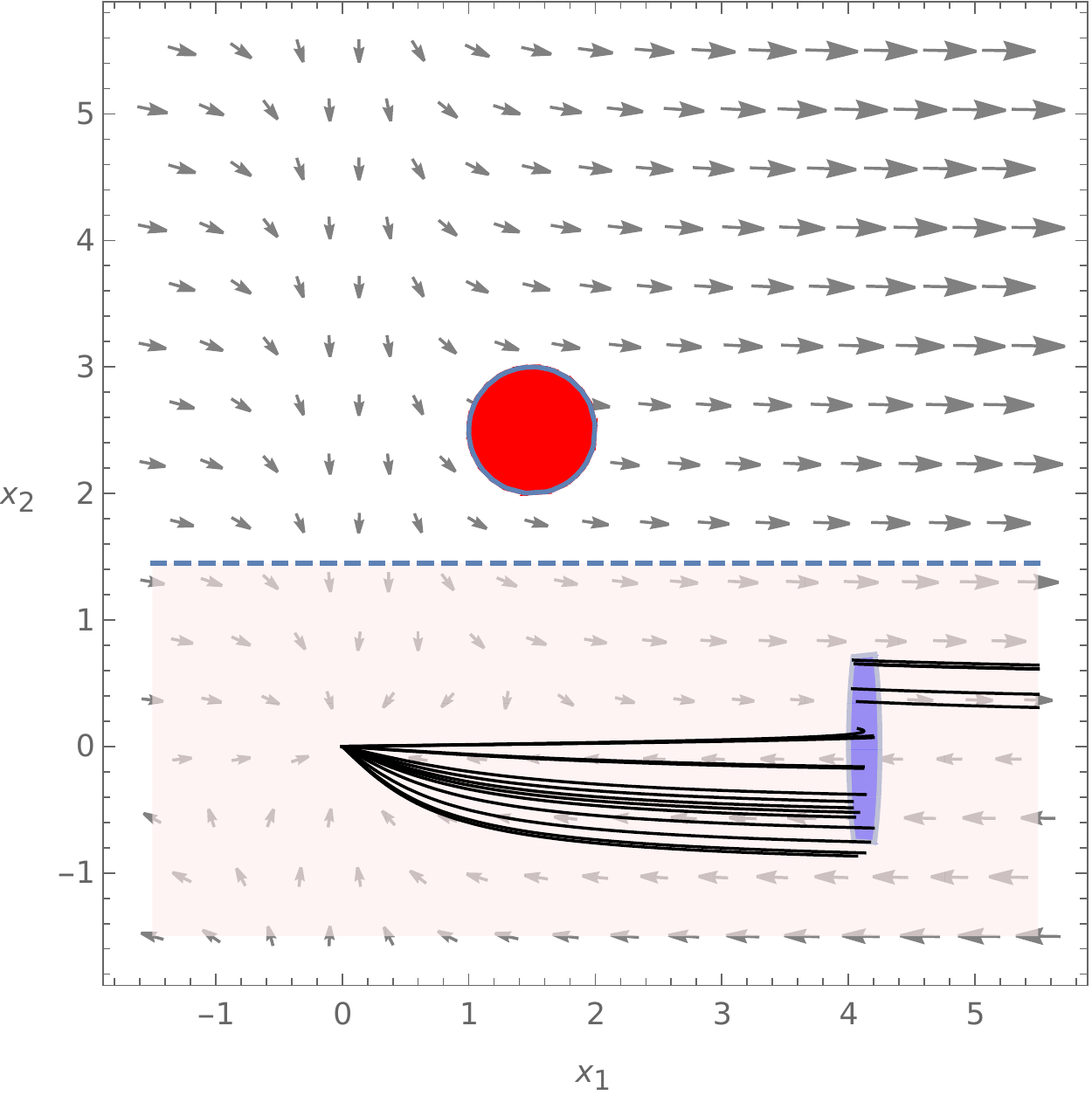}~~~~\label{fig:C1}}& 
			\subfloat[\expname{arch3}]{~~~~\includegraphics[scale=0.38]{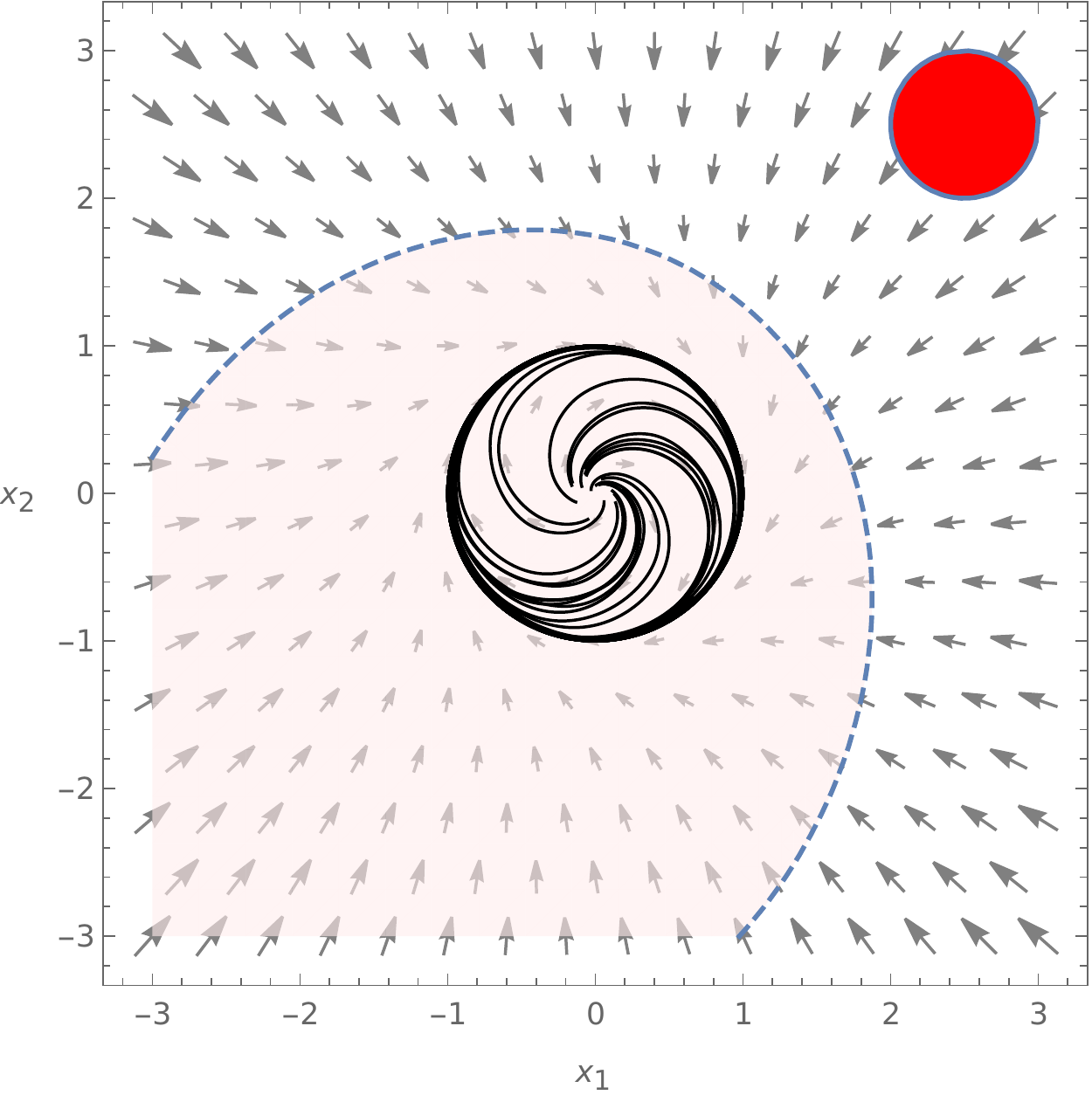}~~~~\label{fig:C5}}& 
			\subfloat[\expname{fitzhugh-nagumo}]{~~~~\includegraphics[scale=0.378]{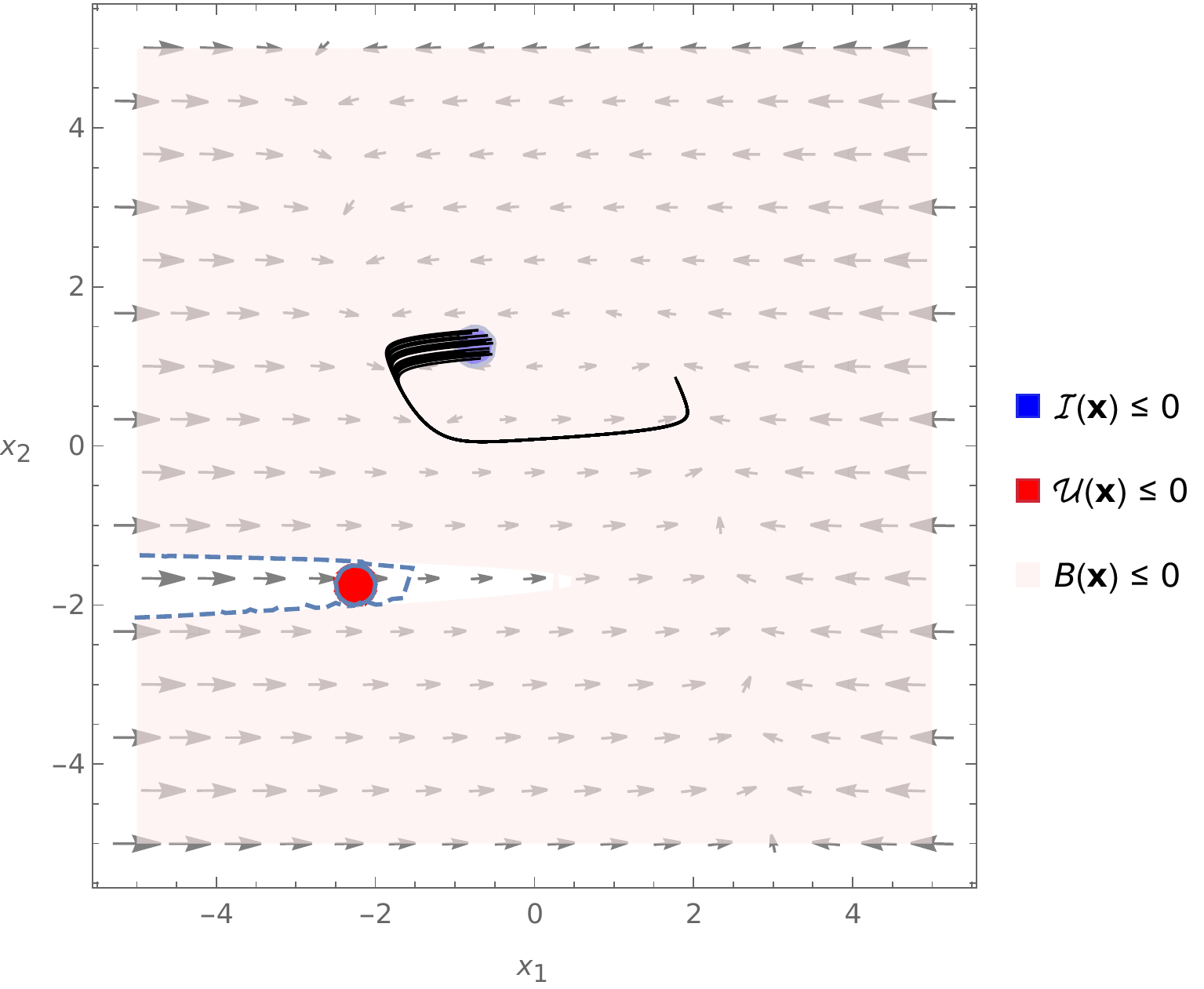}~~~~\label{fig:C12}}\\[-.1cm]
		\end{tabular}
	}
	\caption{Phase portraits of a selected set of examples with the synthesized invariant barrier certificates. The arrows indicate the vector field (hidden in 3D-graphics for a clear presentation) and the solid curves are randomly sampled trajectories.}\label{fig:visualization}
\end{figure}

Table~\ref{tab:results} shows that \toolname{BMI-DC} suffices to synthesize valid barrier certificates in most of the examples within a reasonable number of iterations (i.e., the number of convex sub-problems solved by SDP). This however does not cover all the cases: (1) For the \expname{focus} example, the solution is close enough to a local optimum (after 100 iterations) but yields still an invalid barrier certificate. This problem can be solved (if there exists an invariant barrier certificate as specified) by enforcing the branch-and-bound framework as presented in Section~\ref{sec:bbframework}; (2) For examples \expname{sys-bio1}, \expname{sys-bio2}, and \expname{quadcopter}, neither quantifier elimination in \textsc{Mathematica} nor nonlinear reasoning in \textsc{Z3} can conclude the validity of the synthesized barrier certificates within 15 minutes due to the relatively high system dimensionality (thus marked as \namark; the same applies to \toolname{PENLAB} and \toolname{SOSTOOLS}). The validity for all the other examples is either verified (\cmark) or refuted (\xmark) within 10 seconds. The phase portraits of a selected set of examples and the synthesized invariant barrier certificates are depicted in Fig.~\ref{fig:visualization}.

\paragraph*{\bf Causes of invalid results (\xmark) by \toolname{PENLAB} and \toolname{SOSTOOLS}}
Numerical issues are a common (yet minor) cause of invalid results produced by all the tools in Table~\ref{tab:results}. Whereas the major causes we observed in \toolname{PENLAB} and \toolname{SOSTOOLS} are (1) \toolname{PENLAB} employs non-convex optimization techniques that yield no guarantee on the convergence to local optimums; and (2) \toolname{SOSTOOLS} solves Prajna and Jadbabaie's original, convex barrier-certificate condition~\cite{Prajna04} which is too conservative to recognize the otherwise valid barrier certificates. In fact, most of the invalid results returned by \toolname{SOSTOOLS} have a rather low ``feasibility ratio'' (reported by the underlying SDP solver \toolname{SeDuMi}~\cite{doi:10.1080/10556789908805766}) indicating that \toolname{SOSTOOLS} fails to find barrier certificates adhering to the convex barrier-certificate condition.

\paragraph*{\bf Comparison to \toolname{SOSTOOLS} and \toolname{PENLAB}\footnote{We remark that, even though we perform the comparison under the same problem configurations, it is arguably not a fair comparison in terms of the computation time, as the tools are implemented in different platforms (e.g., \textsc{Mathematica}, \textsc{Matlab}) and rely on different SDP solvers.}}
The comparison in Table~\ref{tab:results} suggests that (1) \emph{Our invariant barrier-certificate condition recognizes more barrier certificates than the original (more conservative) condition as implemented in} \toolname{SOSTOOLS}. In particular, the \expname{lie-high-order} example does admit an inductive invariant in the form of the given template, but none of the existing barrier-certificate conditions~\cite{yang2015exact, zhang2018safety, CAV20BMI} ---concerning Lie derivatives only up to the first order--- recognizes it, since we have $\mathcal{L}_{\ff}^1 B (\xx) = 0$ for some $\xx$ on the boundary of $B$ and hence it requires to exploit the second-order Lie derivative
\footnote{In fact, we have $\LieBound = 2$ for the \expname{lie-high-order} example. For all the other examples in Table~\ref{tab:results}, we either have $\LieBound = 1$ or apply the strengthened consecution condition as described in Remark~\ref{remark:strengthening} with $\mathfrak{R} = 1 < \LieBound$ for efficient synthesis.}; (2) \emph{Our DCP-based synthesis algorithm finds more barrier certificates in less time than directly solving the BMI problems via non-convex optimization techniques as implemented in} \toolname{PENLAB}.

Note that, in our setting, the \emph{volumes} of the invariant sets identified by different approaches are not of primal concern: our goal is to find an invariant that suffices to prove safety of the system instead of a set that ``best'' over- or under-approximates the reachable set (cf.~\cite{DBLP:journals/jns/KordaHM21,DBLP:journals/siamco/MagronGHT19}). However, it would be an interesting future step to investigate the connection between, e.g., robustness, and the volumes of the synthesized invariant sets \`{a} la~\cite{DBLP:journals/siamrev/HenrionLS09,DBLP:journals/automatica/DabbeneHL17}.

We remark that symbolic, monolithic methods based on, e.g., quantifier elimination~\cite{LZZ11} or nonlinear reasoning in SMT, can hardly deal with any of the examples listed in Table~\ref{tab:results} due to the prohibitively high computation complexity. Moreover, it would be desirable to pursue a comparison with the augmented Lagrangian method for solving BMIs as proposed in~\cite{CAV20BMI}, which unfortunately is not yet possible due to the unavailability of the implementation thereof. We will discuss crucial differences to~\cite{CAV20BMI} in Section~\ref{sec:related}.

\begin{figure}[!t]
	\centering
	\resizebox{\textwidth}{!}{
		\begin{tabular}{cc}
			\subfloat[decomposition time for different DC decomposition methods]{
			\includegraphics[width=.5\textwidth]{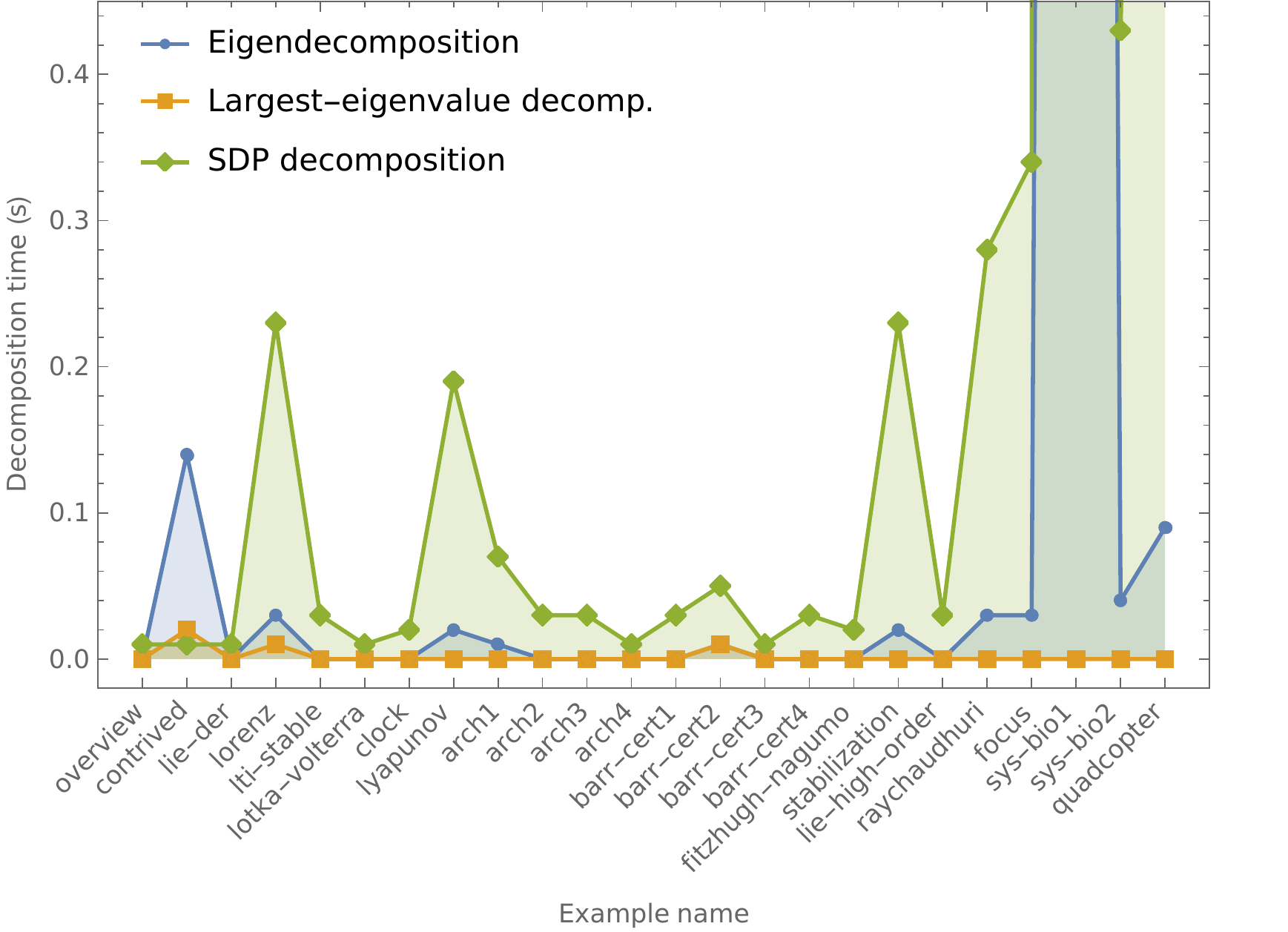}
			\label{decomposition-time}}&
			\subfloat[\#iterations for different DC decomposition methods]{
			\includegraphics[width=.5\textwidth]{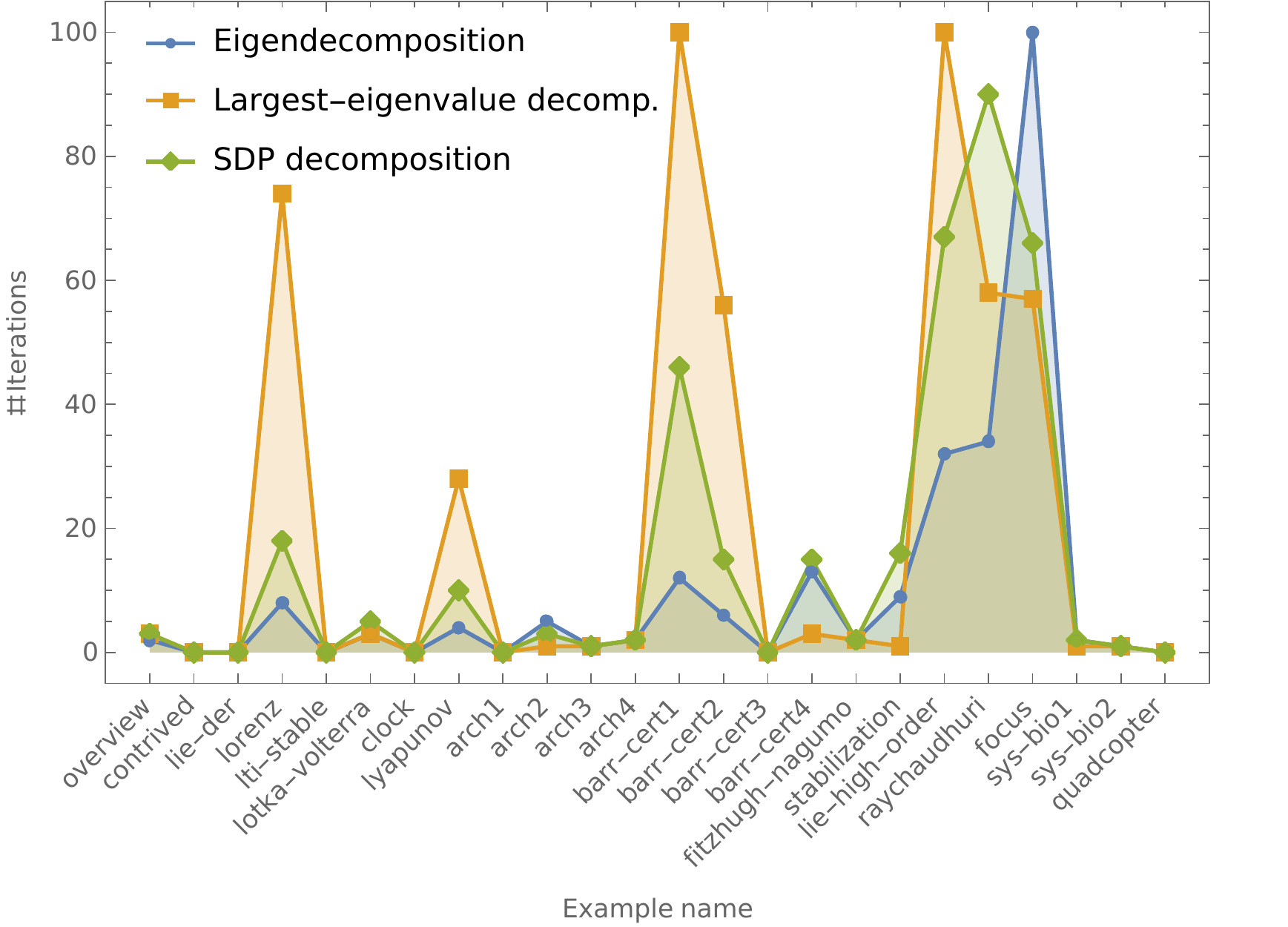}
			\label{fig:decomposition-iteration}}\\[-.1cm]
		\end{tabular}
	}
	\caption{Comparison of the three different DC decomposition methods (see Section~\ref{subsec:dc-decomposition}) in terms of the decomposition time and the number of DCP iterations induced by the decomposition.}\label{fig:decomposition}
\end{figure}

\paragraph*{\bf Comparison between different DC decompositions}
Fig.~\ref{fig:decomposition} depicts a comparison of a naive implementation of the three different DC decomposition methods presented in Section~\ref{subsec:dc-decomposition}. We observe that, in general, (1) the method based on largest eigenvalues enables faster matrix decompositions, but needs more iterations to achieve the desired precisions and yields valid barrier certificates only for 13 out of 24 benchmark examples; (2) the SDP-based method needs a mild amount of iterations (yielding 14/24 valid barrier certificates), but slows down the matrix decompositions (potentially due to the lack of specific sparsity patterns); and (3) the eigendecomposition-based method leads to less number of iterations (yielding 20/24 valid barrier certificates) within a reasonable amount of decomposition time. In summary, there is no clear winner amongst these DC decomposition methods and the implementation can be improved by carefully exploiting the underlying sparsity patterns of the matrices.
\section{Related work}\label{sec:related}

As surveyed in~\cite{DBLP:journals/siglog/Fraenzle19}, the research community has, over the past three decades, extensively addressed the automatic verification of safety-critical hybrid systems. The almost universal undecidability of the unbounded-time reachability problem~\cite{ACH95}, however, confines the sound key-press routines to either semi-decision procedures or approximation schemes, most of which address bounded-time verification by, e.g., computing the finite-time image of a set of initial states.

Invariant generation~\cite{Prajna04,LZZ11}, amongst others, is a well-established approximation scheme that provides a reliable witness for safety (or equivalently, unreachability) of dynamical systems over an infinite time horizon. Invariants can be constructed in various forms, e.g., 
barrier certificates~\cite{Prajna04,Platzer18FM} and differential invariants~\cite{PC08,LZZ11}. With a priori specified templates, the invariant synthesis problem can be reduced to numerical optimizations or constraint solving, as in, e.g.,~\cite{tiwari2003approximate,SSM04a,gulwani2008constraint,kapinski2014simulation}. 

Most pertinently, Prajna and Jadbabaie proposed in their seminal work~\cite{Prajna04} 
a concept coined \emph{barrier certificate} to encode invariants. To enable efficient synthesis using semidefinite programming, the barrier-certificate condition in~\cite{Prajna04} 
strengthens the general condition encoding inductive invariance. 
Since then, significant efforts have been investigated in developing more relaxed (i.e., weaker) forms of barrier-certificate condition that still admit efficient synthesis, thereby leading to, e.g., exponential-type barrier certificates~\cite{Kong13}, Darboux-type barrier certificates~\cite{zeng2016darboux}, general barrier certificates~\cite{Gan17} and vector barrier certificates~\cite{Platzer18FM}. 
Similar barrier-certificate conditions have been explored
to verify systems that address control inputs~\cite{xu2015robustness,ames2016control}, disturbances~\cite{wang2017generating},
and stochastic dynamics~\cite{huang2017probabilistic,jagtap2020formal}.
To attain efficient synthesis, these barrier-certificate conditions share a common property on convexity.
That is, if for some $\aaa_1, \aaa_2 \in \mathbb{R}^m$, $B(\aaa_1, \xx)$ and $B(\aaa_2, \xx)$ both satisfy the barrier-certificate condition, then for any $0 < \mu < 1$, $B(\mu \aaa_1 + (1 - \mu) \aaa_2, \xx)$ must also satisfy the barrier-certificate condition.

However, neither the semantic barrier-certificate condition~\eqref{eqn:semanticBc} encoding the general principle of barrier certificates~\cite{Platzer18FM,Gan17} nor the inductive invariant condition~\eqref{eqn:invariantCondition} is convex. This means, when resorting to convex barrier-certificate conditions, one may miss some potential barrier certificates that suffice as inductive invariants witnessing safety. Therefore, non-convex conditions were suggested~\cite{yang2015exact}, for which the synthesis problem can be reduced to BMI problems solvable via customized schemes, e.g., the augmented Lagrangian method~\cite{CAV20BMI} and the alternating minimization algorithm~\cite{zhang2018safety}. Our synthesis techniques also exploit a BMI reduction, with three crucial differences: (1) our invariant barrier-certificate condition is equivalent to the inductive invariant condition in the sense of Theorem~\ref{thm:inductiveInvariance}, and thus is less conservative than all the aforementioned conditions which consider Lie derivatives only up to the first order; (2) our DCP-based techniques for solving BMIs naturally inherit appealing results on convergence and (weak) completeness, which are not (and can hardly be) provided by the approaches in~\cite{yang2015exact,CAV20BMI,zhang2018safety}; (3) our DCP-based iterative procedure visits only feasible solutions to the original BMI problem, and hence whenever a solution that induces a non-negative objective value is found, we can safely terminate the algorithm and claim a feasible solution to the original BMI problem, which may yield a valid barrier certificate. This is not the case for the approaches in~\cite{yang2015exact,CAV20BMI,zhang2018safety}.

There are recent efforts in synthesizing barrier certificates via machine learning techniques.
Instead of choosing a (polynomial) template and determining the unknown parameters thereof, Zhao et al.~\cite{zhao2020synthesizing} proposes to learn a neural network ---using generated samples from the target system--- 
as a candidate barrier certificate and do posterior verification via, e.g., SMT or interval analysis.
This idea has been further incorporated in a counter-example guided inductive synthesis (CEGIS) framework 
in~\cite{peruffo2021automated, abate2021fossil}. Neural networks in these approaches act as implicit template barrier certificates (with a-priori fixed network structures and activation functions whereas the unknown parameters are the weights to be learnt) which can recognize more complex barrier certificates beyond polynomials. Moreover, applying non-convex barrier-certificate conditions in synthesis does not bring extra overheads to these learning-based approaches.
On the contrary, these approaches cannot guarantee to find a barrier certificate even if there exists one (recognizable by the neural network). Consequently, when the verification fails, one can only resort to supplying the synthesizer with more samples (or heuristically fine-tuning the network and/or the loss function) but no conclusion about the existence of barrier certificates can be drawn. 

Beyond barrier certificates, Wang and Rajamani~\cite{wang2016feasibility} investigated the feasibility problem of general BMI problems with an application to multi-objective nonlinear observer design. The technique of eigendecomposition was also used therein to conduct the DC decomposition. The decomposed concave part, however, is simply ignored and no iterative procedure that exhibits convergence to a local optimum can be provided.

The idea of augmenting a locally-convergent algorithm with a branch-and-bound framework to find the global optimum has been exploited in the realm of optimization~\cite{goh1995global} and control~\cite{tuan2000new}. In contrast, our method is designed for the specific problem of barrier-certificate synthesis, and hence our branch-and-bound algorithm concerns only the parameter space of $\aaa$, i.e., coefficients of the template barrier certificate.

Finally, we refer interested readers to other approaches to solving BMI problems,
e.g., rank minimization~\cite{ibaraki2001rank,orsi2006newton,recht2010guaranteed}, sequential SDP~\cite{correa2004global,eggers2012improving}, as well as methods committed to general non-convex optimizations, e.g., interior point trust-region~\cite{dennis1998trust,leibfritz2002interior,chiu2016method}, successive linearization~\cite{kanzow2005successive} and primal-dual interior point~\cite{yamashita2012local}.

\section{Conclusion}\label{sec:conclusion}
Barrier certificates are a powerful tool to prove time-unbounded safety of hybrid systems. 
We have presented a new condition on barrier certificates ---the invariant barrier-certificate condition, which has been
shown as the weakest possible condition on barrier certificates to attain inductive invariance. We showed that our invariant barrier-certificate condition can be reformulated as an optimization problem subject to bilinear matrix inequalities, which can be solved by our locally-convergent algorithm based on difference-of-convex programming. By incorporating this algorithm into a branch-and-bound framework, we obtained a weak completeness result. 
Experiments on benchmark examples suggested that our invariant barrier-certificate condition recognizes more barrier certificates than existing conditions, and that our DCP-based algorithm is more efficient than directly solving the BMIs via off-the-shelf solvers. 

We stress that our techniques for solving BMIs are of a general nature rather than being confined to barrier-certificate synthesis. Interesting future directions include to extend our method to other synthesis problems, e.g., discovering invariants and/or termination proofs of deterministic/probabilistic programs. 


\paragraph*{\bf Acknowledgements}
The authors would like to thank Hengjun Zhao for the fruitful discussion on differential dynamics requiring high-order Lie derivatives. 

\bibliography{ref}

\newpage
\begin{appendices}

%

\section{Lie derivatives and the trajectory tendency}\label{appendix_lie}

\vspace*{-.2cm}
\begin{figure}[h]
	\centering
	\resizebox{\textwidth}{!}{
		\begin{tabular}{cc}
			\subfloat[first-order Lie derivative and the gradient]{~~~~~~
				\begin{tikzpicture}
					\draw (0, 0) node[inner sep=0]	{\includegraphics[width=.43\textwidth]{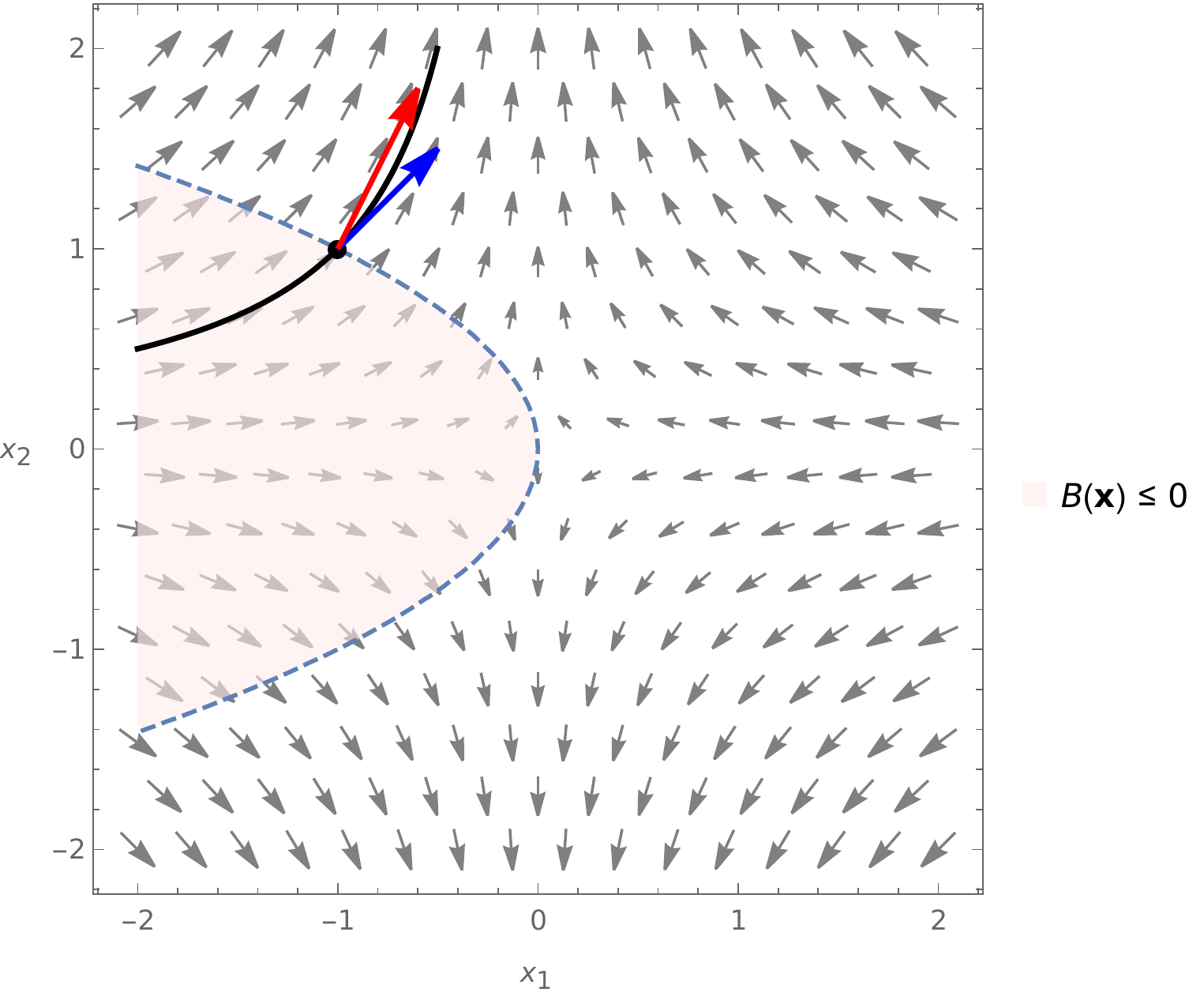}};
					\draw (-0.82, 2.65) node {$\sol$};
					\draw (-0.82, 2.05) node {\textcolor{blue}{$\vv$}};
					\draw (-1.32, 2.35) node {\textcolor{red}{$\uu$}};
				\end{tikzpicture}
				~~\label{fig:lie-illustration1}}&
			\subfloat[demand for the second-order Lie derivative]{~~
				\begin{tikzpicture}
					\draw (0, 0) node[inner sep=0]	{\includegraphics[width=.43\textwidth]{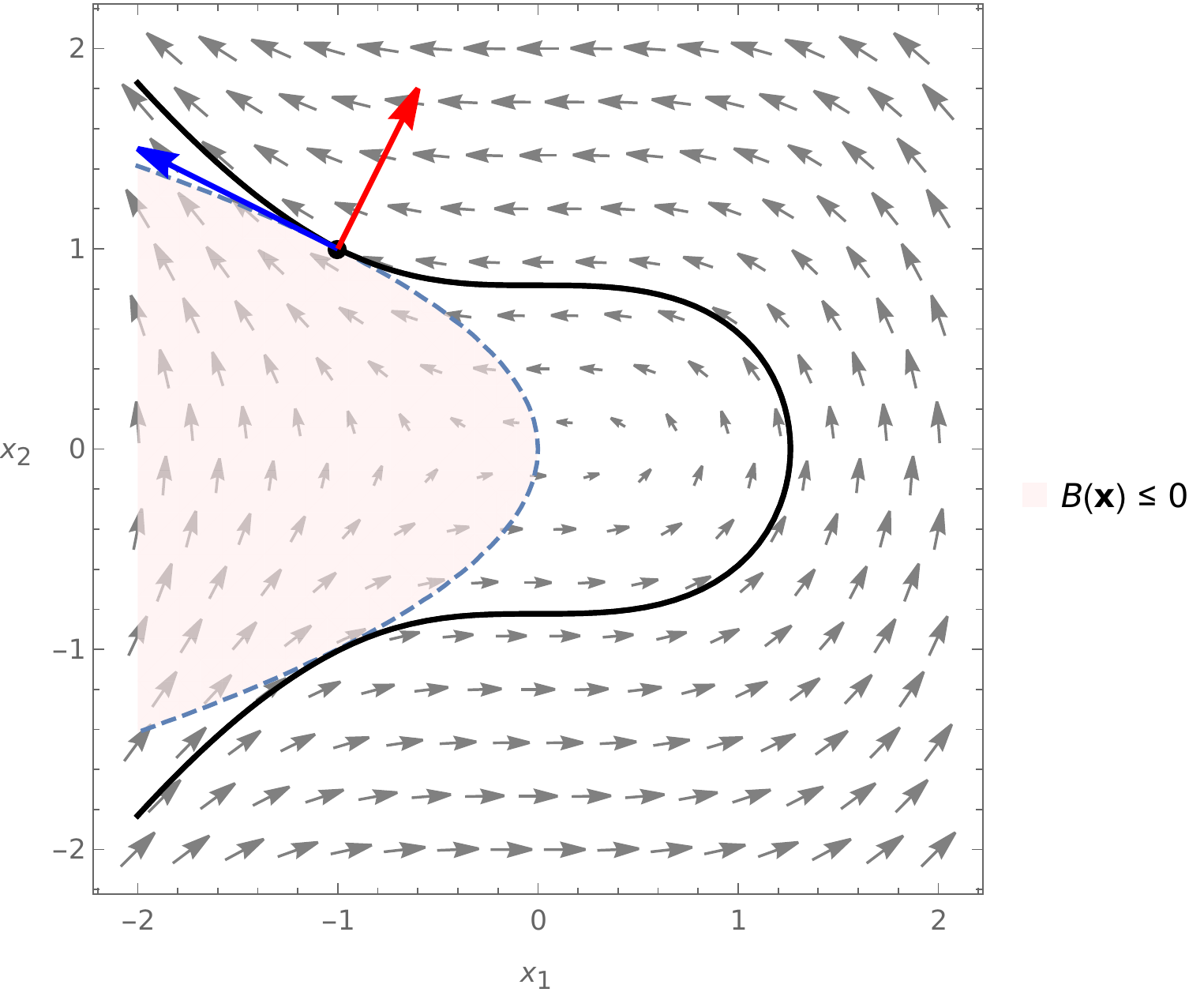}};
					\draw (0.60, 0.82) node {$\sol$};
					\draw (-2.85, 2.14) node {\textcolor{blue}{$\vv$}};
					\draw (-0.86, 2.45) node {\textcolor{red}{$\uu$}};
				\end{tikzpicture}
				~~~~~~\label{fig:lie-illustration2}}\\[-.1cm]
		\end{tabular}
	}
	\caption{An illustration of how Lie derivatives capture the tendency of trajectories in terms of a polynomial function $B(\xx)$. $\sol$: the system trajectory passing through $(-1,1)$; \textcolor{blue}{$\vv$}: the evolution direction per the vector field at $(-1,1)$; \textcolor{red}{$\uu$}: the gradient of $B(\xx)$ at $(-1,1)$.}\label{fig:lie-illustration}
\end{figure}

\begin{example}[Lie derivatives~\textnormal{\cite{LZZ11}}]\label{exmp:lie-illust}
	Let $B(\xx) = x_1 + x_2^2$. Consider the vector field $\ff = (-x_1, x_2)$ as depicted in Fig.~\ref{fig:lie-illustration1}. By Definition~\ref{def:Lie-Der}, we have $\mathcal{L}_{\ff}^0 B(\xx) = x_1 + x_2^2$ and $\mathcal{L}_{\ff}^1 B(\xx) = -x_1 + 2x_2^2$. We exemplify with the point $\xx = (-1,1)$ on the parabola $B(\xx) = x_1 + x_2^2$ that $\mathcal{L}_{\ff}^1 B\big\vert_{(-1,1)} = 3 > 0$ reveals the fact that the system trajectory $\sol$ passing through $(-1,1)$ will escape from the region $B(\xx) \leq 0$. In Fig.~\ref{fig:lie-illustration1}, the vector $\vv = (1,1)$ points to the evolution direction per $\ff = (-x_1, x_2)$, and the vector $\uu = \frac{\partial}{\partial \xx} B\big\vert_{(-1,1)} = (1,2)$ denotes the gradient of $B(\xx)$ at $(-1,1)$. These two vectors together assert that the trajectory $\sol$ will enter the region $B(\xx) > 0$ immediately after passing through $(-1,1)$ since the angle formed by $\uu$ and $\vv$ is less than $\sfrac{\pi}{2}$, that is, the first-order Lie derivative $\mathcal{L}_{\ff}^1 B\big\vert_{(-1,1)} = 3$ is positive. Dually, a negative first-order Lie derivative will witness the crossings of a trajectory from the region $B(\xx) > 0$ to the region $B(\xx) \leq 0$.
	
	However, if the angle between the evolution direction $\vv$ and the gradient $\uu$ is $\sfrac{\pi}{2}$ or the gradient is a zero vector, then it is  impossible to read off the trajectory tendency via the consequent zero first-order Lie derivative. In this case, we resort to non-zero higher-order Lie derivatives: Consider another vector field $\ff' = (-2x_2, x_1^2)$ as depicted in Fig.~\ref{fig:lie-illustration2} with the same function $B(\xx)$. We have $\mathcal{L}_{\ff'}^0 B(\xx) = x_1 + x_2^2$ and $\mathcal{L}_{\ff'}^1 B(\xx) = 2 x_1^2 x_2-2 x_2$, where $\mathcal{L}_{\ff'}^1 B\big\vert_{(-1,1)} = 0$ as the evolution direction $\vv$ is perpendicular to the gradient $\uu$. However, since the second-order Lie derivative $\mathcal{L}_{\ff'}^2 B(\xx) = 2 x_1^4-2 x_1^2-8 x_1 x_2^2$ at $(-1,1)$ is positive, we can conclude that the trajectory passing through $(-1,1)$ will enter the region $B(\xx) > 0$. Notice that, to determine the trajectory tendency, we need to consider Lie derivatives only up to a certain order (as asserted by Theorem~\ref{thm:invariantCondition}), e.g., $2$ in this example.
	\qedTT
\end{example}
%
%

\section{Benchmark examples}\label{appendix_examples}

\begin{example}[\expname{contrived}]
    The vector flow field is: 
    \begin{equation*}
	\dot{\xx} =
    \begin{pmatrix}
        \dot{x}_1 \\
        \dot{x}_2
    \end{pmatrix} 
    =
    \begin{pmatrix}
        -x_1 + x_2 \\
        -x_2 
    \end{pmatrix}
	~.
\end{equation*}

\begin{itemize}
    \item $\init = \{ \xx \in \mathbb{R}^2 \mid (x_1 - 1.125)^2 + (x_2 - 0.625)^2 - 0.0125 \leq 0 \}$. 
    \item $\unsafe = \{ \xx \in \mathbb{R}^2 \mid (x_1 - 0.875)^2 + (x_2 - 0.125)^2 - 0.0125 \leq 0 \}$. 
    \item $\domain = \{ \xx \in \mathbb{R}^2 \mid 0 \leq x_1, x_2 \leq 2 \}$. 
    \item $B(\aaa, \xx)$ includes all monomials up to degree $2$. 
\end{itemize}
\end{example}

\begin{example}[\expname{lie-der}~\textnormal{\cite{LZZ11}}]
    The vector flow field is: 
    \begin{equation*}
	\dot{\xx} =
    \begin{pmatrix}
        \dot{x}_1 \\
        \dot{x}_2
    \end{pmatrix} 
    =
    \begin{pmatrix}
        -2 x_2 \\
        x_1^2
    \end{pmatrix}
	~.
\end{equation*}

\begin{itemize}
    \item $\init = \{ \xx \in \mathbb{R}^2 \mid (x_1 + 1)^2 + (x_2 - 0.5)^2 - 0.16 \leq 0 \}$. 
    \item $\unsafe = \{ \xx \in \mathbb{R}^2 \mid (x_1 + 1)^2 + (x_2 + 0.5)^2 - 0.16 \leq 0 \}$. 
    \item $\domain = \{ \xx \in \mathbb{R}^2 \mid -2 \leq x_1, x_2 \leq 2 \}$. 
    \item $B(\aaa, \xx)$ includes all monomials up to degree $1$. 
\end{itemize}
\end{example}

\begin{example}[\expname{lorenz}~\textnormal{\cite{djaballah2017construction}}]
    The vector flow field is: 
    \begin{equation*}
	\dot{\xx} =
    \begin{pmatrix}
        \dot{x}_1 \\
        \dot{x}_2 \\
        \dot{x}_3
    \end{pmatrix} 
    =
    \begin{pmatrix}
        10.0 (-x_1 + x_2) \\
        -x_2 + x_1 (28.0 - x_3) \\
        x_1 x_2 - \frac{8}{3} x_3
    \end{pmatrix}
	~.
\end{equation*}

\begin{itemize}
    \item $\init = \{ \xx \in \mathbb{R}^3 \mid (x_1 + 14.5)^2 + (x_2 + 14.5)^2 + (x_3 - 12.5)^2 - 0.25 \leq 0 \}$. 
    \item $\unsafe = \{ \xx \in \mathbb{R}^3 \mid (x_1 + 16.5)^2 + (x_2 + 14.5)^2 + (x_3 - 2.5)^2 - 0.25 \leq 0 \}$. 
    \item $\domain = \{ \xx \in \mathbb{R}^3 \mid -20 \leq x_1, x_2, x_3 \leq 20 \}$. 
    \item $B(\aaa, \xx)$ includes all monomials up to degree $2$. 
\end{itemize}
\end{example}

\begin{example}[\expname{lti-stable}~\textnormal{\cite{DBLP:conf/cav/GaoKDRSAK19}}]
    The vector flow field is: 
    \begin{equation*}
	\dot{\xx} =
    \begin{pmatrix}
        \dot{x}_1 \\
        \dot{x}_2
    \end{pmatrix} 
    =
    \begin{pmatrix}
        -0.1 x_1 - 10 x_2 \\
        4 x_1 - 2 x_2 
    \end{pmatrix}
	~.
\end{equation*}

\begin{itemize}
    \item $\init = \{ \xx \in \mathbb{R}^2 \mid (x_1 - 1.125)^2 + (x_2 - 0.625)^2 - 0.125^2 \leq 0 \}$. 
    \item $\unsafe = \{ \xx \in \mathbb{R}^2 \mid (x_1 + 1.5)^2 + (x_2 + 1.25)^2 - 0.25^2 \leq 0 \}$. 
    \item $\domain = \{ \xx \in \mathbb{R}^2 \mid -2 \leq x_1, x_2 \leq 2 \}$. 
    \item $B(\aaa, \xx)$ includes all monomials up to degree $2$. 
\end{itemize}
\end{example}

\begin{example}[\expname{lotka-volterra}~\textnormal{\cite{goubault2014finding}}]
    The vector flow field is: 
    \begin{equation*}
	\dot{\xx} =
    \begin{pmatrix}
        \dot{x}_1 \\
        \dot{x}_2 \\
        \dot{x}_3
    \end{pmatrix} 
    =
    \begin{pmatrix}
        x_1 (1 - x_3) \\
        x_2 (1 - 2 x_3) \\
        x_3 (-1 + x_1 + x_2)
    \end{pmatrix}
	~.
\end{equation*}

\begin{itemize}
    \item $\init = \{ \xx \in \mathbb{R}^3 \mid (x_1 - 1)^2 + (x_2 - 1)^2 + x_3^2 - 0.64 \leq 0 \}$. 
    \item $\unsafe = \{ \xx \in \mathbb{R}^3 \mid x_1^2 + (x_2 + 1)^2 - 0.25 \leq 0 \}$. 
    \item $\domain = \{ \xx \in \mathbb{R}^3 \mid -2 \leq x_1, x_2, x_3 \leq 2 \}$. 
    \item $B(\aaa, \xx) = a x_2$. 
\end{itemize}
\end{example}

\begin{example}[\expname{clock}~\textnormal{\cite{RatschanS05}}]
    The vector flow field is: 
    \begin{equation*}
	\dot{\xx} =
    \begin{pmatrix}
        \dot{x}_1 \\
        \dot{x}_2
    \end{pmatrix} 
    =
    \begin{pmatrix}
        -x_1 + 2 x_1^2 x_2 \\
        -x_2 
    \end{pmatrix}
	~.
\end{equation*}

\begin{itemize}
    \item $\init = \{ \xx \in \mathbb{R}^2 \mid (8 x_1 - 33)^2 + x_2^2 - 1 \leq 0 \}$. 
    \item $\unsafe = \{ \xx \in \mathbb{R}^2 \mid (x_1 - 1.5)^2 + (x_2 - 2.5)^2 - 0.25 \leq 0 \}$. 
    \item $\domain = \{ \xx \in \mathbb{R}^2 \mid -1.5 \leq x_1, x_2 \leq 5.5 \}$. 
    \item $B(\aaa, \xx)$ includes all monomials up to degree $1$. 
\end{itemize}
\end{example}

\begin{example}[\expname{lyapunov}~\textnormal{\cite{ratschan2010providing}}]
    The vector flow field is: 
    \begin{equation*}
	\dot{\xx} =
    \begin{pmatrix}
        \dot{x}_1 \\
        \dot{x}_2 \\
        \dot{x}_3
    \end{pmatrix} 
    =
    \begin{pmatrix}
        -x_2 \\
        -x_3 \\
        -x_1 - 2 x_2 - x_3 + x_1^3 
    \end{pmatrix}
	~.
\end{equation*}

\begin{itemize}
    \item $\init = \{ \xx \in \mathbb{R}^3 \mid (x_1 - 0.25)^2 + (x_2 - 0.25)^2 + (x_3 - 0.25)^2 - 0.25 \leq 0 \}$. 
    \item $\unsafe = \{ \xx \in \mathbb{R}^3 \mid (x_1 - 1.5)^2 + (x_2 + 1.5)^2 + (x_3 + 1.5)^2 - 0.25 \leq 0 \}$. 
    \item $\domain = \{ \xx \in \mathbb{R}^3 \mid -2 \leq x_1, x_2, x_3 \leq 2 \}$. 
    \item $B(\aaa, \xx)$ includes all monomials up to degree $2$. 
\end{itemize}
\end{example}

\begin{example}[\expname{arch1}~\textnormal{\cite{sogokon2016non}}]
    The vector flow field is: 
    \begin{equation*}
	\dot{\xx} =
    \begin{pmatrix}
        \dot{x}_1 \\
        \dot{x}_2
    \end{pmatrix} 
    =
    \begin{pmatrix}
        -x_1 + 2 x_1^3 x_2^2 \\
        -x_2 
    \end{pmatrix}
	~.
\end{equation*}

\begin{itemize}
    \item $\init = \{ \xx \in \mathbb{R}^2 \mid x_1^2 + (x_2 - 0.5)^2 - 0.04 \leq 0 \}$. 
    \item $\unsafe = \{ \xx \in \mathbb{R}^2 \mid (x_1 + 1.5)^2 + (x_2 + 1.5)^2 - 0.25 \leq 0 \}$. 
    \item $\domain = \{ \xx \in \mathbb{R}^2 \mid -2 \leq x_1, x_2 \leq 2 \}$. 
    \item $B(\aaa, \xx)$ includes all monomials up to degree $2$. 
\end{itemize}
\end{example}

\begin{example}[\expname{arch2}~\textnormal{\cite{sogokon2016non}}]
    The vector flow field is: 
    \begin{equation*}
	\dot{\xx} =
    \begin{pmatrix}
        \dot{x}_1 \\
        \dot{x}_2
    \end{pmatrix} 
    =
    \begin{pmatrix}
        x_1^2 + x_2^2 - 1\\
        5(x_1 x_2 - 1)
    \end{pmatrix}
	~.
\end{equation*}

\begin{itemize}
    \item $\init = \{ \xx \in \mathbb{R}^2 \mid (x_1 + 0.5)^2 + (x_2 + 0.5)^2 - 0.25 \leq 0 \}$. 
    \item $\unsafe = \{ \xx \in \mathbb{R}^2 \mid (x_1 + 1.5)^2 + (x_2 + 1.5)^2 - 0.25 \leq 0 \}$. 
    \item $\domain = \{ \xx \in \mathbb{R}^2 \mid -2 \leq x_1, x_2 \leq 2 \}$. 
    \item $B(\aaa, \xx)$ includes all monomials up to degree $2$. 
\end{itemize}
\end{example}

\begin{example}[\expname{arch3}~\textnormal{\cite{sogokon2016non}}]
    The vector flow field is: 
    \begin{equation*}
	\dot{\xx} =
    \begin{pmatrix}
        \dot{x}_1 \\
        \dot{x}_2
    \end{pmatrix} 
    =
    \begin{pmatrix}
        x_1 - x_1^3 + x_2 - x_1 x_2^2 \\
        -x_1 + x_2 - x_1^2 x_2 - x_2^3
    \end{pmatrix}
	~.
\end{equation*}

\begin{itemize}
    \item $\init = \{ \xx \in \mathbb{R}^2 \mid x_1^2 + x_2^2 - 0.04 \leq 0 \}$. 
    \item $\unsafe = \{ \xx \in \mathbb{R}^2 \mid (x_1 - 2.5)^2 + (x_2 - 2.5)^2 - 0.25 \leq 0 \}$. 
    \item $\domain = \{ \xx \in \mathbb{R}^2 \mid -3 \leq x_1, x_2 \leq 3 \}$. 
    \item $B(\aaa, \xx)$ includes all monomials up to degree $2$. 
\end{itemize}
\end{example}

\begin{example}[\expname{arch4}~\textnormal{\cite{sogokon2016non}}]
    The vector flow field is: 
    \begin{equation*}
	\dot{\xx} =
    \begin{pmatrix}
        \dot{x}_1 \\
        \dot{x}_2
    \end{pmatrix} 
    =
    \begin{pmatrix}
        -2 x_1 + x_1^2 + x_2 \\
        x_1 - 2 x_2 + x_2^2
    \end{pmatrix}
	~.
\end{equation*}

\begin{itemize}
    \item $\init = \{ \xx \in \mathbb{R}^2 \mid x_1^2 + x_2^2 - 0.1^2 \leq 0 \}$. 
    \item $\unsafe = \{ \xx \in \mathbb{R}^2 \mid (x_1 - 0.75)^2 + (x_2 - 0.75)^2 - 0.25^2 \leq 0 \}$. 
    \item $\domain = \{ \xx \in \mathbb{R}^2 \mid -0.5 \leq x_1, x_2 \leq 1 \}$. 
    \item $B(\aaa, \xx)$ includes all monomials up to degree $1$. 
\end{itemize}
\end{example}

\begin{example}[\expname{barr-cert1}~\textnormal{\cite{Prajna04}}]
    The vector flow field is: 
    \begin{equation*}
	\dot{\xx} =
    \begin{pmatrix}
        \dot{x}_1 \\
        \dot{x}_2
    \end{pmatrix} 
    =
    \begin{pmatrix}
        x_2 \\
        -x_1 + \frac{1}{3} x_1^3 - x_2 
    \end{pmatrix}
	~.
\end{equation*}

\begin{itemize}
    \item $\init = \{ \xx \in \mathbb{R}^2 \mid (x_1 - 1.5)^2 + x_2^2 - 0.25 \leq 0 \}$. 
    \item $\unsafe = \{ \xx \in \mathbb{R}^2 \mid (x_1 + 1)^2 + (x_2 + 1)^2 - 0.16 \leq 0 \}$. 
    \item $\domain = \{ \xx \in \mathbb{R}^2 \mid -4 \leq x_1, x_2 \leq 4 \}$. 
    \item $B(\aaa, \xx)$ includes all monomials up to degree $2$. 
\end{itemize}
\end{example}

\begin{example}[\expname{barr-cert2}~\textnormal{\cite{djaballah2017construction}}]
    The vector flow field is: 
    \begin{equation*}
	\dot{\xx} =
    \begin{pmatrix}
        \dot{x}_1 \\
        \dot{x}_2
    \end{pmatrix} 
    =
    \begin{pmatrix}
        -x_1 + x_1 x_2 \\
        -x_2 
    \end{pmatrix}
	~.
\end{equation*}

\begin{itemize}
    \item $\init = \{ \xx \in \mathbb{R}^2 \mid (x_1 - 1.125)^2 + (x_2 - 0.625)^2 - 0.125^2 \leq 0 \}$. 
    \item $\unsafe = \{ \xx \in \mathbb{R}^2 \mid (x_1 - 0.875)^2 + (x_2 - 0.125)^2 - 0.075^2 \leq 0 \}$. 
    \item $\domain = \{ \xx \in \mathbb{R}^2 \mid 0 \leq x_1, x_2 \leq 1.5 \}$. 
    \item $B(\aaa, \xx)$ includes all monomials up to degree $2$. 
\end{itemize}
\end{example}

\begin{example}[\expname{barr-cert3}~\textnormal{\cite{zhang2018safety}}]
    The vector flow field is: 
    \begin{equation*}
	\dot{\xx} =
    \begin{pmatrix}
        \dot{x}_1 \\
        \dot{x}_2
    \end{pmatrix} 
    =
    \begin{pmatrix}
        -x_1 + x_1 x_2 \\
        -x_2 
    \end{pmatrix}
	~.
\end{equation*}

\begin{itemize}
    \item $\init = \{ \xx \in \mathbb{R}^2 \mid (x_1 + 1)^2 + (x_2 + 1)^2 - 0.25 \leq 0 \}$. 
    \item $\unsafe = \{ \xx \in \mathbb{R}^2 \mid x_1^2 + (x_2 - 1)^2 - 0.25 \leq 0 \}$. 
    \item $\domain = \{ \xx \in \mathbb{R}^2 \mid -2 \leq x_1, x_2 \leq 2 \}$. 
    \item $B(\aaa, \xx)$ includes all monomials up to degree $1$. 
\end{itemize}
\end{example}

\begin{example}[\expname{barr-cert4}~\textnormal{\cite{zhang2018safety}}]
    The vector flow field is: 
    \begin{equation*}
	\dot{\xx} =
    \begin{pmatrix}
        \dot{x}_1 \\
        \dot{x}_2
    \end{pmatrix} 
    =
    \begin{pmatrix}
        -x_1 + 2 x_1^2 x_2 \\
        -x_2 
    \end{pmatrix}
	~.
\end{equation*}

\begin{itemize}
    \item $\init = \{ \xx \in \mathbb{R}^2 \mid 9 x_1^2 + (2 x_2 - 2.25)^2 - 0.75^2 \leq 0 \}$. 
    \item $\unsafe = \{ \xx \in \mathbb{R}^2 \mid (x_1 - 2)^2 + (x_2 - 2)^2 - 0.5^2 \leq 0 \}$. 
    \item $\domain = \{ \xx \in \mathbb{R}^2 \mid -1 \leq x_1, x_2 \leq 3 \}$. 
    \item $B(\aaa, \xx)$ includes all monomials up to degree $2$. 
\end{itemize}
\end{example}

\begin{example}[\expname{fitzhugh-nagumo}~\textnormal{\cite{DBLP:conf/cdc/SassiGS14}}]
    The vector flow field is: 
    \begin{equation*}
	\dot{\xx} =
    \begin{pmatrix}
        \dot{x}_1 \\
        \dot{x}_2
    \end{pmatrix} 
    =
    \begin{pmatrix}
        -1/3 x_1^3 + x_1 - x_2 + 0.875 \\
        0.08 (x_1 - 0.8 x_2 + 0.7)
    \end{pmatrix}
	~.
\end{equation*}

\begin{itemize}
    \item $\init = \{ \xx \in \mathbb{R}^2 \mid (x_1 + 0.75)^2 + (x_2 -1.25)^2 - 0.25^2 \leq 0 \}$. 
    \item $\unsafe = \{ \xx \in \mathbb{R}^2 \mid (x_1 + 2.25)^2 + (x_2 + 1.75)^2 - 0.25^2 \leq 0 \}$. 
    \item $\domain = \{ \xx \in \mathbb{R}^2 \mid -5 \leq x_1, x_2 \leq 5 \}$. 
    \item $B(\aaa, \xx)$ includes all monomials up to degree $2$. 
\end{itemize}
\end{example}

\begin{example}[\expname{stabilization}~\textnormal{\cite{DBLP:conf/hybrid/SassiS15}}]
    The vector flow field is: 
    \begin{equation*}
	\dot{\xx} =
    \begin{pmatrix}
        \dot{x}_1 \\
        \dot{x}_2 \\
        \dot{x}_3
    \end{pmatrix} 
    =
    \begin{pmatrix}
        -x_1 + x_2 - x_3 \\
        -x_1 (x_3 + 1) - x_2 \\
        0.76524 x_1 - 4.7037 x_3
    \end{pmatrix}
	~.
\end{equation*}

\begin{itemize}
    \item $\init = \{ \xx \in \mathbb{R}^3 \mid x_1^2 + x_2^2 + x_3^2 - 1 \leq 0 \}$. 
    \item $\unsafe = \{ \xx \in \mathbb{R}^3 \mid -x_1^2 - x_2^2 + 3 \leq 0 \}$. 
    \item $\domain = \{ \xx \in \mathbb{R}^3 \mid -2 \leq x_1, x_2, x_3 \leq 2 \}$. 
    \item $B(\aaa, \xx)$ includes all monomials up to degree $2$. 
\end{itemize}
\end{example}

\begin{example}[\expname{lie-high-order}]
	The vector flow field is: 
	\begin{equation*}
		\dot{\xx} =
		\begin{pmatrix}
			\dot{x}_1 \\
			\dot{x}_2 
		\end{pmatrix} 
		=
		\begin{pmatrix}
			x_1 \\
			x_2
		\end{pmatrix}
		~.
	\end{equation*}
	
	\begin{itemize}
		\item $\init = \{ \xx \in \mathbb{R}^2 \mid (x_1-1.125)^2+(x_2-0.625)^2-0.0125 \leq 0 \}$. 
		\item $\unsafe = \{ \xx \in \mathbb{R}^2 \mid (x_1-0.875)^2+(x_2-0.125)^2-0.0125 \leq 0 \}$. 
		\item $\domain = \{ \xx \in \mathbb{R}^2 \mid -2 \leq x_1, x_2 \leq 2 \}$. 
		\item $B(\aaa, \xx) = x_1^2+ a_1 x_2^2 + a_2 x_1+ a_3 x_2+a_4$. 
	\end{itemize}
\end{example}

\begin{example}[\expname{raychaudhuri}~\textnormal{\cite{ferragut2015seeking}}]
    The vector flow field is: 
    \begin{equation*}
	\dot{\xx} =
    \begin{pmatrix}
        \dot{x}_1 \\
        \dot{x}_2 \\
        \dot{x}_3 \\
        \dot{x}_4
    \end{pmatrix} 
    =
    \begin{pmatrix}
        -0.5 x_1^2 - 2 (x_2^2 + x_3^2 - x_4^2) \\
        -x_1 x_2 - 1 \\
        -x_1 x_3 \\
        -x_1 x_4
    \end{pmatrix}
	~.
\end{equation*}

\begin{itemize}
    \item $\init = \{ \xx \in \mathbb{R}^4 \mid x_1^2 + (x_2 + 1)^2 - 0.1 \leq 0 \}$. 
    \item $\unsafe = \{ \xx \in \mathbb{R}^4 \mid (x_1 + 1)^2 + x_2^2 - 0.1 \leq 0 \}$. 
    \item $\domain = \{ \xx \in \mathbb{R}^4 \mid -1.5 \leq x_1, \ldots, x_4 \leq 1.5 \}$. 
    \item $B(\aaa, \xx) = a_1 x_1^2 + a_2 x_1 x_2 + a_3 x_2^2 + a_4 x_1 + a_5 x_2 + a_6$. 
\end{itemize}
\end{example}

\begin{example}[\expname{focus}~\textnormal{\cite{ratschan2006constraints}}]
    The vector flow field is: 
    \begin{equation*}
	\dot{\xx} =
    \begin{pmatrix}
        \dot{x}_1 \\
        \dot{x}_2
    \end{pmatrix} 
    =
    \begin{pmatrix}
        x_1 - x_2 \\
        x_1 + x_2 
    \end{pmatrix}
	~.
\end{equation*}

\begin{itemize}
    \item $\init = \{ \xx \in \mathbb{R}^2 \mid (x_1 - 2.75)^2 + (5 x_2 - 10)^2 - 0.25^2 \leq 0 \}$. 
    \item $\unsafe = \{ \xx \in \mathbb{R}^2 \mid x_1 - 2 \leq 0 \}$. 
    \item $\domain = \{ \xx \in \mathbb{R}^2 \mid 1.5 \leq x_1, x_2 \leq 3.5 \}$. 
    \item $B(\aaa, \xx)$ includes all monomials up to degree $4$. 
\end{itemize}
\end{example}

\begin{example}[\expname{sys-bio1}~\textnormal{\cite{klipp2008systems}}]
    The vector flow field is: 
    \begin{equation*}
	\dot{\xx} =
    \begin{pmatrix}
        \dot{x}_1 \\
        \dot{x}_2 \\
        \dot{x}_3 \\
        \dot{x}_4 \\
        \dot{x}_5 \\
        \dot{x}_6 \\
        \dot{x}_7 
    \end{pmatrix} 
    =
    \begin{pmatrix}
        -0.4 x_1 + 5 x_3 x_4 \\
        0.4 x_1 - x_2 \\
        x_2 - 5 x_3 x_4 \\
        5 x_5 x_6 - 5 x_3 x_4 \\
        -5 x_5 x_6 + 5 x_3 x_4 \\
        0.5 x_7 - 5 x_5 x_6 \\
        -0.5 x_7 + 5 x_5 x_6
    \end{pmatrix}
	~.
\end{equation*}

\begin{itemize}
    \item $\init = \{ \xx \in \mathbb{R}^7 \mid \sum_{i=1}^{7}(x_i - 1)^2 - 0.01^2 \leq 0 \}$. 
    \item $\unsafe = \{ \xx \in \mathbb{R}^7 \mid \sum_{i=1}^{7}(x_i - 1.9)^2 - 0.1^2 \leq 0 \}$. 
    \item $\domain = \{ \xx \in \mathbb{R}^7 \mid -2 \leq x_1, \ldots, x_7 \leq 2 \}$. 
    \item $B(\aaa, \xx)$ includes all monomials up to degree $2$. 
\end{itemize}
\end{example}

\begin{example}[\expname{sys-bio2}~\textnormal{\cite{klipp2008systems}}]
    The vector flow field is: 
    \begin{equation*}
	\dot{\xx} =
    \begin{pmatrix}
        \dot{x}_1 \\
        \dot{x}_2 \\
        \dot{x}_3 \\
        \dot{x}_4 \\
        \dot{x}_5 \\
        \dot{x}_6 \\
        \dot{x}_7 \\
        \dot{x}_8 \\
        \dot{x}_9 
    \end{pmatrix} 
    =
    \begin{pmatrix}
        3 x_3 - x_1 x_6 \\
        x_4 - x_2 x_6 \\
        x_1 x_6 - 3 x_3 \\
        x_2 x_6 - x_4 \\
        3 x_3 + 5 x_1 - x_5 \\
        5 x_5 + 3 x_3 + x_4 - x_6 (x_1 + x_2 + 2 x_8 + 1) \\
        5 x_4 + x_2 - 0.5 x_7 \\
        5 x_7 - 2 x_6 x_8 + x_9 - 0.2 x_8 \\
        2 x_6 x_8 - x_9
    \end{pmatrix}
	~.
\end{equation*}

\begin{itemize}
    \item $\init = \{ \xx \in \mathbb{R}^9 \mid \sum_{i=1}^{9}(x_i - 1)^2 - 0.01^2 \leq 0 \}$. 
    \item $\unsafe = \{ \xx \in \mathbb{R}^9 \mid \sum_{i=1}^{9}(x_i - 1.9)^2 - 0.1^2 \leq 0 \}$. 
    \item $\domain = \{ \xx \in \mathbb{R}^9 \mid -2 \leq x_1, \ldots, x_9 \leq 2 \}$. 
    \item $B(\aaa, \xx)$ includes all monomials up to degree $1$. 
\end{itemize}
\end{example}

\begin{example}[\expname{quadcopter}~\textnormal{\cite{DBLP:conf/cav/GaoKDRSAK19}}]
    The vector flow field is: 
    \begin{equation*}
	\dot{\xx} =
    \begin{pmatrix}
        \dot{x}_1 \\
        \dot{x}_2 \\
        \dot{x}_3 \\
        \dot{x}_4 \\
        \dot{x}_5 \\
        \dot{x}_6 \\
        \dot{x}_7 \\
        \dot{x}_8 \\
        \dot{x}_9 \\
        \dot{x}_{10} \\
        \dot{x}_{11} \\
        \dot{x}_{12} 
    \end{pmatrix} 
    =
    \begin{pmatrix}
        x_4 \\
        x_5 \\
        x_6 \\
        -7253.4927 x_1 + 1936.3639 x_{11} - 1338.7624 x_4 + 1333.3333 x_8 \\
        -1936.3639 x_{10} - 7253.4927 x_2 - 1338.7624 x_5 - 1333.3333 x_7 \\
        -769.2308 x_3 - 770.2301 x_6 \\
        x_{10} \\
        x_{11} \\
        x_{12} \\
        9.81 x_2 \\
        -9.81 x_1 \\
        -16.3541 x_{12} - 15.3846 x_9
    \end{pmatrix}
	~.
\end{equation*}

\begin{itemize}
    \item $\init = \{ \xx \in \mathbb{R}^{12} \mid \sum_{i=1}^{12} x_i^2 - 0.01 \leq 0 \}$. 
    \item $
    \begin{aligned}[t]
    \unsafe = \{ \xx \in \mathbb{R}^{12} \ & \mid (2 x_1 - 0.5)^2 + (2 x_2 - 0.5)^2 + (2 x_3 - 0.5)^2 + (x_4 - 1)^2 \\
    &+ (x_5 - 1)^2 + (x_6 - 1)^2 + (x_7 - 1)^2 + (x_8 + 1)^2 + (x_9 - 1)^2 \\
    &+ (x_{10} - 1)^2 + (x_{11} + 1)^2 + (x_{12} - 1)^2 - 0.25 \leq 0 \}.
    \end{aligned} $
    \item $\domain = \{ \xx \in \mathbb{R}^{12} \mid -2 \leq x_1, \ldots, x_{12} \leq 2 \}$. 
    \item $B(\aaa, \xx)$ includes all monomials up to degree $1$. 
\end{itemize}
\end{example}


\end{appendices}

\end{document}